# Taxonomy assignment approach determines the efficiency of identification of metabarcodes in marine nematodes

**Oleksandr Holovachov[1], Quiterie Haenel[2], Sarah J. Bourlat[3] and Ulf Jondelius[1]**

[1]Department of Zoology, Swedish Museum of Natural History, Stockholm, Sweden

[2]Zoological Institute, University of Basel, Switzerland

[3]Department of Marine Sciences, University of Gothenburg, Sweden

Abstract: Precision and reliability of barcode-based biodiversity assessment can be affected at several steps during acquisition and analysis of the data. Identification of barcodes is one of the crucial steps in the process and can be accomplished using several different approaches, namely, alignment-based, probabilistic, tree-based and phylogeny-based. Number of identified sequences in the reference databases affects the precision of identification. This paper compares the identification of marine nematode barcodes using alignment-based, tree-based and phylogeny-based approaches. Because the nematode reference dataset is limited in its taxonomic scope, barcodes can only be assigned to higher taxonomic categories, families. Phylogeny-based approach using Evolutionary Placement Algorithm provided the largest number of positively assigned metabarcodes and was least affected by erroneous sequences and limitations of reference data, comparing to alignment-based and tree-based approaches.

# 1. Introduction

Metabarcoding studies based on high throughput sequencing of amplicons from marine samples have reshaped our understanding of the biodiversity of marine microscopic eukaryotes, revealing a much higher diversity than previously known [1]. Early metabarcoding of the slightly larger sediment-dwelling meiofauna have mainly focused on scoring relative diversity of taxonomic groups [1-3]. The next step in metabarcoding: identification of species, is limited by the available reference database, which is sparse for most marine taxa, and by the matching algorithms. In this paper we are evaluating to what extent OTUs of marine nematodes can be assigned to family level taxa using publicly available reference sequences, and which of three matching strategies, alignment-based, tree-based or phylogeny-based that is the most effective.

The reference datasets for marine nematodes are sparsely populated, as correctly pointed out in [4]. The most recent check of NCBI GenBank (February 2017) reveals that less than 180 genera and about 170 identified species of marine nematodes are included, comparing to over 530 described genera and almost 4750 described species (based on [5] with updates). This summarized number of records in GenBank does not take into consideration which genes are represented (mostly near complete or partial 18S and partial 28S rDNA), but gives the total number of entries. Not all of these entries include sequences suitable to be used as references for metabarcoding. Since completeness of the reference databases for marine nematodes is insufficient to assign all anonymous metabarcodes (operational taxonomic units, OTUs) to species level [6], one has to consider if they can be assigned to taxonomic categories above species level, and if this type of data can be used in research.

Assignment of OTUs to nematode genera faces the same problem as assignment of OTUs to species – limited representation of identified taxa in reference databases (see above). Identification to the family level of those OTUs that cannot be assigned to any particular species or genus is the next best option. It provides enough information to group nematode OTUs into trophic [7-8] and functional [9] groups and apply ecological metrics, such as Maturity Index [10], used to evaluate the complexity and functioning of nematode communities [11]. This approach has already been applied in metabarcoding studies of terrestrial nematode communities from the Arctic and the tropics [12-13].

Although, it would be possible to generate new barcodes for marine nematodes from our study sites to supplement existing reference datasets, the purpose of the present paper is to follow the typical scenario when metabarcoding projects rely on existing databases and do not publish new reference sequences.

Identification of OTUs can be done using a number of currently available approaches and applications, several of which will be tested and compared below. In general, all taxonomy assignment methods can be grouped into four categories: alignment-based, probabilistic, tree-based and phylogeny-based.

*Alignment-based* approaches utilize various measures of similarity between query and reference sequences based solely on their alignment. They are implemented in VAMPS [14], Taxonerator [15] and CREST [16], or can be performed directly through BLASTN [17] function of the NCBI server (https://blast.ncbi.nlm.nih.gov/Blast.cgi). The performances of CREST and BLASTN are evaluated in details in this publication. On the other hand, since VAMPS is specifically designed for procaryotic organisms, while Taxonerator uses the same routine as BLASTN, neither one is included in this comparison.

*Probabilistic* approaches rely on likelihood estimates of OTU placement and include the UTAX algorithm of the USEARCH software package [18] and Statistical Assignment Package (SAP) [19]. For technical reasons, none of these tools are included in this comparison: 1) exact details of the UTAX algorithm have not been published, and thus the results produced by this approach are difficult to evaluate; 2) standalone version of SAP could not be successfully installed, while the web server (http://services.birc.au.dk/sap/server) was not stable in use and consistently returned error messages.

The *tree-based* approach evaluates similarity between query and reference sequences by analyzing the position of each individual OTU relative to reference sequence on the cladogram and the bootstrap support that it receives. This approach includes the following bioinformatic steps: multiple sequence alignment of short query reads with reference sequences is done *de-novo* using any available multiple sequence alignment tool; the dataset is usually trimmed to the barcode size; the cladogram is built using one of the phylogeny inference algorithms, most commonly Neighbour Joining followed by bootstrapping [20-25].

*Phylogeny-based* identification of query sequences is performed in three stages. During the preparation stage, a manually curated reference alignment is created using full length sequences of the gene that includes the barcoding region. A reference phylogeny is estimated based on this alignment. Taxonomic assignment of the query barcodes is then done by using the reference tree as constraint and testing placement of query reads across all nodes in the reference topology, with placement likelihood calculated for every combination. The highest scoring placements are retained for evaluation. This approach is implemented in MLTreeMap [26], pplacer [27] and Evolutionary Placement Algorithm (EPA) [28]. Of the three, only Evolutionary Placement Algorithm is used in this paper, since "there was no clear difference in accuracy between EPA and pplacer" (cited from [27]) in comparative tests performed in [28]. MLTreeMap is designed for taxonomy assignment of barcodes into higher-level taxonomic categories (Phylum and above) and was not suitable for our purpose.

# 2. Materials and methods

## 2.1. Sampling sites, sampling, extraction and fixation

Samples used in this study were collected in two ecologically distinct locations along the west coast of Sweden. Coarse shell sand was sampled at 7-8 m depth with a bottom dredge along the north-eastern side of the Hållö island near Smögen (N 58° 20.32-20.38' E 11° 12.73-12.68'). Soft mud was collected using a Warén dredge at 53 m depth in the Gullmarn Fjord near Lysekil (N 58° 15.73' E 11° 26.10'), in the so-called "Telekabeln" site. Samples from both sites were extracted using two different techniques each. Material for metabarcoding was preserved in 96% ethanol and stored at -20°C, material for morphology-based identification was preserved in 4% formaldehyde.

Meiofauna from the coarse sand from Hållö was extracted using two variations of the flotation (decanting and sieving) technique. In the first case, fresh water was used to induce osmotic shock in meiofaunal organisms and force them to detach from the substrate. 200 ml volume of sediment was placed in a large volume of fresh water, thoroughly mixed to suspend meiofauna and sediment. The supernatant was sieved through 1000 µm sieve in order to separate and discard macrofaunal fraction. The filtered sample was then sieved through a 45 µm sieve to collect meiofauna, which was preserved either for sequencing or morphological identification. The sieving step was repeated three times. Ten replicates were preserved for molecular studies and two replicates were preserved for morphology-based observations. In the second case, a 7.2% solution of $MgCl_2$ was used to anaesthetize nematodes and other organisms to detach them from the substrate. Meiofauna was decanted through 125 µm sieve. Similarly, ten replicates were preserved for molecular studies and two replicates were preserved for morphology-based observations.

Meiofauna from the mud samples was also extracted using two different methods: flotation and siphoning. For the flotation, fresh water was used to induce osmotic shock in meiofaunal organisms. 2.4 l volume of sediment was placed in a large volume of fresh water, thoroughly mixed to suspend meiofauna and sediment. The supernatant was sieved through 1000 µm sieve in order to separate and discard macrofaunal fraction. The filtered sample was then sieved through a 70 µm sieve to collect meiofauna. The last procedure was repeated three times. Meiofauna was collected, divided into 12 subsamples and preserved: six subsamples were preserved for molecular studies and six subsamples were preserved for morphology-based observations. For siphoning, a total volume of 12 l of sediment was transferred to a plastic container, covered with 20 cm of seawater, and left to settle overnight. Meiofauna was then collected through siphoning off the top layer of sediment and passing it through a 125µm sieve from which samples were taken. One sample was fixed in 96% ethanol, and split into six equal subsamples for molecular studies. The second sample was also split into six subsamples and preserved for morphology-based observations.

## 2.2. Morphology-based analysis of samples

In order to estimate nematode diversity it is usually recommended to either count and identify all nematode individuals in the entire sample, or in a subsample of predetermined volume. The alternative, least time consuming and most commonly used option is to count a predetermined number (usually 100 or 200) of randomly picked nematodes from the sample. Unfortunately, this latter approach can be imprecise for samples with high species diversity. Moreover, since nematodes are affected by Stokes law, that causes uneven distribution of specimens of different size along the bottom of the counting dish, it is difficult to obtain randomized data with this approach. Therefore, we opted to count and identify all nematodes for all samples (or subsamples). The amount of time required for this task limited the effort to two replicates for each site and extraction method, eight in total. We appreciate that counting nematodes in only two replicates per sample is not enough to quantitatively evaluate the composition of nematode communities, it is nevertheless satisfactory to provide the list of species and genera for each sampling site and extraction method for the purpose of this publication.

All nematode specimens were identified and counted for two replicates each from Hållö flotation with $MgCl_2$, Hållö flotation with freshwater and Telekabeln siphoning. Telekabeln flotation with freshwater were subsampled by taking 1/10 of the entire sample. Specimens from formaldehyde-preserved samples were transferred to pure glycerine using modified Seinhorst's rapid method [29] and mounted on glass slides using the paraffin wax ring method. All nematode specimens were identified to genus, and, when possible, to species level and placed in the classification system published in [5] and accepted in WoRMS [30] and NeMys [31] reference databases. Please note that this classification is in many cases different from the nematode classification used in GenBank [32], SILVA [33] and GBIF (www.gbif.org).

## 2.3. Sequencing procedures

DNA extractions from the samples preserved in 96% ethanol were performed on about 10g of sediment using the PowerMax® Soil DNA Isolation Kit, (MO BIO Laboratories), according to manufacturer's instructions. Primers were designed for the 18S rRNA gene including Illumina

MiSeq overhang adapter sequences for compatibility with Illumina index and sequencing adapters. The 18S rRNA marker was amplified using PCR primers modified from [2] yielding a ≈370 bp fragment that includes the V1-V2 hypervariable domains of 18S rRNA (Supplementary Figure 1). Illumina MiSeq library preparation was done using the dual PCR amplification method [33]. All subsequent sequencing and bioinformatic analysis steps are fully described in [6].

## 2.4. Preliminary taxonomic assignment using QIIME

Preliminary taxonomic assignment was done using the QIIME [35] script *assign_taxonomy.py* against the SILVA database [33] release 111. Default settings in QIIME used for preliminary sorting of OTUs grouped query sequences into two groups based on similarity level: to phyla at 80% similarity and to species at 97% similarity. The output for each query sequence included closest match but did not give similarity level, making it impossible to evaluate these assignments. Only two OTUs were positively identified using QIIME to species level: *Viscosia viscosa* (TS6.SSU58722) and *Chromadora nudicapitata* (HF2.SSU192072). Six more OTUs were identified to the genus level: *Enoplus* sp. (HE3.SSU110275), *Enoploides* (HE3.SSU124287), *Symplocostoma* sp. (HE5.SSU188855), *Calomicrolaimus* (HF9.SSU20251), *Odontophora* sp. (HF1.SSU779114) and *Sabatieria* sp. (TF6.SSU48167).

The original output from the QIIME analysis included 145 OTUs assigned to the phylum Nematoda. Four of them were incorrectly placed among nematodes due to errors in the reference database derived from SILVA – they group with Arthropoda (HE1.SSU866120, HE6.SSU382930, HF6.SSU331569) and Phoronida (TS6.SSU559982) in all other analyses and were excluded. Two more sequences cluster with nematodes but appear to have long insertions within conserved regions (HE6.SSU358113 and TF5.SSU411806). Both of them were found only in one sample each, further supporting the idea that they are derived from an erroneous amplification product, and were removed from any further analysis. The final list of nematode OTUs includes 139 query sequences.

## 2.5. Taxonomy assignment of nematode OTUs using alignment-based methods

All 139 nematode OTUs were manually analysed using BLASTN 2.5.0+ [17] against the nucleotide collection of the NCBI database (http://blast.ncbi.nlm.nih.gov/Blast.cgi) on August 22, 2016 with the following settings: *optimize for highly similar sequences* (megablast), *exclude uncultured/environmental sample sequences*, *max target sequences – 100*, sorted by max score. Closest matches were evaluated. If the top match sequence was still labelled as "uncultured", "unidentified" or "environmental", the next best match was evaluated. Assignment to the family level was based on the top hit with at least 90% identity score, with 100% sequence cover, as well as assignment consistency (e.g. top hits assigned to the same family) following the observations described in [36], and contrary to the approach used in [37].

LCAClassifier function of the CREST web server (http://apps.cbu.uib.no/crest) was used to assign taxonomy to 139 OTUs using built-in silvamod database [16] on August 25, 2016. Three different scores of LCA relative range were tested separately: 2%, 5% and 10%, but only the results based on LCA range of 2% were retained for further analysis and comparison.

## 2.6. Taxonomy assignment of nematode OTUs using tree-based approach

According to published tests [38], the tree-based approach does not allow grouping of sequences into well supported monophyletic clades equivalent in their taxonomic composition to nematode orders, but most of the marine nematode families are well resolved and supported. The reference sequence dataset was based on the "filtered" alignment from [38] that was updated with newly published sequences of marine nematodes. The final reference dataset is composed of 305 sequences representing the majority of marine nematode families as well as selected freshwater and terrestrial families, some species of which are known to inhabit the marine environment, plus three outgroup taxa (Supplementary Table 1). The same set of sequences was used for the taxonomy placement using a phylogeny-based approach (Section 2.7).

The reference dataset was trimmed to the barcoding region and aligned with query sequences using ClustalW [39] algorithm at default settings implemented in MEGA ver. 7 [40]. Phylogenetic tree was built using Maximum Likelihood phylogeny inference with RAxML ver. HPC2 [41] at default settings with 1000 bootstrap replicates via CIPRES portal [42]. Two independent analyses were performed: in the first case, all 139 query sequences (cumulative reference dataset) were aligned with the reference dataset and analysed at once; in the second case, 139 query sequences were split in 14 groups of 10 or nine (partitioned query dataset), each group was separately aligned with the complete reference dataset and analysed. It was done to verify if the number and composition of query sequences have any impact on the effectiveness of the tree-based taxonomy assignment approach.

## 2.7. Taxonomy assignment of nematode OTUs using phylogeny-based approach

Alignments from [43-44] were combined together and supplemented with other sequences of marine nematodes available in GenBank. In order to minimize any potential errors and inconsistencies, both at the tree-building stage, alignment stage and placement stage, all sequences used for generating reference alignment and reference tree were selected to be as complete as possible, with the exception of taxa for which no alternative option was available. Secondary structure annotation was manually added to all non-annotated sequences using the JAVA-based editor 4SALE [45], and all sequences were manually aligned to maximize apparent positional homology of nucleotides. The resulting alignment includes representatives of all families of marine nematodes for which sequence data is available, as well as selected freshwater, terrestrial and animal parasitic taxa (Supplementary Table 1). The reference tree was built using RAxML ver. HPC2 [41] via the CIPRES portal [42] with Maximum likelihood inference of the partitioned dataset. The GTR+Γ nucleotide substitution model was used for non-paired sites, whereas the RNA7A [46] substitution model was used for paired sites. Bootstrap ML analysis was performed using the rapid bootstrapping option with 1000 iterations.

Query sequences were aligned to a fixed reference alignment (created in the previous step) using either mothur ver. 1.36.1 [47] or PaPaRa [48] under default settings. Taxonomy predictions for query sequences were than generated with Evolutionary Placement Algorithm [28] implemented in RAxML [41] using the following command: *raxmlHPC-PTHREADS -T 2 -f v -s alignment_file -t reference_tree -m GTRCAT -n output.* Taxonomic assignments to family level taxonomic categories were based either on high likelihood of single placement (above the 95% threshold), or on high cumulative likelihood of multiple placements all of which are within a single monophyletic clade equal to a family (see Section 4.4 for explanation).

### 2.8. Image processing

Trees were visualized using FigTree [49] and iTOL [50]. All clades with bootstrap support lower than 70% were collapsed in the final illustrations. Secondary structure of the barcoding region of 18S rRNA (Supplementary Figure 1) was visualized using VARNA [51].

## 3. Results

### 3.1. Morphology-based analysis of samples

The nematode fauna in the coarse sand from the Hållö site included 107 different nematode species belonging to 86 genera and 33 families (Supplementary Table 2). Of these, flotation using $MgCl_2$ recovered 88 species from 73 genera and 26 families, while flotation using $H_2O$ recovered 101 species from 83 genera and 33 families. The differences in nematode fauna extracted using two variations of the same method are limited to rare species of different size classes (from less than 0.5 mm to over 2 mm). Relative abundance of these rare species does not exceed 0.14% (0.01– 0.14%, with the average of 0.03%). The list of nematodes from Hållö site includes four species new to the fauna of Sweden *(Bolbonema brevicolle, Bradylaimus pellita, Desmodora granulata, Odontophora villoti*) and five species new to science (from the genera *Adelphos, Paramesonchium, Leptolaimus* and *Diplopeltoides*).

Mud sediments from the Telekabeln site were inhabited by 113 different nematode species, belonging to 77 genera and 33 families (Supplementary Table 3). Of these, siphoning recovered 81 species from 62 genera and 29 families, while flotation using $H_2O$ recovered 102 species from 70 genera and 32 families. The differences in nematode fauna extracted using two different methods include both rare and uncommon species of various size classes (from less than 0.5 mm to over 2 mm). Relative abundance of these rare species does not exceed 2.02% (0.01–2.02%, with the average of 0.29%). The list of nematodes from the Telekabeln samples includes seven species new to the fauna of Sweden *(Campylaimus rimatus, C. amphidialis, C. tkatchevi, C. orientalis, Diplopeltoides asetosus, D. linkei, D. nudus*) and one species new to science (from the genus *Diplopeltoides*).

## 3.2. Taxonomy placement of OTUs using alignment-based approaches

**3.2.1. BLASTN.** Out of 139 queried anonymous OTUs, 52 could be assigned to family-level categories based on the following criteria: 90% or more identity score and 100% sequence cover, as well as assignment consistency (Supplementary Table 4). In one case BLASTN search produced conflicting results – two top hits with the same identity score and sequence cover that belonged to different families, and still fall within the threshold limit. This is the barcode TF1.SSU676746 that showed 90% identity and 100% sequence cover to *Haliplectus* sp. (family Haliplectidae) and *Prodesmodora* sp. (family Desmodoridae). It was considered unassigned. Similar examples were seen in BLASTN results of other OTUs that did not reach the threshold. These examples shows that considering only one top hit when assigning taxonomy to anonymous OTUs using alignment-based approaches may sometimes lead to questionable or dubious identification.

**3.2.2**. **CREST.** Only 26 out of 139 queried anonymous OTUs were assigned to families using LCAClassifier of CREST under default parameters (Supplementary Table 5), and following built-in classification. In two cases, OTUs were placed outside Nematoda: HE3.SSU118424 was placed within Copepoda (Phylum Arthropoda) and TS1.SSU284163 was placed in Scolecida (Phylum Annelida). The first OTU was positively assigned to the family Oxystominidae (Phylum Nematoda) using tree-based and phylogeny-based approaches (see Sections 3.3 and 3.4); the second OTU was unassigned in all other analyses.

## 3.3. Taxonomy placement of OTUs using tree-based approaches

**3.3.1. Cumulative query dataset.** Tree-based taxonomy assignment of the cumulative query dataset produced 54 well supported placements (Figure 1; Supplementary Table 6) that fulfilled the following criteria: OTU must cluster within the monophyletic clade that has high bootstrap support (≥70%) and is at or below family-level. The remaining 85 OTUs could not be placed in clades satisfying these criteria, and are thus treated as unidentified.

**3.3.2. Partitioned query dataset.** The results of taxonomic assignment using a tree-based approach of the partitioned query dataset produced somewhat different results comparing to the cumulative query dataset – 67 OTUs were placed in monophyletic clades equivalent to family-level categories with sufficient support (Supplementary Table 6). Of these, taxonomic placement of only 47 OTUs matched the identification produced using the cumulative query dataset, and identifications of 20 OTUs were new. Seven OTUs were not assigned using a partitioned query dataset but positively identified using a cumulative query dataset.

## 3.4. Taxonomy placement of OTUs using phylogeny-based approaches

**3.4.1. EPA/mothur.** Phylogeny-based taxonomy assignment using mothur-based alignment and Evolutionary Placement Algorithm produced 105 well supported placements with single or accumulated likelihood of 0.95 or more (Figure 2; Supplementary Table 7). There are ten additional cases when the positive identity can not be attained because OTUs are placed either within a paraphyletic assemblage (family Desmodoridae or Linhomoeidae) or closely related monophyletic clade (Draconematidae or Siphonolaimidae respectively).

**3.4.2. EPA/PaPaRa.** The results produced using PaPaRa-based alignment and Evolutionary Placement Algorithm are exactly the same as obtained using mothur-based alignment and described in the section 3.4.1 (Supplementary Table 7), even though visual comparison of alignments produced by mothur and by PaPaRa revealed some differences.

## 3.5. Comparison of different taxonomy assignment approaches

Among the three different taxonomy assignment approaches tested (each with two variations), the Evolutionary Placement Algorithm (both variations) placed the largest number of query OTUs into family level taxonomic categories (105 out of 139), while CREST implementation of alignment-based assignment was the least efficient (26 out of 139). Despite such a broad success rate, most of the identified OTUs were assigned to the same families (Supplementary Table 8), with the following exceptions:

1) HF1.SSU759758 was placed in the family Camacolaimidae using Tree-based and Phylogeny-based approaches, in the family Leptolaimidae using CREST, and unassigned using BLASTN;

2) HF5.SSU995414 was placed in the family Rhabdolaimidae using BLASTN, in the family Ironidae using CREST and both variations of EPA, and unassigned using Tree-based approach;

3) TF1.SSU698227 was placed in the family Teratocephalidae using BLASTN and in the family Benthimermithidae using both variations of EPA, unassigned in other cases;

4) TF1.SSU700188 was placed in the family Linhomoeidae using BLASTN and in the family Cyartonematidae using using Tree-based and Phylogeny-based approaches, unassigned using CREST;

5) TF6.SSU47996 was placed in the family Oncholaimidae using BLASTN and in the family Enchelidiidae in all other cases.

## 3.6. Comparison between barcode-based and morphology-based identification

The Evolutionary Placement Algorithm (phylogeny based approach) provided the largest number of positively identified OTUs and will be compared with the faunistic lists created by identifying nematode specimens using morphological characters. Since species-level identification can not be achieved for most of the OTUs, the results of barcode-based and morphology-based identifications can only be compared as the number of identified OTUs/morphospecies per family (Figure 3; Supplementary Table 9). Among families with available reference sequences barcode-based identification failed to identify the families Phanodermatidae, Leptosomatidae, Trefusiidae, Epsilonematidae, Draconematidae, Monoposthiidae and Sphaerolaimidae. On the other hand, barcode-based identification also uncovered several taxa that were overlooked during morphology-based identification, such as the families Achromadoridae, Mermithidae and Benthimermithidae – the last two are internal parasites of invertebrates during part of their life cycle and were most likely overlooked, because examination of meiofauna for internal parasites was not attempted. In all other cases, the efficiency of either barcode-based or morphology-based identification varied considerably, even within the same taxon across different samples (Figure 3). Nevertheless, *Pearson correlation coefficient* revealed moderate positive correlation ($\rho$ = 0.7296967138) between the number of assigned OTUs and identified morphospecies in each family/extraction/sample (Supplementary Figure 2).

# 4. Discussion

## 4.1 General notes

Three different taxonomy assignment approaches (with two modifications each) tested in this project provide some variation in the number of positively identified OTUs, however, the assigned identities of those OTUs that were identified was consistent with very few exceptions (Section 3.5). These discrepancies can possibly be caused by several different factors. Placement of one of the OTUs (HF1.SSU759758) either in the family Camacolaimidae (Tree-based and Phylogeny-based approaches) or in the family Leptolaimidae (CREST) is likely a result of outdated classification of the phylum Nematoda used in the SILVA-derived reference database implemented in CREST, comparing to the nematode classification used in WoRMS and in this publication (Section 4.6). Conflicting results of the assignment of TF1.SSU698227 either in the family Teratocephalidae (BLASTN) or in the family Benthimermithidae (EPA) can be due to poor representation of reference dataset in this part of the nematode tree. The remaining conflicting placements of HF5.SSU995414 (Rhabdolaimidae *versus* Ironidae), TF1.SSU700188 (Linhomoeidae *versus* Cyartonematidae) and TF6.SSU47996 (Oncholaimidae *versus* Enchelidiidae) are possibly caused by the fact that overall sequence similarity used by BLASTN does not necessarily reflect common phylogenetic history, which is the basis of the tree-based and phylogeny-based assignment approaches. Differences in the individual success rates of each taxonomy assignment approach will be discussed below (Sections 4.2-4.4).

## 4.2 Alignment-based approach

Alignment-based approaches tested in this publication include manual analysis using BLASTN 2.5.0+ [17] against the nucleotide collection of the NCBI database and LCAClassifier function of the CREST against built-in silvamod database [16]. Both tested approaches have their own advantages and disadvantages. NCBI implementation of BLASTN allows visual examination of multiple top hits in the output and individual evaluation of these top hits, manual application of variable similarity threshold if it has been predetermined empirically, and, if necessary, correction of classification. Taxonomy assignment using CREST is less flexible and has the following limitations: (1) similarity thresholds used in CREST are based on the procaryotic 16S rRNA analysis and do not account for the differences in the variability of rRNA within and between different taxa [36]; (2) classification of the phylum Nematoda that is used in the CREST database is different from the most recent and widely accepted classification scheme published in WoRMS; (3) results of the taxonomy assignment in the output files can not be verified and, if necessary, updated.

Strictly speaking, alignment-based assignment approaches should not be used to place OTUs to supraspecific taxa without critical evaluation of the results. First of all, similarity scores used in BLASTN search results do not reflect phylogenetic affinities of analysed taxa, and do not account for the fact that the level of variability of the 5′ barcoding region of 18S rRNA (Supplementary Figure 1) is different in various nematode taxa [36]. Too narrow similarity thresholds can exclude potentially identifiable sequences, while too broad thresholds can lead to misidentifications. Dell'Anno et al. [4] is an example where broad similarity threshold resulted in incorrect assignment of several nematode OTUs from deep-sea samples to nematode species known to inhabit fresh-water and soil and never found in the marine environment (e. g. *Anaplectus porosus, Anaplectus* sp., *Pakira orae* amd *Tylolaimophorus* sp.).

## 4.3 Tree-based approach

Phylogenetic hypotheses used to infer relationships of taxa are usually thoroughly described and rigorously evaluated, undergo comparison and testing using different alignment and tree-building algorithms. Phylogenetic trees used to identify unknown barcodes are less so [20-21]. Barcodes are by definition relatively short in length, hypervariable sites flanked by conserved regions. Hypervariable domains V1 and V2, which are part of the barcoding region of the 18S rRNA used in this publication, are the culprit that causes poor alignment and hence has negative effect on the quality of the resulting phylogeny. Different alignment and phylogeny-inference algorithms may provide competing phylogenetic hypotheses [38], and, as a result, different placements of OTUs in the cladogram. Taxon composition and sequence quality (exclusion of incorrectly identified species, low quality and short sequences) of the reference dataset is also crucial [38], as it determines which taxa can be identified and which taxa can not. Even the number and composition of OTUs, have strong effect on the final phylogenetic tree, and, as a result, on the outcome of the taxonomy assignment, as shown in Section 3.3. The latter is caused by the need to align *de-novo* the combined datasets that includes reference and query sequences – presence of unidentified sequencing errors among query OTUs can have negative effect on the alignment and phylogeny inference, even if all reference sequences are of high quality. This effect is global, i.e. by affecting the entire alignment and tree topology and bootstrap, erroneous sequences can potentially cause other OTUs to be misidentified or unidentified. In conclusion, successful use of tree-based approaches to assign taxonomy to anonymous OTUs is highly dependent not only on the quality and completeness of the reference dataset and alignment and phylogeny inference algorithms, but also on the quality and diversity of query sequences.

## 4.4 Phylogeny-based approach

Phylogeny-based approaches allow the estimation of the most likely position of each OTUs within the constrained phylogenetic tree, estimation of the rank of its taxonomic placement in supraspecific categories if these are well resolved and supported in the reference phylogeny, and can even work with paraphyletic taxa. Moreover, since the reference alignment and reference phylogeny are constrained during phylogeny-based taxonomy assignment procedures, the quality of query sequences has no impact on the result, i.e. the presence of erroneous sequences among query OTUs (chimaeras) has no effect on the identification of other query OTUs. The outcome of the analysis solely depends on the quality of the reference alignment and reference phylogeny. Even minor differences in the alignment of OTUs against the reference alignment noted above (Section 3.4.2) had no effect on the results. An additional advantage of the phylogeny-based taxonomy assignment approach implemented in Evolutionary Placement Algorithm is the possibility to use cumulative likelihood scores when assigning taxa to clades equivalent to supraspecific taxonomic categories (Supplementary Figure 3).

## 4.5 Metabarcoding *versus* morphology-based identification

Morphology-based identification procedures are strongly biased by the expertise and experience of the researcher performing the identification, as well as the state of the knowledge on the diversity of particular groups of nematodes. Metabarcoding, on the other hand, should be able to better estimate the diversity of poorly known groups of nematodes, or groups for which taxonomic expertise is not available at the moment, as well as unidentifiable specimens (eggs, juveniles, damaged specimens, etc). Moreover, metabarcoding can reveal taxa that are physically hidden and can not be observed by the researcher during sorting and identification, such as internal parasites – similarly to the results obtained by [52], barcode-based identification revealed the presence of endoparasitic nematodes from the families Mermithidae and Benthimermithidae in our samples. They had been overlooked during morphology-based identification, likely being juveniles within bodies of other invertebrates.

The number of OTUs identified by metabarcoding is strongly influenced by the clustering procedures of the raw sequence data, and, depending on the threshold used, will give different results. Assuming that the OTUs produced through metabarcoding are equivalent to currently recognized morphospecies, the only reason it would not be able to correctly estimate the number of species in the sample is if there are issues with amplification of the barcoding gene. The genus *Halalaimus* is a good example of a problematic taxon in this case – only one *Halalaimus* OTU (TS5.SSU874117) was recovered with metabarcoding, and only from the Telekabeln site. Morphology-based identification recovered at least two different *Halalaimus* species in Hållö site and more than eight species in the Telekabeln site, some of which were relatively common. GenBank hosts a number of *Halalaimus* sequences, confirming that the genus is sufficiently diverse genetically, and that our single *Halalaimus* OTU is unlikely to encompass multiple morphospecies, but is rather a result of amplification problems.

## 4.6 Reference databases

Taxonomy-assignment procedures described in the literature [16, 35] often rely on various releases of the SILVA database [33], which in turn is based on the sequence data published in GenBank or EMBL. These databases can be "built-in" (CREST), and completely inaccessible for the user, or "pre-made" and hard to modify (QIIME). Presence of erroneously identified sequences of nematodes and other organisms in GenBank and SILVA databases has been mentioned multiple times [36, 38, 53-54]. If the reference database is not checked for errors prior to the analysis, the results produced by any taxonomy-assignment algorithm should be evaluated using available data on geographical or ecological distribution of species, in order to avoid mistakes.

As mentioned above, the SILVA database in itself does not always follow the most recent accepted classification for certain groups of organisms. As a result, placing some of the OTUs into nematode families based on SILVA classification turned out to be incorrect. For example, genera *Paracyatholaimus* and *Preacanthonchus* were placed in the family Chromadoridae using QIIME, while they do belong to the family Cyatholaimidae. Same examples are *Enoploides* placed in Enoplidae instead of Thoracostomopsidae, *Calyptronema* in Oncholaimidae instead of Enchelidiidae, *Achromadora* in Chromadoridae instead of Achromadoridae, *Camacolaimus* in Leptolaimidae instead of Camacolaimidae, and some others. Output from CREST [16] only gives the name of the supraspecific taxon for those cases where an anonymous OTU can not be identified to species level. This prevents proper evaluation of the assignment results and correction of assignments based on erroneous reference sequence or incorrect classification. We do not expect any database to be able to quickly reflect changes in nematode classification, but we expect end users of these databases to be aware of the need to verify and, if necessary, to update the output of any taxonomy-assignment procedure that they may use.

Another disadvantage of taxonomy-assignment software that uses built-in databases and offers only top-pick assignments in the output files (QIIME, CREST), is that a substantial number of OTUs are matched with environmental samples, labelled in such databases with the words "environmental" (e.g. "environmental sample"), "uncultured" (e.g. "uncultured eucaryote") and "unidentified" ("unidentified nematode"). They themselves are OTUs generated during previous metabarcoding projects and identified not by looking at actual morphological vouchers but by using one of the multiple taxonomy-assignment methods. Moreover, by giving only one top "hit" assignment, such software eliminates the possibility to see if the "second best" hit is based on sequence data from the physically observed and identified morphological voucher, and its similarity score, preventing the researcher from making educated decisions on the taxonomic identity of an OTU.

# 5. Conclusions and future prospects

The identification of OTUs is obviously a key step in metabarcoding and it is essential that the most effective method is used (as opposed to the fastest or simplest). Ideally the barcoding sequences should be assigned taxonomic names that provide a link to all biological knowledge that may exist in relation to the organism. Misidentification will compromise the results in studies of e.g. biogeography, community structure, habitat state or presence of certain important species (invasive, rare, indicators, etc).

Identification of OTUs should be at the appropriate taxonomic level, which is determined by the available reference sequences and the purpose of the study. In the case of marine nematodes we were able to assign our barcode sequences to family-level taxa to a high degree despite the very incomplete reference database. The relevance of family-level metabarcoding data in ecological studies remains poorly tested and requires extensive comparison with data obtained using classical approaches.

The full potential of metabarcoding is realised when sequences are identified to species level. This conveys the most information and permits more robust inferences. A prerequisite for this is taxonomic groundwork in the form of complete curated reference databases with sequences of reliably identified specimens.

We found the phylogeny-based taxonomy assignment approach to be the most efficient and the least error-prone. The alignment-based approach is less reliable because the similarity thresholds it depends on do not account for inter-and intra-taxon variations in barcode sequence, while tree-based approaches can be affected by the quality of the input OTU data. If phylogeny-based taxonomy assignment methods become widely used in nematode metabarcoding, it is imperative to create and maintain high quality reference alignments and reference phylogenetic trees to be used by researchers worldwide.

## Ethics statement

There are no particular ethical aspects specific to this publication. It did not involve: (1) experiments on animals; (2) collection of protected species; (3) research on human subjects; and (4) collection of personal data.

## Data accessibility

The data supporting this article are available in the electronic supplementary material.

## Compelling interests

The authors have no competing interests.

## Authors contributions

O.H. conceived and designed the study. U.J. and O.H. performed fieldwork. Q.H. and S.J.B. performed molecular analyses. O.H. performed morphology-based identification and taxonomy-assignment analyses. Q.H., O.H., U.J. and S.J.B. contributed reagents, materials and analysis tools. O.H. wrote the manuscript with input from Q.H., S.J.B. and U.J.

All authors gave final approval for publication.

## Acknowledgements

We would like to thank the Genomics Core facility platform at the Sahlgrenska Academy, University of Gothenburg. Sampling was conducted using vessel ("Oscar von Sydow") and facilities of the Sven Lovén Centre for Marine Infrastructure in Kristineberg.

## Funding

This work was in part supported by the project "Systematics of Swedish free-living nematodes of the orders Desmodorida and Araeolaimida" (Swedish Taxonomy Initiative,

ArtDatabanken, Sweden) awarded to OH and by the Swedish Research Council project (2012-3446) "Biodiversity genomics: Species identification pipelines for analyzing marine invertebrate larval stages, community structure, and trophic interactions" awarded to SJB.

## References

1. Leray, M., & Knowlton, N. (2016). Censusing marine eukaryotic diversity in the twenty-first century. Philosophical Transactions of the Royal Society B: Biological Sciences 371: 20150331 (doi: 10.1098/rstb.2015.0331).
2. Fonseca, V.G., Carvalho, G.R., Sung, W., Johnson, H.F., Power, D.M., Neill, S.P., Packer, M., Blaxter, M.L., Lambshead, P.J.D., Thomas, W.K. & Creer, S. (2010). Second-generation environmental sequencing unmasks marine metazoan biodiversity. Nature Communications 1: 98 (doi: 10.1038/ncomms1095).
3. Fonseca, V.G., Carvalho, G.R., Nichols, B., Quince, C., Johnson, H.F., Neill, S.P., Lambshead, P.J.D., Thomas, W.K., Power, D.M. & Creer, S. (2014). Metagenetic analysis of patterns of distribution and diversity of marine meiobenthic eukaryotes. Global Ecology and Biogeography 23: 1293-1302 (doi: 10.1111/geb.12223
4. Dell'Anno, A., Carugati, L., Corinaldesi, C., Riccioni, G., Danovaro, R. (2015). Unveiling the biodiversity of deep-sea nematodes through metabarcoding: are we ready to bypass the classical taxonomy? PLoS ONE 10: e0144928 (doi: 0.1371/journal.pone.0144928)
5. Schmidt-Rhaesa, A. (2014). Handbook of Zoology. Gastrotricha, Cycloneurali and Gnathifera. Volume 2. nematoda. De Gruyter, 759 pp.
6. Haenel, Q., Holovachov, O., Jondelius, U., Sundberg, P. & Bourlat, S.J. (2017). NGS-based biodiversity and community structure analysis of meiofaunal eukaryotes in shell sand from Hållö island, Smögen, and soft mud from Gullmarn Fjord, Sweden. Biodiversity Data Journal. Submitted
7. Jensen, P. (1987). Feeding ecology of free-living aquatic nematodes. Marine Ecology – Progress Series 35: 187-196.
8. Yeates, G.W., Bongers, T., de Goede, R.G.M., Freckman, D.W. & Georgieva, S.S. (1993). Feeding habits in soil nematode families and genera – an outline for soil ecologists. Journal of Nematology 25: 315-331.
9. Bongers, T. & Bongers, M. (1998). Functional diversity of nematodes. Applied Soil Ecology 10: 239-251.
10. Bongers, T. (1999). The Maturity Index, the evolution of nematode life history traits, adaptive radiation and cp-scaling. Plant and Soil 212: 13-22.

11. Ahmed, M., Sapp, M., Prior ,T., Karssen ,G. & Back, M.A. (2016). Technological advancements and their importance for nematode identification. SOIL 2: 257-270.
12. Kerfahi, D., Tripathi, B.M., Porazinska, D.L., Park, J., Go, R. & Adams, J.M. (2016). Do tropical rain forest soils have greater nematode diversity than High Arctic tundra? A metagenetic comparison of Malaysia and Svalbard. Global Ecology and Biogeography 25: 716-728 (doi: 10.111/geb.12448).
13. Kerfahi, D., Park, J., Tripathi, B.M., Singh, D., Porazinska, D.L., Moroenyane, I. & Adams J.M. (2016). Molecular methods reveal controls on nematode community and unexpectedly high nematode diversity, in Svalbard high Arctic tundra. Polar Biology. (doi: 10.1007/s00300-016-1999-6).
14. Sogin, M.L., Morrison, H.G., Huber, J.A., Welch, D.M., Huse, S.M., Neal, P.R., Arrieta, J.M. & Herndl, G.J. (2006). Microbial diversity in the deep sea and the underexplored "rare biosphere". Proceedings of the National Academy of Sciences 103: 12115-12120.
15. Jones, M., Ghoorah, A. & Blaxter, M. (2011). jMOTU and Taxonerator: turning DNA barcode sequences into annotated operational taxonomic units. PLoS One 6: e19259 (doi:10.1371/journal.pone.0019259).
16. Lanzén, A., Jørgensen, S., Huson, D., Gorfer, M., Grindhaug, S.H., Jonassen, I., Øvreås, L. & Urich, T. (2012). CREST – classification resources for environmental sequence tags. PLoS ONE 7: e49334 (doi: 10.1371/journal.pone.0049334).
17. Madden, T. (2002). Chapter 16. The BLAST sequence analysis tool. The NCBI Handbook.
18. Edgar, R.C. (2010). Search and clustering orders of magnitude faster than BLAST, Bioinformatics 26, 2460-2461 (doi: 10.1093/bioinformatics/btq461)
19. Munch, K., Boomsma, W., Huelsenbeck, J., Willerslev, E. & Nielsen, R. (2008) .Statistical assignment of DNA sequences using Bayesian phylogenetics. Systematic Biology 57: 750--757 (doi: 10.1080/10635150802422316).
20. Morise, H., Miyazaki, E., Yoshimitsu, S. & Eki, T. (2012). Profiling nematode communities in unmanaged flowerbed and agricultural field soils in Japan by DNA barcode sequencing. PLoS ONE 7: e51785 (doi: 10.1371/journal.pone.0051785).
21. Sapkota, R. & Nicolaisen, M. (2015). High-throughput sequencing of nematode communities from total soil DNA extractions. BMC Ecology 15: 3 (doi: 10.1186/s12898-014-0034-4).
22. Bhadury, P., Austen, M., Bilton, D., Lambshead, P., Rogers, A. & Smerdon, G. (2006). Development and evaluation of a DNA-barcoding approach for the rapid identication of nematodes. Marine Ecology Progress Series 320: 1 9 (doi: 10.3354/meps320001).
23. Bhadury, P. & Austen, M. (2010). Barcoding marine nematodes: an improved set of

nematode 18S rRNA primers to overcome eukaryotic co-interference. Hydrobiologia 641: 245-251 (doi: 10.1007/s10750-009-0088-z)

24. De Ley, P., Tandingan De Ley, I., Morris, K., Abebe, E., Mundo-Ocampo, M., Yoder, M., Heras, J., Waumann, D., Rocha-Olivares, A., Burr, A.H.J., Baldwin, J.G. & Thomas, W.K. (2005). An integrated approach to fast and informative morphological vouchering of nematodes for applications in molecular barcoding. Philosophical Transactions of the Royal Society B: Biological Sciences 360: 1945-1958 (doi:10.1098/rstb.2005.1726).
25. Derycke, S., Vanaverbeke, J., Rigaux, A., Backeljau, T. & Moens, T. (2010). Exploring the use of cytochrome oxidase c subunit 1 (COI) for DNA barcoding of free-living marine nematodes. PLoS ONE 5: e13716 (doi: 10.1371/journal.pone.0013716).
26. Stark, M., Berger, S.A., Stamatakis, A. & Mering C. (2010). MLTreeMap – accurate Maximum Likelihood placement of environmental DNA sequences into taxonomic and functional reference phylogenies. BMC Genomics 11: 461 (doi: 10.1186/1471-2164-11-461)
27. Matsen, F.A., Kodner, R.B. & Armbrust, E.V. (2010). pplacer: linear time maximum-likelihood and Bayesian phylogenetic placement of sequences onto a xed reference tree. BMC Bioinformatics 11: 538 (doi: 10.1186/1471-2105-11-538).
28. Berger, S.A., Krompass, D. & Stamatakis A. (2011). Performance, accuracy, and web server for evolutionary placement of short sequence reads under maximum likelihood. Systematic Biology 60: 291-302 (doi: 10.1093/sysbio/syr010)
29. De Grisse, A.T. (1969). Redescription ou modifications de quelques techniques utilisées dans l'etude des nematodes phytoparasitaires. Mededelingen Rijksfakulteit Landbouwwetenschappen Gent 34: 351-369.
30. WoRMS Editorial Board (2016). World Register of Marine Species. Available from http://www.marinespecies.org at VLIZ. Accessed 2016-09-29 (doi:10.14284/170).
31. Guilini, K., Bezerra, T.N., Eisendle-Flöckner, U., Deprez, T., Fonseca, G., Holovachov, O., Leduc, D., Miljutin, D., Moens, T., Sharma, J., Smol, N., Tchesunov, A., Mokievsky, V., Vanaverbeke, J., Vanreusel, A., Venekey, V. & Vincx, M. (2016). NeMys: World Database of Free-Living Marine Nematodes. Accessed at http://nemys.ugent.be on 2016-09-29
32. Benson, D.A., Cavanaugh, M., Clark, K., Karsch-Mizrachi, I., Lipman, D.J., Ostell, J. & Sayers, E.W. (2013). GenBank. Nucleic Acids Res. 41: D36-42 (doi: 10.1093/nar/gks1195).
33. Quast, C., Pruesse, E., Yilmaz, P., Gerken, J., Schweer, T., Yarza, P., Peplies, J. & Glockner, F.O. (2012). The SILVA ribosomal RNA gene database project: improved data processing and web-based tools. Nucleic Acids Research 41: D590-D596 (doi: 10.1093/nar/gks1219).
34. Bourlat, S.J., Haenel, Q., Finnman, J. & Leray, M. (2016). Preparation of amplicon libraries

for metabarcoding of marine eukaryotes using Illumina MiSeq: the Dual-PCR method. In: Bourlat, S.J. editor. Marine Genomics - Methods and protocols: Springer, 197-208.

35. Caporaso, J.G., Kuczynski, J., Stombaugh, J., Bittinger, K., Bushman, F.D., Costello, E.K., Fierer, N., Peña, A.G., Goodrich, J.K., Gordon, J.I., Huttley, G.A., Kelley, S.T., Knights, D., Koenig, J.E., Ley, R.E., Lozupone, C.A., McDonald, D., Muegge, B.D., Pirrung, M., Reeder, J., Sevinsky, J.R., Turnbaugh, P.J., Walters, W.A., Widmann, J., Yatsunenko, T., Zaneveld, J. & Knight, R. (2010) QIIME allows analysis of high-throughput community sequencing data. Nature Methods 7: 335-336 (doi: 10.1038/nmeth.f.303).

36. Holovachov, O. (2016). Metabarcoding of marine nematodes – evaluation of similarity scores used in alignment-based taxonomy assignment approach. Biodiversity Data Journal 4: e10647 (doi: 10.3897/BDJ.4.e10647).

37. Rzeznik-Originac, J., Kalenitchenko, D., Mariette, J., Bodiou, J.-Y., Le Bris, N. & Derelle, E. (2017). Comparison of meiofaunal diversity by combined morphological and molecular approaches in a shallow Mediterranean sediment. Marine Biology 164: 40 (doi: 10.1007/s00227-017-3074-4).

38. Holovachov, O. (2016). Metabarcoding of marine nematodes – evaluation of reference datasets used in tree-based taxonomy assignment approach. Biodiversity Data Journal 4: e10021 (doi: 10.3897/BDJ.4.e10021).

39. Larkin, M.A., Blackshields, G., Brown, N.P., Chenna, R., McGettigan, P.A., McWilliam, H., Valentin, F., Wallace, I.M., Wilm, A., Lopez, R., Thompson, J.D., Gibson, T.J. & Higgins, D.G. (2007). Clustal W and Clustal X version 2.0. Bioinformatics 23: 2947-2948 (doi: 10.1093/bioinformatics/btm404).

40. Tamura, K., Stecher, G., Peterson, D., Filipski, A. & Kumar, S. (2013). MEGA6: Molecular Evolutionary Genetics Analysis Version 6.0. Molecular Biology and Evolution 30: 2725-2729 (doi: 10.1093/molbev/mst197)

41. Stamatakis, A. (2014). RAxML version 8: a tool for phylogenetic analysis and post-analysis of large phylogenies. Bioinformatics 30: 1312-1313 (doi: 10.1093/ bioinformatics/btu033).

42. Miller, M.A., Pfeier, W. & Schwartz, T. (2010). Creating the CIPRES Science Gateway for inference of large phylogenetic trees. Proceedings of the Gateway Computing Environments Workshop. New Orleans, LA.

43. Holovachov, O., Rodrigues, C.F., Zbinden, M. & Duperron, S. (2013). *Trophomera conchicola* sp. n. (Nematoda: Benthimermithidae) from chemosymbiotic bivalves *Idas modiolaeiformis* and *Lucionoma kazani* (Mollusca: Mytilidae and Lucinidae) in Eastern Mediterranean. Russian Journal of Nematology 21: 1-12.

44. Holovachov, O., Boström, S., Tandingan De Ley, I., Robinson, C., Mundo-Ocampo, M. &

Nadler, S.A. (2013). Morphology, molecular characterisation and systematic position of the genus *Cynura* Cobb, 1920 (Nematoda: Plectida). Nematology 15: 611-627.

45. Seibel, P. N., Müller, T., Dandekar, T., Schultz, J. & Wolf, M. (2006). 4SALE – A tool for synchronous RNA sequence and secondary structure alignment and editing. BMC Bioinformatics 7:498.
46. Higgs, P.G. (2000). RNA secondary structure: physical and computational aspects. Quarterly Review in Biophysics 33:199-253.
47. Schloss, P.D., Westcott, S.L., Ryabin, T., Hall, J.R., Hartmann, M., Hollister, E.B., Lesniewski, R.A., Oakley, B.B., Parks, D.H., Robinson, C.J., Sahl, J.W., Stres. B., Thallinger, G.G., Van Horn, D.J. & Weber, C.F. (2009). Introducing mothur: Open-source, platform-independent, community-supported software for describing and comparing microbial communities. Applied Environmental Microbiology 75: 7537-7541
48. Berger S.A. & Stamatakis A. (2011). Aligning short reads to reference alignments and trees. Bioinformatics 27: 2068-2075 (doi:10.1093/bioinformatics/btr320)
49. Rambaut, A. (2015). FigTree. http://tree.bio.ed.ac.uk/software/gtree/
50. Letunic, I. & Bork, P. (2016). Interactive tree of life (iTOL) v3: an online tool for the display and annotation of phylogenetic and other trees. Nucleic Acids Research: gkw290 (doi: 10.1093/nar/gkw290)
51. Darty, K., Denise, A., & Ponty, Y. (2009). VARNA: Interactive drawing and editing of the RNA secondary structure. Bioinformatics 25: 1974-1975.
52. Lindeque, P.K., Parry, H.E., Harmer, R.A., Somerfield, P.J. & Atkinson, A. (2013). Next generation sequencing reveals the hidden diversity of zooplankton assemblages. PLoS ONE 8: e81327 (doi: 10.1371/journal.pone.0081327).
53. Buhay J.E. (2009). "COI-Like" sequences are becoming problematic in molecular systematic and DNA barcoding studies. Journal of Crustacean Biology 29: 96-110.
54. Schnell, I.B., Sollmann, R., Calvignac-Spencer, S., Siddall, M.E., Yu, D.W., Wilting, A. & Gilbert, M.T.P. (2015). iDNA from terrestrial haematophagous leeches as a wildlife surveying and monitoring tool – prospects, pitfalls and avenues to be developed. Frontiers in Zoology 12: 24.
55. Wuyts, J., Van de Peer, Y., Winkelmans, T. & De Wachter, R. (2002). The European database on small subunit ribosomal RNA. Nucleic Acids Research 30: 183-185.
56. Neefs, J.M., Van de Peer, Y., Hendriks, L. & De Wachter, R. (1990). Compilation of small ribosomal subunit RNA sequences. Nucleic Acids Research 18 Suppl: 2237-2317.

## Figure legends

**Figure 1.** Cladogram based on tree-based taxonomy assignment approach using complete query dataset. Families that include positively assigned OTUs are colour coded, remaining reference taxa are shaded in grey.

**Figure 2.** Cladogram based on phylogeny-based taxonomy assignment approach. Families that include positively assigned OTUs are colour coded, remaining reference taxa are shaded in grey.

**Figure 3.** Comparison of the total number of taxa identified using phylogeny-based taxonomy assignment approach (OTUs, red) and morphology-based identification (morphospecies, green) for each nematode family in each sample (sampling site/extraction method) based on Supplementary Table 9 (excluding families without reference sequence data).

## Supplementary files

**Supplementary Figure 1.** Barcoding region used in this study – generalized secondary structure model of the 90% consensus sequence (excluding rare insertions) based on combined data from secondary-structure-based alignment of reference sequences and mothur-based alignment of OTU sequences (444 sequences in total). Helices (6-10, 10/e1, 11-14) are according to [55]. Variable regions V1 and V2 are labelled according to [56].

**Supplementary Figure 2.** Graphical comparison of the number of taxa identified using morphology-based identification (X axis) *versus* phylogeny-based taxonomy assignment approach (Y axis) for each nematode family in each sample (sampling site/extraction method) based on Supplementary Table 9 (excluding families without reference sequence data). Size of "bubbles" corresponds to the total number of families that show this result for all sites and extraction methods.

**Supplementary Figure 3.** Examples of cumulative placement of two different OTUs within the family Thoracostomopsidae that provide reliable identification of anonymous OTUs to the family-level taxonomic categories (branch length not representative). A: Part of the reference phylogeny with labelled clades (*i463-i477*) and bootstrap support. B: Placements of HE3.SSU124287 in clades

*i464* (probability 0.333433), *i465* (probability 0.333132) and *i466* (probability 0.333435) gives cumulative probability higher than the required threshold of 0.95. C: Placements of HF9.SSU17250 in clades *i464* (probability 0.924697) and *i466* (probability 0.037675) gives cumulative probability higher than the required threshold of 0.95.

**Supplementary Table 1.** GenBank accession numbers and classification of sequences used in reference datasets for the tree-based and phylogeny-based taxonomy assignment algorithms.

**Supplementary Table 2.** Taxonomic composition and relative abundance (% of the total number of specimens) of nematode species in Hållö site.

**Supplementary Table 3.** Taxonomic composition and relative abundance (% of the total number of specimens) of nematode species in Telekabeln site.

**Supplementary Table 4.** Results of alignment-based taxonomy assignment using BLASTN 2.5.0+ against the nucleotide collection of the NCBI database.

**Supplementary Table 5.** Results of alignment-based taxonomy assignment using LCAClassifier of CREST against built-in reference database.

**Supplementary Table 6.** Results of tree-based taxonomy assignment using complete and partitioned query dataset.

**Supplementary Table 7.** Results of phylogeny-based taxonomy assignment using mothur and PaPaRa-based alignments.

**Supplementary Table 8.** Comparison of the results produced by different taxonomy assignment approaches and distribution of different OTUs among study sites (presence vs. absence).

**Supplementary Table 9.** Number of taxa (species/OTUs) per family recovered in morphology-based identification and best scoring taxonomy assignment approach (EPA) grouped per each site and extraction method.

**Supplementary Data 1.** Nematode OTU sequences used in this study.

**Figure 1.** Cladogram based on tree-based taxonomy assignment approach using complete query dataset. Families that include positively assigned OTUs are colour coded, remaining reference taxa are shaded in grey.

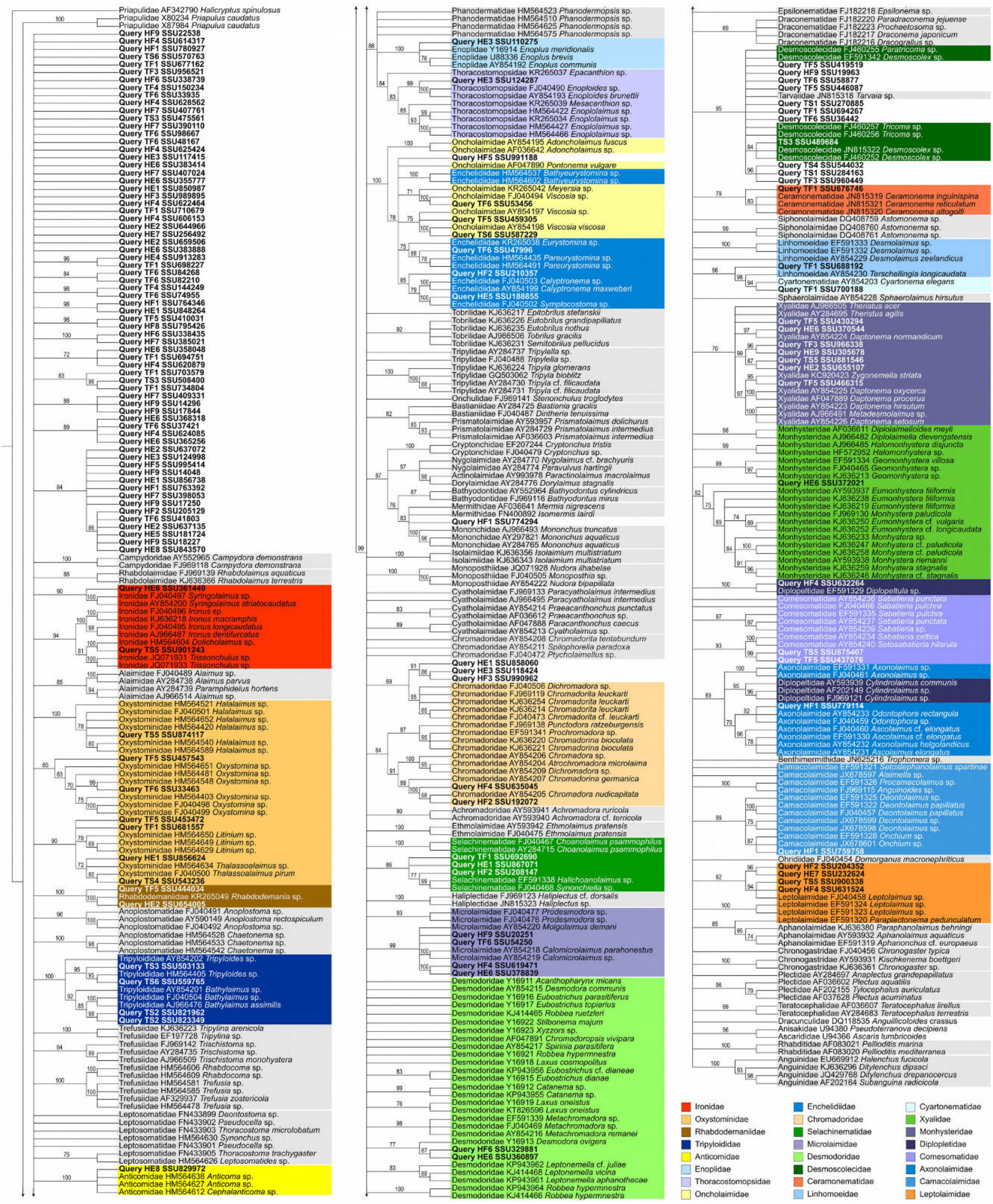

**Figure 2.** Cladogram based on phylogeny-based taxonomy assignment approach. Families that include positively assigned OTUs are colour coded, remaining reference taxa are shaded in grey.

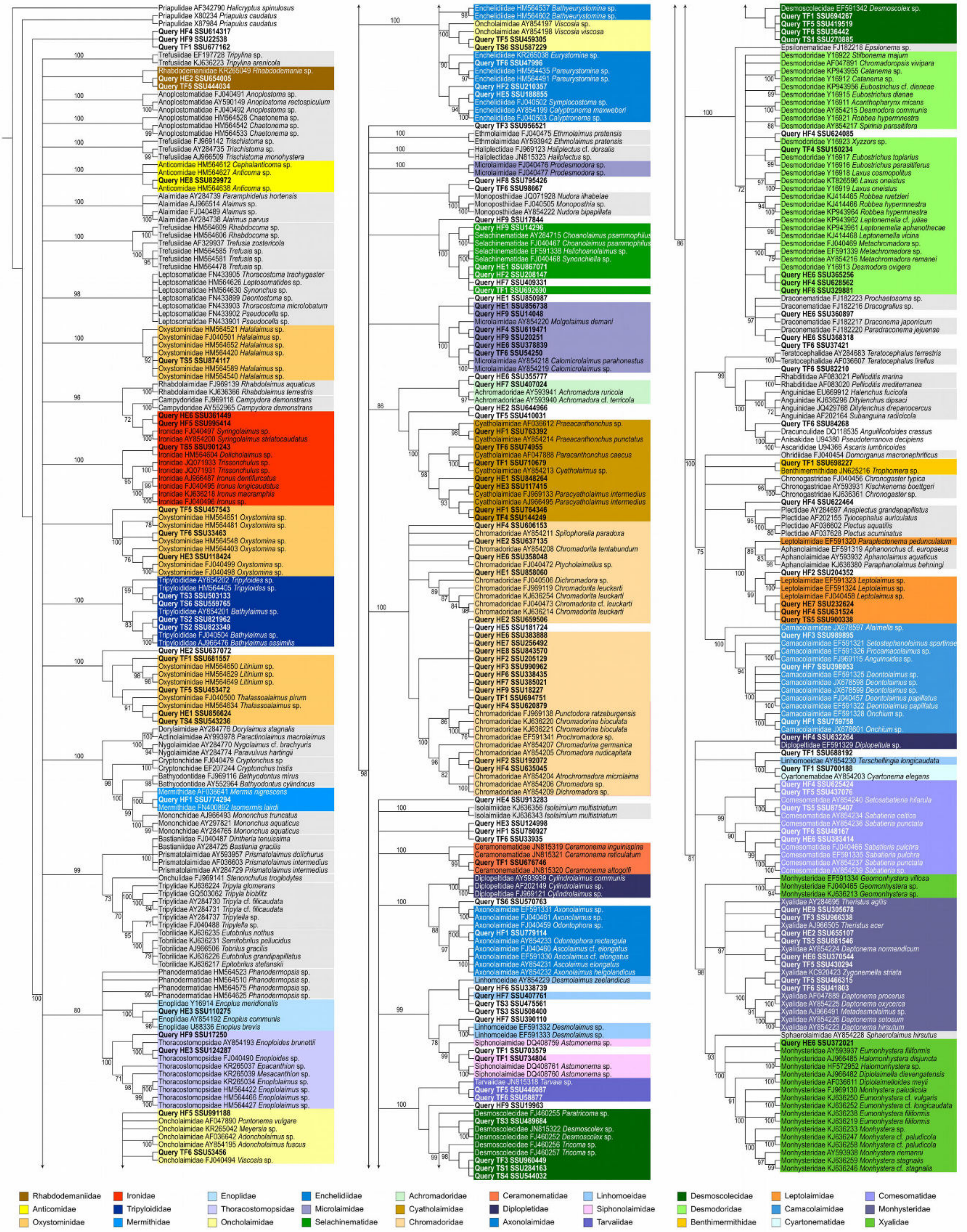

**Figure 3.** Comparison of the total number of taxa identified using phylogeny-based taxonomy assignment approach (OTUs, red) and morphology-based identification (morphospecies, green) for each nematode family in each sample (sampling site/extraction method) based on Supplementary Table 9 (excluding families without reference sequence data).

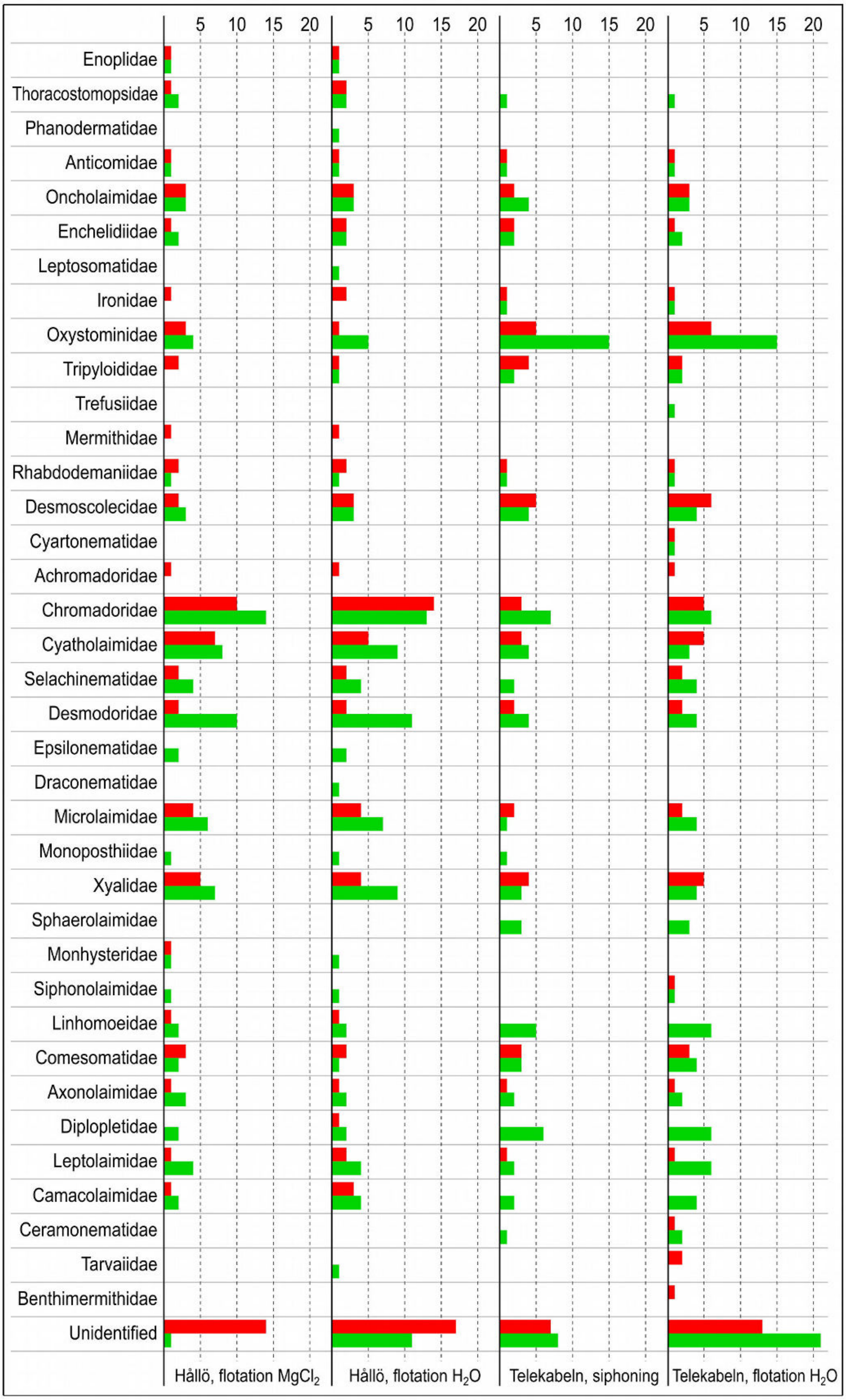

**Supplementary Figure 1.** Barcoding region used in this study – generalized secondary structure model of the 90% consensus sequence (excluding rare insertions) based on combined data from secondary-structure-based alignment of reference sequences and mothur-based alignment of OTU sequences (444 sequences in total). Helices (6-10, 10/e1, 11-14) are according to [55]. Variable regions V1 and V2 are labelled according to [56].

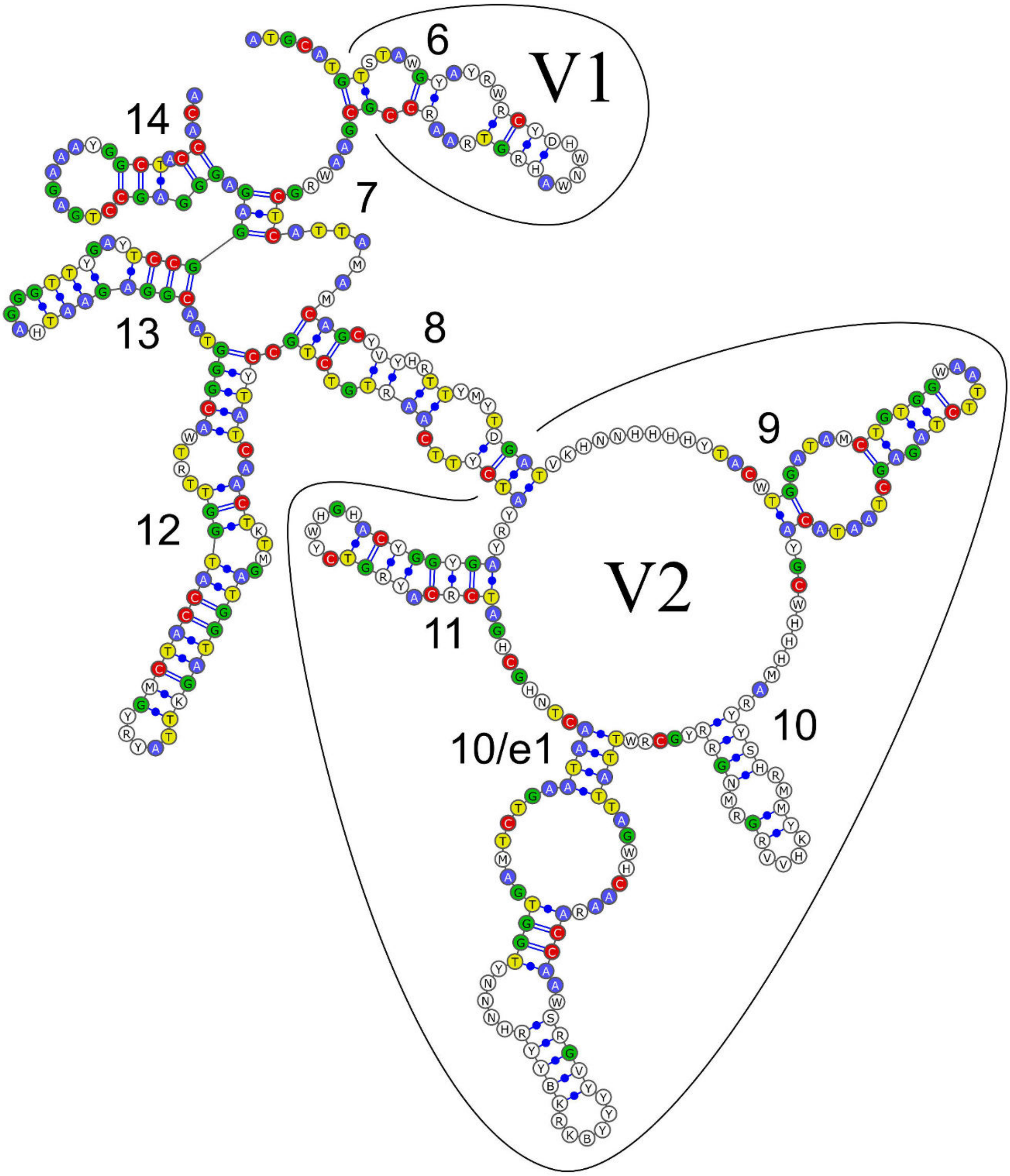

**Supplementary Figure 2.** Graphical comparison of the number of taxa identified using morphology-based identification (X axis) *versus* phylogeny-based taxonomy assignment approach (Y axis) for each nematode family in each sample (sampling site/extraction method) based on Supplementary Table 9 (excluding families without reference sequence data). Size of "bubbles" correspond to the number of families separately for each site and extraction method.

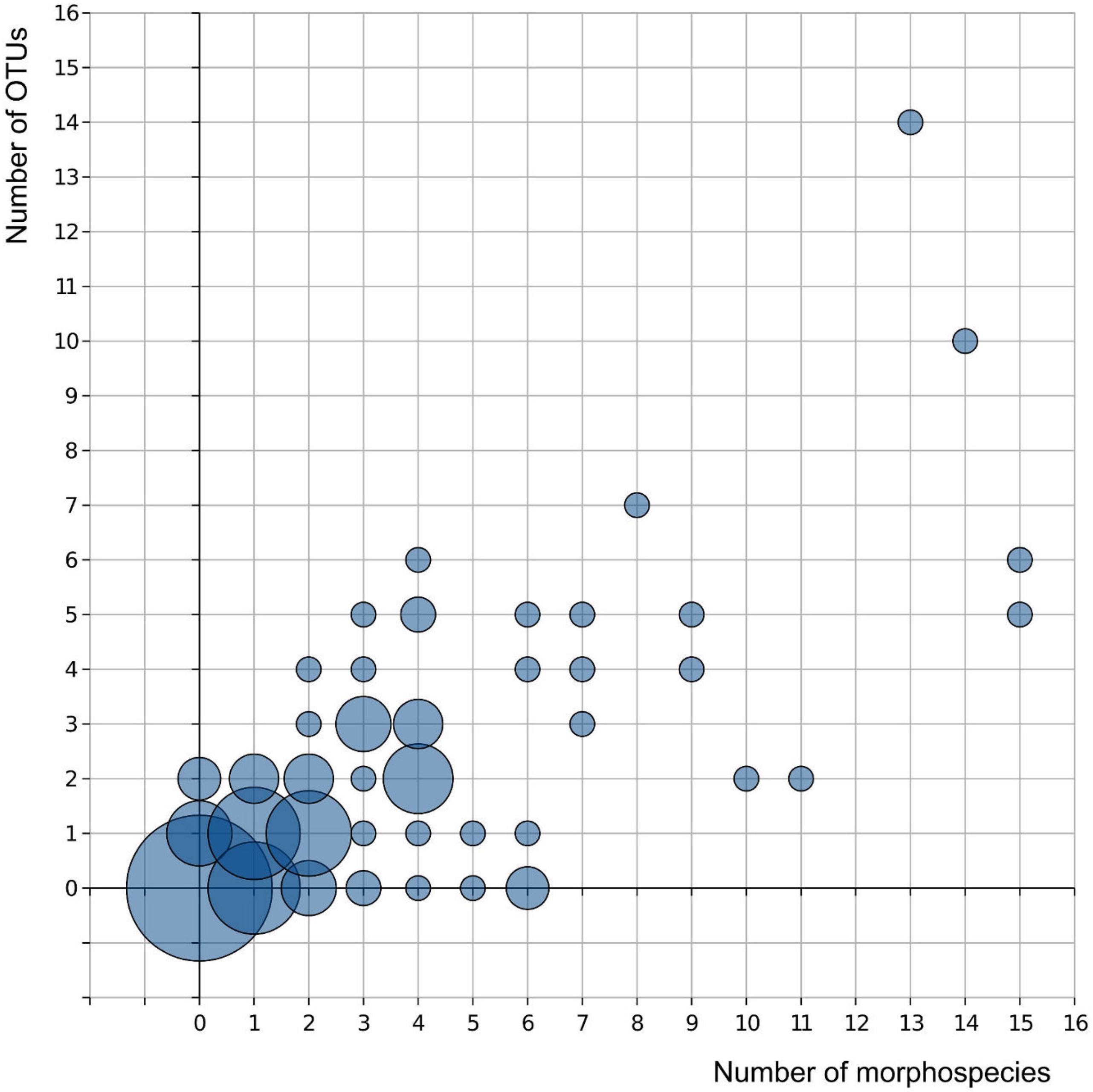

**Supplementary Figure 3.** Examples of cumulative placement of two different OTUs within the family Thoracostomopsidae that provide reliable identification of anonymous OTUs to the family-level taxonomic categories (branch length not representative). A: Part of the reference phylogeny with labelled clades (*i463-i477*) and bootstrap support. B: Placements of HE3.SSU124287 in clades *i464* (probability 0.333433), *i465* (probability 0.333132) and *i466* (probability 0.333435) gives cumulative probability higher than the required threshold of 0.95. C: Placements of HF9.SSU17250 in clades *i464* (probability 0.924697) and *i466* (probability 0.037675) gives cumulative probability higher than the required threshold of 0.95.

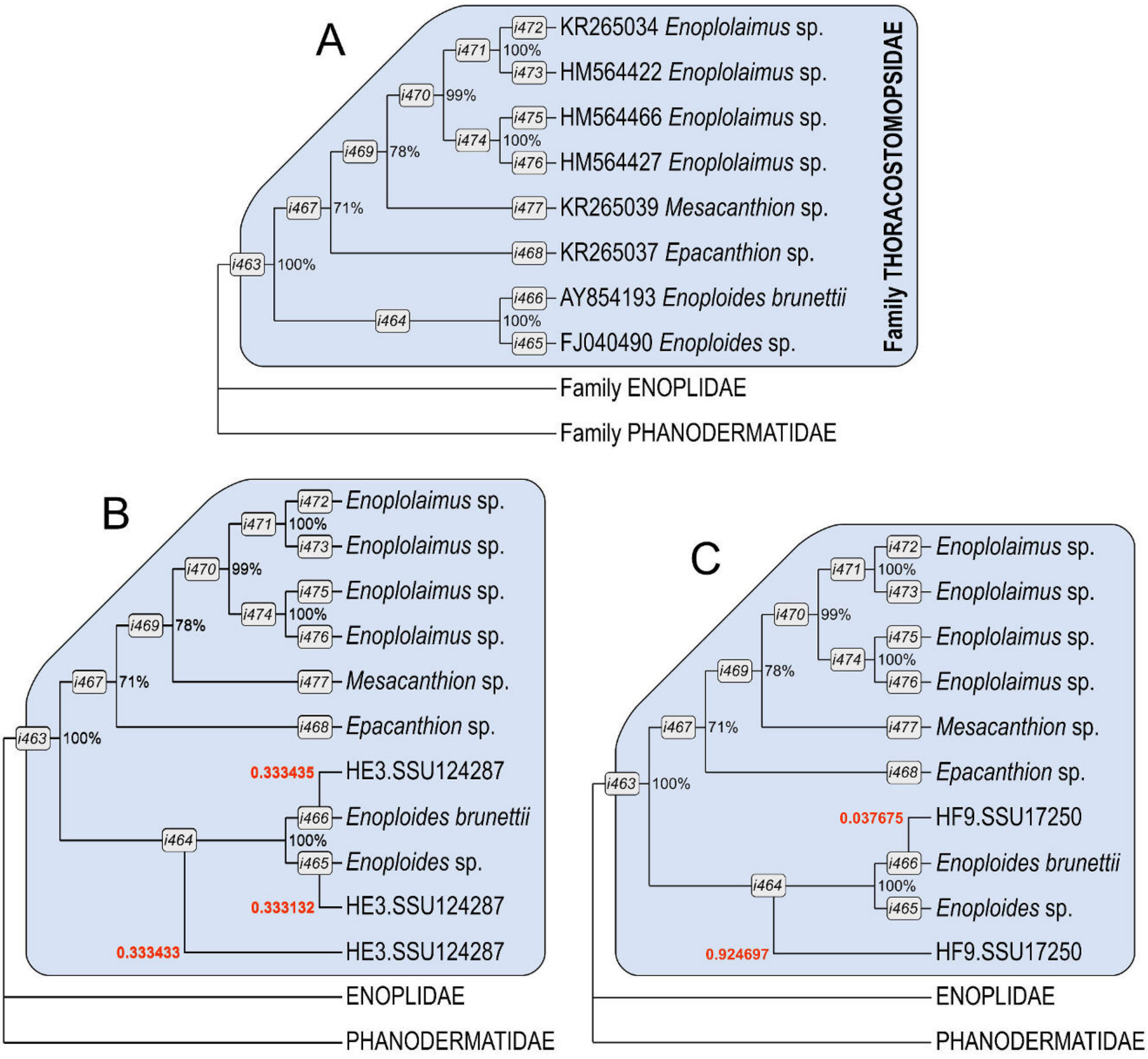

**Supplementary Table 1.** GenBank accession numbers and classification of sequences used in reference datasets for the tree-based and phylogeny-based taxonomy assignment algorithms.

| | **Acc. number** | **Family** | **Genus** | **Species** |
|---|---|---|---|---|
| 1. | AF202164 | Anguinidae | *Subanguina* | *radicicola* |
| 2. | EU669912 | Anguinidae | *Halenchus* | *fucicola* |
| 3. | JQ429768 | Anguinidae | *Ditylenchus* | *drepanocercus* |
| 4. | KJ636296 | Anguinidae | *Ditylenchus* | *dipsaci* |
| 5. | AF083020 | Rhabditidae | *Pellioditis* | *mediterranea* |
| 6. | AF083021 | Rhabditidae | *Pellioditis* | *marina* |
| 7. | U94366 | Ascarididae | *Ascaris* | *lumbricoides* |
| 8. | DQ118535 | Dracunculidae | *Anguillicoloides* | *crassus* |
| 9. | U94380 | Anisakidae | *Pseudoterranova* | *decipiens* |
| 10. | AY284683 | Teratocephalidae | *Teratocephalus* | *terrestris* |
| 11. | AF036607 | Teratocephalidae | *Teratocephalus* | *lirellus* |
| 12. | AF202155 | Plectidae | *Tylocephalus* | *auriculatus* |
| 13. | AF037628 | Plectidae | *Plectus* | *acuminatus* |
| 14. | AF036602 | Plectidae | *Plectus* | *aquatilis* |
| 15. | AY284697 | Plectidae | *Anaplectus* | *grandepapillatus* |
| 16. | AY593931 | Chronogastridae | *Kischkenema* | *boettgeri* |
| 17. | FJ040456 | Chronogastridae | *Chronogaster* | *typica* |
| 18. | KJ636361 | Chronogastridae | *Chronogaster* | sp. |
| 19. | EF591319 | Aphanolaimidae | *Aphanonchus* | cf. *europaeus* |
| 20. | AY593932 | Aphanolaimidae | *Aphanolaimus* | *aquaticus* |
| 21. | KJ636380 | Aphanolaimidae | *Paraphanolaimus* | *behningi* |
| 22. | EF591320 | Leptolaimidae | *Paraplectonema* | *pedunculatum* |
| 23. | EF591323 | Leptolaimidae | *Leptolaimus* | sp. |
| 24. | EF591324 | Leptolaimidae | *Leptolaimus* | sp. |
| 25. | FJ040458 | Leptolaimidae | *Leptolaimus* | sp. |
| 26. | FJ040454 | Ohridiidae | *Domorganus* | *macronephriticus* |
| 27. | EF591321 | Camacolaimidae | *Setostephanolaimus* | *spartinae* |
| 28. | JX678597 | Camacolaimidae | *Alaimella* | sp. |
| 29. | FJ969115 | Camacolaimidae | *Anguinoides* | sp. |
| 30. | JX678601 | Camacolaimidae | *Onchium* | sp. |
| 31. | EF591328 | Camacolaimidae | *Onchium* | sp. |
| 32. | EF591322 | Camacolaimidae | *Deontolaimus* | *papillatus* |
| 33. | FJ040457 | Camacolaimidae | *Deontolaimus* | *papillatus* |
| 34. | JX678599 | Camacolaimidae | *Deontolaimus* | sp. |
| 35. | JX678598 | Camacolaimidae | *Deontolaimus* | sp. |
| 36. | EF591325 | Camacolaimidae | *Deontolaimus* | sp. |
| 37. | EF591326 | Camacolaimidae | *Procamacolaimus* | sp. |

| | Acc. number | Family | Genus | Species |
|---|---|---|---|---|
| 38. | JN625216 | Benthimermithidae | *Trophomera* | sp. |
| 39. | FJ040460 | Axonolaimidae | *Ascolaimus* | cf. *elongatus* |
| 40. | EF591330 | Axonolaimidae | *Ascolaimus* | cf. *elongatus* |
| 41. | AY854231 | Axonolaimidae | *Ascolaimus* | *elongatus* |
| 42. | AY854232 | Axonolaimidae | *Axonolaimus* | *helgolandicus* |
| 43. | EF591331 | Axonolaimidae | *Axonolaimus* | sp. |
| 44. | FJ040461 | Axonolaimidae | *Axonolaimus* | sp. |
| 45. | AY854233 | Axonolaimidae | *Odontophora* | *rectangula* |
| 46. | FJ040459 | Axonolaimidae | *Odontophora* | sp. |
| 47. | EF591335 | Comesomatidae | *Sabatieria* | *pulchra* |
| 48. | FJ040466 | Comesomatidae | *Sabatieria* | *pulchra* |
| 49. | AY854240 | Comesomatidae | *Setosabatieria* | *hilarula* |
| 50. | AY854237 | Comesomatidae | *Sabatieria* | *punctata* |
| 51. | AY854236 | Comesomatidae | *Sabatieria* | *punctata* |
| 52. | AY854234 | Comesomatidae | *Sabatieria* | *celtica* |
| 53. | AY854239 | Comesomatidae | *Sabatieria* | sp. |
| 54. | AY593939 | Diplopeltidae | *Cylindrolaimus* | *communis* |
| 55. | FJ969121 | Diplopeltidae | *Cylindrolaimus* | sp. |
| 56. | AF202149 | Diplopeltidae | *Cylindrolaimus* | sp. |
| 57. | EF591329 | Diplopeltidae | *Diplopeltula* | sp. |
| 58. | EF591334 | Monhysteridae | *Geomonhystera* | *villosa* |
| 59. | FJ040465 | Monhysteridae | *Geomonhystera* | sp. |
| 60. | KJ636213 | Monhysteridae | *Geomonhystera* | sp. |
| 61. | HF572952 | Monhysteridae | *Halomonhystera* | sp. |
| 62. | AJ966485 | Monhysteridae | *Halomonhystera* | *disjuncta* |
| 63. | AY593938 | Monhysteridae | *Monhystera* | *riemanni* |
| 64. | FJ969130 | Monhysteridae | *Monhystera* | *paludicola* |
| 65. | KJ636247 | Monhysteridae | *Monhystera* | cf. *paludicola* |
| 66. | KJ636258 | Monhysteridae | *Monhystera* | cf. *paludicola* |
| 67. | KJ636246 | Monhysteridae | *Monhystera* | cf. *stagnalis* |
| 68. | KJ636259 | Monhysteridae | *Monhystera* | *stagnalis* |
| 69. | KJ636233 | Monhysteridae | *Monhystera* | sp. |
| 70. | KJ636250 | Monhysteridae | *Eumonhystera* | cf. *vulgaris* |
| 71. | KJ636238 | Monhysteridae | *Eumonhystera* | *filiformis* |
| 72. | AY593937 | Monhysteridae | *Eumonhystera* | *filiformis* |
| 73. | KJ636219 | Monhysteridae | *Eumonhystera* | *filiformis* |
| 74. | KJ636252 | Monhysteridae | *Eumonhystera* | cf. *longicaudata* |
| 75. | AJ966482 | Monhysteridae | *Diplolaimella* | *dievengatensis* |
| 76. | AF036611 | Monhysteridae | *Diplolaimelloides* | *meyli* |
| 77. | AJ966505 | Xyalidae | *Theristus* | *acer* |
| 78. | AY284695 | Xyalidae | *Theristus* | *agilis* |

| | Acc. number | Family | Genus | Species |
|---|---|---|---|---|
| 79. | AF047889 | Xyalidae | *Daptonema* | *procerus* |
| 80. | AY854226 | Xyalidae | *Daptonema* | *setosum* |
| 81. | AY854225 | Xyalidae | *Daptonema* | *oxycerca* |
| 82. | AY854223 | Xyalidae | *Daptonema* | *hirsutum* |
| 83. | AY854224 | Xyalidae | *Daptonema* | *normandicum* |
| 84. | KC920423 | Xyalidae | *Zygonemella* | *striata* |
| 85. | AJ966491 | Xyalidae | *Metadesmolaimus* | sp. |
| 86. | AY854228 | Sphaerolaimidae | *Sphaerolaimus* | *hirsutus* |
| 87. | EF591333 | Linhomoeidae | *Desmolaimus* | sp. |
| 88. | EF591332 | Linhomoeidae | *Desmolaimus* | sp. |
| 89. | AY854229 | Linhomoeidae | *Desmolaimus* | *zeelandicus* |
| 90. | AY854230 | Linhomoeidae | *Terschellingia* | *longicaudata* |
| 91. | DQ408760 | Siphonolaimidae | *Astomonema* | sp. |
| 92. | DQ408761 | Siphonolaimidae | *Astomonema* | sp. |
| 93. | DQ408759 | Siphonolaimidae | *Astomonema* | sp. |
| 94. | JN815318 | Tarvaiidae | *Tarvaia* | sp. |
| 95. | JN815319 | Ceramonematidae | *Ceramonema* | *inguinispina* |
| 96. | JN815320 | Ceramonematidae | *Ceramonema* | *altogolfi* |
| 97. | JN815321 | Ceramonematidae | *Ceramonema* | *reticulatum* |
| 98. | FJ460256 | Desmoscolecidae | *Tricoma* | sp. |
| 99. | FJ460257 | Desmoscolecidae | *Tricoma* | sp. |
| 100. | FJ460255 | Desmoscolecidae | *Paratricoma* | sp. |
| 101. | FJ460252 | Desmoscolecidae | *Desmoscolex* | sp. |
| 102. | EF591342 | Desmoscolecidae | *Desmoscolex* | sp. |
| 103. | JN815322 | Desmoscolecidae | *Desmoscolex* | sp. |
| 104. | AY854203 | Cyartonematidae | *Cyartonema* | *elegans* |
| 105. | FJ182217 | Draconematidae | *Draconema* | *japonicum* |
| 106. | FJ182220 | Draconematidae | *Paradraconema* | *jejuense* |
| 107. | FJ182216 | Draconematidae | *Dracograllus* | sp. |
| 108. | FJ182223 | Draconematidae | *Prochaetosoma* | sp. |
| 109. | FJ182218 | Epsilonematidae | *Epsilonema* | sp. |
| 110. | AY854217 | Desmodoridae | *Spirinia* | *parasitifera* |
| 111. | EF591339 | Desmodoridae | *Metachromadora* | sp. |
| 112. | AY854216 | Desmodoridae | *Metachromadora* | *remanei* |
| 113. | FJ040469 | Desmodoridae | *Metachromadora* | sp. |
| 114. | Y16911 | Desmodoridae | *Acanthopharynx* | *micans* |
| 115. | AF047891 | Desmodoridae | *Chromadoropsis* | *vivipara* |
| 116. | AY854215 | Desmodoridae | *Desmodora* | *communis* |
| 117. | Y16913 | Desmodoridae | *Desmodora* | *ovigera* |
| 118. | Y16923 | Desmodoridae | *Xyzzors* | sp. |
| 119. | KJ414468 | Desmodoridae | *Leptonemella* | *vicina* |

| | Acc. number | Family | Genus | Species |
|---|---|---|---|---|
| 120. | KP943962 | Desmodoridae | *Leptonemella* | cf. *juliae* |
| 121. | KP943961 | Desmodoridae | *Leptonemella* | *aphanothecae* |
| 122. | Y16915 | Desmodoridae | *Eubostrichus* | *dianae* |
| 123. | Y16917 | Desmodoridae | *Eubostrichus* | *topiarius* |
| 124. | Y16916 | Desmodoridae | *Eubostrichus* | *parasitiferus* |
| 125. | KP943956 | Desmodoridae | *Eubostrichus* | cf. *dianeae* |
| 126. | Y16918 | Desmodoridae | *Laxus* | *cosmopolitus* |
| 127. | KT826596 | Desmodoridae | *Laxus* | *oneistus* |
| 128. | Y16919 | Desmodoridae | *Laxus* | *oneistus* |
| 129. | Y16922 | Desmodoridae | *Stilbonema* | *majum* |
| 130. | KJ414465 | Desmodoridae | *Robbea* | *ruetzleri* |
| 131. | KJ414466 | Desmodoridae | *Robbea* | *hypermnestra* |
| 132. | Y16921 | Desmodoridae | *Robbea* | *hypermnestra* |
| 133. | KP943964 | Desmodoridae | *Robbea* | *hypermnestra* |
| 134. | KP943955 | Desmodoridae | *Catanema* | sp. |
| 135. | Y16912 | Desmodoridae | *Catanema* | sp. |
| 136. | AY854218 | Microlaimidae | *Calomicrolaimus* | *parahonestus* |
| 137. | AY854219 | Microlaimidae | *Calomicrolaimus* | sp. |
| 138. | AY854220 | Microlaimidae | *Molgolaimus* | *demani* |
| 139. | FJ040477 | Microlaimidae | *Prodesmodora* | sp. |
| 140. | FJ040476 | Microlaimidae | *Prodesmodora* | sp. |
| 141. | JN815323 | Haliplectidae | *Haliplectus* | sp. |
| 142. | FJ969123 | Haliplectidae | *Haliplectus* | cf. *dorsalis* |
| 143. | AY284715 | Selachinematidae | *Choanolaimus* | *psammophilus* |
| 144. | FJ040467 | Selachinematidae | *Choanolaimus* | *psammophilus* |
| 145. | FJ040468 | Selachinematidae | *Synonchiella* | sp. |
| 146. | EF591338 | Selachinematidae | *Halichoanolaimus* | sp. |
| 147. | AY593942 | Ethmolaimidae | *Ethmolaimus* | *pratensis* |
| 148. | FJ040475 | Ethmolaimidae | *Ethmolaimus* | *pratensis* |
| 149. | AY593941 | Achromadoridae | *Achromadora* | *ruricola* |
| 150. | AY593940 | Achromadoridae | *Achromadora* | cf. *terricola* |
| 151. | AY854205 | Chromadoridae | *Chromadora* | *nudicapitata* |
| 152. | AY854206 | Chromadoridae | *Chromadora* | sp. |
| 153. | AY854209 | Chromadoridae | *Dichromadora* | sp. |
| 154. | FJ040506 | Chromadoridae | *Dichromadora* | sp. |
| 155. | FJ969119 | Chromadoridae | *Chromadorita* | *leuckarti* |
| 156. | FJ969138 | Chromadoridae | *Punctodora* | *ratzeburgensis* |
| 157. | EF591341 | Chromadoridae | *Prochromadora* | sp. |
| 158. | FJ040473 | Chromadoridae | *Chromadorita* | cf. *leuckarti* |
| 159. | KJ636254 | Chromadoridae | *Chromadorita* | *leuckarti* |
| 160. | KJ636214 | Chromadoridae | *Chromadorita* | *leuckarti* |

| | Acc. number | Family | Genus | Species |
|---|---|---|---|---|
| 161. | AY854208 | Chromadoridae | *Chromadorita* | *tentabundum* |
| 162. | AY854207 | Chromadoridae | *Chromadorina* | *germanica* |
| 163. | KJ636220 | Chromadoridae | *Chromadorina* | *bioculata* |
| 164. | KJ636221 | Chromadoridae | *Chromadorina* | *bioculata* |
| 165. | AY854204 | Chromadoridae | *Atrochromadora* | *microlaima* |
| 166. | AY854211 | Chromadoridae | *Spilophorella* | *paradoxa* |
| 167. | FJ040472 | Chromadoridae | *Ptycholaimellus* | sp. |
| 168. | AJ966495 | Cyatholaimidae | *Paracyatholaimus* | *intermedius* |
| 169. | FJ969133 | Cyatholaimidae | *Paracyatholaimus* | *intermedius* |
| 170. | AF036612 | Cyatholaimidae | *Praeacanthonchus* | sp. |
| 171. | AF047888 | Cyatholaimidae | *Paracanthonchus* | *caecus* |
| 172. | AY854214 | Cyatholaimidae | *Praeacanthonchus* | *punctatus* |
| 173. | AY854213 | Cyatholaimidae | *Cyatholaimus* | sp. |
| 174. | JQ071928 | Monoposthiidae | *Nudora* | *ilhabelae* |
| 175. | FJ040505 | Monoposthiidae | *Monoposthia* | sp. |
| 176. | AY854222 | Monoposthiidae | *Nudora* | *bipapillata* |
| 177. | AY284776 | Dorylaimidae | *Dorylaimus* | *stagnalis* |
| 178. | AY993978 | Actinolaimidae | *Paractinolaimus* | *macrolaimus* |
| 179. | AY284774 | Nygolaimidae | *Paravulvus* | *hartingii* |
| 180. | AY284770 | Nygolaimidae | *Nygolaimus* | cf. *brachyuris* |
| 181. | KJ636343 | Isolaimiidae | *Isolaimium* | *multistriatum* |
| 182. | KJ636356 | Isolaimiidae | *Isolaimium* | *multistriatum* |
| 183. | EF207244 | Cryptonchidae | *Cryptonchus* | *tristis* |
| 184. | FJ040479 | Cryptonchidae | *Cryptonchus* | sp. |
| 185. | FJ969116 | Bathyodontidae | *Bathyodontus* | *mirus* |
| 186. | AY552964 | Bathyodontidae | *Bathyodontus* | *cylindricus* |
| 187. | AY284765 | Mononchidae | *Mononchus* | *aquaticus* |
| 188. | AY297821 | Mononchidae | *Mononchus* | *aquaticus* |
| 189. | AJ966493 | Mononchidae | *Mononchus* | *truncatus* |
| 190. | FN400892 | Mermithidae | *Isomermis* | *lairdi* |
| 191. | AF036641 | Mermithidae | *Mermis* | *nigrescens* |
| 192. | AY284729 | Prismatolaimidae | *Prismatolaimus* | *intermedius* |
| 193. | AF036603 | Prismatolaimidae | *Prismatolaimus* | *intermedius* |
| 194. | AY593957 | Prismatolaimidae | *Prismatolaimus* | *dolichurus* |
| 195. | AY284725 | Bastianiidae | *Bastiania* | *gracilis* |
| 196. | FJ040487 | Bastianiidae | *Dintheria* | *tenuissima* |
| 197. | FJ969141 | Onchulidae | *Stenonchulus* | *troglodytes* |
| 198. | AY284731 | Tripylidae | *Tripyla* | cf. *filicaudata* |
| 199. | AY284730 | Tripylidae | *Tripyla* | cf. *filicaudata* |
| 200. | KJ636224 | Tripylidae | *Tripyla* | *glomerans* |
| 201. | GQ503062 | Tripylidae | *Tripyla* | *bioblitz* |

| | Acc. number | Family | Genus | Species |
|---|---|---|---|---|
| 202. | AY284737 | Tripylidae | *Tripylella* | sp. |
| 203. | FJ040488 | Tripylidae | *Tripylella* | sp. |
| 204. | AJ966506 | Tobrilidae | *Tobrilus* | *gracilis* |
| 205. | KJ636235 | Tobrilidae | *Eutobrilus* | *nothus* |
| 206. | KJ636226 | Tobrilidae | *Eutobrilus* | *grandipapillatus* |
| 207. | KJ636217 | Tobrilidae | *Epitobrilus* | *stefanskii* |
| 208. | KJ636231 | Tobrilidae | *Semitobrilus* | *pellucidus* |
| 209. | AF047890 | Oncholaimidae | *Pontonema* | *vulgare* |
| 210. | AF036642 | Oncholaimidae | *Adoncholaimus* | sp. |
| 211. | AY854195 | Oncholaimidae | *Adoncholaimus* | *fuscus* |
| 212. | AY854198 | Oncholaimidae | *Viscosia* | *viscosa* |
| 213. | AY854197 | Oncholaimidae | *Viscosia* | sp. |
| 214. | FJ040494 | Oncholaimidae | *Viscosia* | sp. |
| 215. | KR265042 | Oncholaimidae | *Meyersia* | sp. |
| 216. | FJ040502 | Enchelidiidae | *Symplocostoma* | sp. |
| 217. | AY854199 | Enchelidiidae | *Calyptronema* | *maxweberi* |
| 218. | FJ040503 | Enchelidiidae | *Calyptronema* | sp. |
| 219. | KR265038 | Enchelidiidae | *Eurystomina* | sp. |
| 220. | HM564491 | Enchelidiidae | *Pareurystomina* | sp. |
| 221. | HM564435 | Enchelidiidae | *Pareurystomina* | sp. |
| 222. | HM564537 | Enchelidiidae | *Bathyeurystomina* | sp. |
| 223. | HM564602 | Enchelidiidae | *Bathyeurystomina* | sp. |
| 224. | U88336 | Enoplidae | *Enoplus* | *brevis* |
| 225. | Y16914 | Enoplidae | *Enoplus* | *meridionalis* |
| 226. | AY854192 | Enoplidae | *Enoplus* | *communis* |
| 227. | AY854193 | Thoracostomopsidae | *Enoploides* | *brunettii* |
| 228. | FJ040490 | Thoracostomopsidae | *Enoploides* | sp. |
| 229. | KR265034 | Thoracostomopsidae | *Enoplolaimus* | sp. |
| 230. | HM564466 | Thoracostomopsidae | *Enoplolaimus* | sp. |
| 231. | HM564427 | Thoracostomopsidae | *Enoplolaimus* | sp. |
| 232. | HM564422 | Thoracostomopsidae | *Enoplolaimus* | sp. |
| 233. | KR265037 | Thoracostomopsidae | *Epacanthion* | sp. |
| 234. | KR265039 | Thoracostomopsidae | *Mesacanthion* | sp. |
| 235. | HM564625 | Phanodermatidae | *Phanodermopsis* | sp. |
| 236. | HM564510 | Phanodermatidae | *Phanodermopsis* | sp. |
| 237. | HM564575 | Phanodermatidae | *Phanodermopsis* | sp. |
| 238. | HM564523 | Phanodermatidae | *Phanodermopsis* | sp. |
| 239. | HM564638 | Anticomidae | *Anticoma* | sp. |
| 240. | HM564627 | Anticomidae | *Anticoma* | sp. |
| 241. | HM564612 | Anticomidae | *Cephalanticoma* | sp. |
| 242. | FN433905 | Leptosomatidae | *Thoracostoma* | *trachygaster* |

| | Acc. number | Family | Genus | Species |
|---|---|---|---|---|
| 243. | FN433903 | Leptosomatidae | *Thoracostoma* | *microlobatum* |
| 244. | FN433902 | Leptosomatidae | *Pseudocella* | sp. |
| 245. | FN433901 | Leptosomatidae | *Pseudocella* | sp. |
| 246. | FN433899 | Leptosomatidae | *Deontostoma* | sp. |
| 247. | HM564626 | Leptosomatidae | *Leptosomatides* | sp. |
| 248. | HM564630 | Leptosomatidae | *Synonchus* | sp. |
| 249. | HM564581 | Trefusiidae | *Trefusia* | sp. |
| 250. | AF329937 | Trefusiidae | *Trefusia* | *zostericola* |
| 251. | HM564585 | Trefusiidae | *Trefusia* | sp. |
| 252. | HM564478 | Trefusiidae | *Trefusia* | sp. |
| 253. | HM564606 | Trefusiidae | *Rhabdocoma* | sp. |
| 254. | HM564609 | Trefusiidae | *Rhabdocoma* | sp. |
| 255. | AJ966509 | Trefusiidae | *Trischistoma* | *monohystera* |
| 256. | AY284735 | Trefusiidae | *Trischistoma* | sp. |
| 257. | FJ969142 | Trefusiidae | *Trischistoma* | sp. |
| 258. | KJ636223 | Trefusiidae | *Tripylina* | *arenicola* |
| 259. | EF197728 | Trefusiidae | *Tripylina* | sp. |
| 260. | AJ966476 | Tripyloididae | *Bathylaimus* | *assimilis* |
| 261. | FJ040504 | Tripyloididae | *Bathylaimus* | sp. |
| 262. | AY854201 | Tripyloididae | *Bathylaimus* | sp. |
| 263. | HM564405 | Tripyloididae | *Tripyloides* | sp. |
| 264. | AY854202 | Tripyloididae | *Tripyloides* | sp. |
| 265. | FJ040491 | Anoplostomatidae | *Anoplostoma* | sp. |
| 266. | FJ040492 | Anoplostomatidae | *Anoplostoma* | sp. |
| 267. | AY590149 | Anoplostomatidae | *Anoplostoma* | *rectospiculum* |
| 268. | HM564528 | Anoplostomatidae | *Chaetonema* | sp. |
| 269. | HM564542 | Anoplostomatidae | *Chaetonema* | sp. |
| 270. | HM564533 | Anoplostomatidae | *Chaetonema* | sp. |
| 271. | KR265049 | Rhabdodemaniidae | *Rhabdodemania* | sp. |
| 272. | FJ040501 | Oxystominidae | *Halalaimus* | sp. |
| 273. | HM564589 | Oxystominidae | *Halalaimus* | sp. |
| 274. | HM564540 | Oxystominidae | *Halalaimus* | sp. |
| 275. | HM564652 | Oxystominidae | *Halalaimus* | sp. |
| 276. | HM564521 | Oxystominidae | *Halalaimus* | sp. |
| 277. | HM564420 | Oxystominidae | *Halalaimus* | sp. |
| 278. | HM564481 | Oxystominidae | *Oxystomina* | sp. |
| 279. | HM564548 | Oxystominidae | *Oxystomina* | sp. |
| 280. | HM564403 | Oxystominidae | *Oxystomina* | sp. |
| 281. | HM564651 | Oxystominidae | *Oxystomina* | sp. |
| 282. | FJ040499 | Oxystominidae | *Oxystomina* | sp. |
| 283. | FJ040498 | Oxystominidae | *Oxystomina* | sp. |

| | Acc. number | Family | Genus | Species |
|---|---|---|---|---|
| 284. | FJ040500 | Oxystominidae | *Thalassoalaimus* | *pirum* |
| 285. | HM564634 | Oxystominidae | *Thalassoalaimus* | sp. |
| 286. | HM564650 | Oxystominidae | *Litinium* | sp. |
| 287. | HM564649 | Oxystominidae | *Litinium* | sp. |
| 288. | HM564629 | Oxystominidae | *Litinium* | sp. |
| 289. | AY284738 | Alaimidae | *Alaimus* | *parvus* |
| 290. | AJ966514 | Alaimidae | *Alaimus* | sp. |
| 291. | FJ040489 | Alaimidae | *Alaimus* | sp. |
| 292. | AY284739 | Alaimidae | *Paramphidelus* | *hortens* |
| 293. | FJ040496 | Ironidae | *Ironus* | sp. |
| 294. | AJ966487 | Ironidae | *Ironus* | *dentifurcatus* |
| 295. | FJ040495 | Ironidae | *Ironus* | *longicaudatus* |
| 296. | KJ636218 | Ironidae | *Ironus* | *macramphis* |
| 297. | HM564604 | Ironidae | *Dolicholaimus* | sp. |
| 298. | JQ071931 | Ironidae | *Trissonchulus* | sp. |
| 299. | JQ071933 | Ironidae | *Trissonchulus* | sp. |
| 300. | AY854200 | Ironidae | *Syringolaimus* | *striatocaudatus* |
| 301. | FJ040497 | Ironidae | *Syringolaimus* | sp. |
| 302. | KJ636366 | Rhabdolaimidae | *Rhabdolaimus* | *terrestris* |
| 303. | FJ969139 | Rhabdolaimidae | *Rhabdolaimus* | *aquaticus* |
| 304. | AY552965 | Campydoridae | *Campydora* | *demonstrans* |
| 305. | FJ969118 | Campydoridae | *Campydora* | *demonstrans* |
| 306. | X87984 | Priapulidae | *Priapulus* | *caudatus* |
| 307. | X80234 | Priapulidae | *Priapulus* | *caudatus* |
| 308. | AF342790 | Priapulidae | *Halicryptus* | *spinulosus* |

**Supplementary Table 2.** Taxonomic composition and relative abundance (% of the total number of specimens) of nematode species in Hållö site.

| | Classification and identity | Flotation with $MgCl_2$ | Flotation with $H_2O$ |
|---|---|---|---|
| | **ORDER ENOPLIDA** | | |
| | **Family Enoplidae** | | |
| 1 | *Enoplus* sp. | 0.46 | 0.43 |
| | **Family Thoracostomopsidae** | | |
| 2 | Enoplolaiminae gen. sp. 1 | 0.33 | 0.29 |
| 3 | Enoplolaiminae gen. sp. 2 | 0.30 | 0.25 |
| | **Family Phanodermatidae** | | |
| 4 | Phanodermatidae gen. sp. | – | 0.03 |
| | **Family Anticomidae** | | |
| 5 | *Anticoma* sp. | 1.07 | 0.66 |
| | **Family Oncholaimidae** | | |
| 6 | *Viscosia* sp. | 3.73 | 1.96 |
| 7 | *Pontonema* sp. | 0.15 | 0.03 |
| 8 | Oncholaimidae gen. sp. | 0.41 | 0.31 |
| | **Family Enchelidiidae** | | |
| 9 | *Symplocostoma* sp. | 0.78 | 0.22 |
| 10 | *Pareurystomina* sp. | 0.08 | 0.03 |
| | **Family Leptosomatidae** | | |
| 11 | Leptosomatidae gen. sp. | – | 0.01 |
| | **Family Oxystominidae** | | |
| 12 | *Nemanema* sp. | – | 0.04 |
| 13 | *Oxystomina* sp. | 0.04 | 0.08 |
| 14 | *Thalassoalaimus* sp. | 0.08 | 0.16 |
| 15 | *Halalaimus* sp. 1 | 0.23 | 0.66 |
| 16 | *Halalaimus* sp. 2 | – | 0.01 |
| 17 | *Maldivea* sp. | 0.08 | – |
| | **Family Tripyloididae** | | |
| 18 | *Bathylaimus* sp. | – | 0.01 |
| | **Family Xennellidae** | | |
| 19 | *Xennella* sp. | 0.28 | 0.05 |
| | **ORDER TRIPLONCHIDA** | | |
| | **Family Rhabdodemaniidae** | | |
| 20 | *Rhabdodemania* sp. | 2.36 | 0.83 |
| | **ORDER DESMOSCOLECIDA** | | |
| | **Family Desmoscolecidae** | | |
| 21 | *Desmoscolex* spp. | 0.32 | 0.46 |
| 22 | *Tricoma* (*Tricoma*) spp. | 0.25 | 0.39 |

| | Classification and identity | Flotation with $MgCl_2$ | Flotation with $H_2O$ |
|---|---|---|---|
| 23 | *Tricoma* (*Quadricoma*) spp. | 3.05 | 3.49 |
| | **Family Meyliidae** | | |
| 24 | *Gerlachius* sp. | – | 0.01 |
| 25 | *Meylia* sp. | – | 0.01 |
| | **ORDER CHROMADORIDA** | | |
| | **Family Chromadoridae** | | |
| 26 | *Chromadora* sp. | 0.04 | 0.07 |
| 27 | *Chromadorita* sp. | 1.53 | 1.39 |
| 28 | *Chromadorella* cf. *filiformis* | 0.68 | 0.46 |
| 29 | *Neochromadora* sp. 1 | 2.33 | 1.81 |
| 30 | *Neochromadora* sp. 2 | 0.24 | 0.04 |
| 31 | *Neochromadora* sp. 3 | 0.16 | 0.10 |
| 32 | *Dichromadora* sp. 1 | 0.33 | 0.87 |
| 33 | *Dichromadora* sp. 2 | 0.04 | – |
| 34 | *Euchromadora* sp. | 1.03 | 0.36 |
| 35 | *Innocuanema tentabundum* | 1.19 | 1.30 |
| 36 | *Ptycholaimellus* sp. | 0.40 | 0.31 |
| 37 | *Spilophorella* sp. | 0.17 | 0.25 |
| 38 | *Actinonema* sp. | 0.45 | 0.14 |
| 39 | Chromadoridae gen. spp. | 0.28 | 0.16 |
| | **Family Cyatholaimidae** | | |
| 40 | *Pomponema* sp. | 0.16 | 0.05 |
| 41 | *Paracanthonchus* sp. 1 | 0.80 | 0.23 |
| 42 | *Paracanthonchus* cf. *spectabilis* | 0.04 | 0.16 |
| 43 | *Paracanthonchus* cf. *longus* | 1.89 | 2.26 |
| 44 | *Preacanthonchus* cf. *inglisi* | 0.80 | 0.55 |
| 45 | *Paracyatholaimoides* sp. | 0.23 | 1.98 |
| 46 | *Marylinnia* sp | 0.04 | 0.16 |
| 47 | *Longicyatholaimus* sp. | – | 0.14 |
| 48 | Cyatholaimidae gen. sp. | 0.17 | 0.04 |
| | **Family Selachinematidae** | | |
| 49 | *Halichoanolaimus* sp. | 1.92 | 1.17 |
| 50 | *Latronema* sp. | 0.12 | 0.09 |
| 51 | *Gammanema* sp. | 0.21 | 0.13 |
| 52 | *Choniolaimus* sp. | 0.04 | 0.01 |
| | **ORDER DESMODORIDA** | | |
| | **Family Desmodoridae** | | |
| 53 | *Desmodora* cf. *communis* | 1.14 | 0.59 |
| 54 | *Desmodora pontica* | 15.53 | 14.26 |
| 55 | *Desmodora granulata* | – | 0.01 |
| 56 | *Desmodorella schulzi* | 10.49 | 10.22 |

| | Classification and identity | Flotation with $MgCl_2$ | Flotation with $H_2O$ |
|---|---|---|---|
| 57 | *Bradylaimus pellita* | 2.74 | 2.09 |
| 58 | *Bolbonema brevicolle* | 0.28 | 0.38 |
| 59 | *Chromaspirina parapontica* | 10.42 | 20.34 |
| 60 | *Spirinia* sp. 1 | 0.35 | 0.27 |
| 61 | *Spirinia* sp. 2 | 0.27 | 0.27 |
| 62 | *Leptonemella* sp. | 0.04 | 0.03 |
| 63 | *Adelphos* sp. | 0.04 | 0.01 |
| | **Family Epsilonematidae** | | |
| 64 | *Epsilonema/Metepsilonema* sp. | 0.11 | 0.35 |
| 65 | *Perepsilonema* sp. | 1.08 | 4.97 |
| | **Family Draconematidae** | | |
| 66 | *Draconema cephalatum* | – | 0.01 |
| | **Family Microlaimidae** | | |
| 67 | *Ixonema powelli* | 0.21 | 0.95 |
| 68 | *Microlaimus* sp. 1 | 8.24 | 5.81 |
| 69 | *Microlaimus* sp. 2 | 1.97 | 2.03 |
| 70 | *Microlaimus* sp. 3 | 1.42 | 2.00 |
| 71 | *Microlaimus* sp. 4 | 0.04 | – |
| 72 | *Microlaimus* sp. 5 | – | 0.04 |
| 73 | *Microlaimus acanthus* | – | 0.01 |
| 74 | *Pseudoncholaimus dentatus* | 0.13 | 0.08 |
| | **Family Monoposthiidae** | | |
| 75 | *Monoposthia costata* | 0.31 | 0.23 |
| | **ORDER MONHYSTERIDA** | | |
| | **Family Xyalidae** | | |
| 76 | *Daptonema* sp. 1 | 1.45 | 0.83 |
| 77 | *Daptonema* sp. 2 | – | 0.03 |
| 78 | *Theristus* sp. 1 | 1.39 | 1.43 |
| 79 | *Theristus* sp. 2 | 0.04 | 0.03 |
| 80 | *Gonionchus* sp. | 0.11 | 0.07 |
| 81 | *Echinotheristus* sp. | 0.20 | 0.14 |
| 82 | *Amphimonhystera* sp. | 0.04 | 0.09 |
| 83 | *Sphaerotheristus* sp. | – | 0.01 |
| 84 | Xyalidae gen. sp. | 0.04 | 0.07 |
| | **Family Monhysteridae** | | |
| 85 | Monhysteridae gen. sp. | 0.04 | 0.01 |
| | **Family Siphonolaimidae** | | |
| 86 | *Siphonolaimus* sp. | 0.12 | 0.17 |
| | **Family Linhomoeidae** | | |
| 87 | *Disconema suecicum* | 0.12 | 0.07 |
| 88 | Linchomoeidae gen. spp. | 0.37 | 0.20 |

| | Classification and identity | Flotation with $MgCl_2$ | Flotation with $H_2O$ |
|---|---|---|---|
| | **ORDER ARAEOLAIMIDA** | | |
| | **Family Comesomatidae** | | |
| 89 | *Sabatieria* sp. | 9.21 | 3.56 |
| 90 | *Paramesonchium* sp. n. | 0.12 | – |
| | **Family Axonolaimidae** | | |
| 91 | *Axonolaimus helgolandicus* | 0.27 | 0.10 |
| 92 | *Ascolaimus elongatus* | 0.04 | – |
| 93 | *Odontophora villoti* | 1.35 | 2.85 |
| | **Family Diplopeltidae** | | |
| 94 | *Southerniella* sp. | 0.08 | 0.07 |
| 95 | *Diplopeltula* sp. | 0.12 | 0.05 |
| | **ORDER PLECTIDA** | | |
| | **Family Leptolaimidae** | | |
| 96 | *Leptolaimus pellucidus* | 0.08 | 0.05 |
| 97 | *Leptolaimus* sp. | 0.04 | 0.03 |
| 98 | *Leptolaimus* sp. n. | 0.40 | 0.23 |
| 99 | *Manunema* sp. | 0.04 | 0.01 |
| | **Family Camacolaimidae** | | |
| 100 | *Stephanolaimus elegans* | 0.12 | 0.03 |
| 101 | *Dagda bipapillata* | – | 0.01 |
| 102 | *Deontolaimus* sp. | 0.04 | 0.04 |
| 103 | *Onchium* sp. | – | 0.01 |
| | **Family Diplopeltoididae** | | |
| 104 | *Diplopeltoides* sp. n. 1 | 0.04 | – |
| 105 | *Diplopeltoides* sp. n. 2 | – | 0.01 |
| | **Family Tarvaiidae** | | |
| 106 | *Tarvaia* sp. | – | 0.03 |
| | **Family Tubolaimoididae** | | |
| 107 | *Tubolaimoides* sp. | – | 0.01 |
| | | | |
| 108 | Nematoda indet. | 0.04 | 0.14 |

**Supplementary Table 3.** Taxonomic composition and relative abundance (% of the total number of specimens) of nematode species in Telekabeln site.

| | Classification and identity | Siphoning | Flotation with $H_2O$ |
|---|---|---|---|
| | **ORDER ENOPLIDA** | | |
| | **Family Thoracostomopsidae** | | |
| 1 | Enoplolaiminae gen. sp. | 3.45 | 0.08 |
| | **Family Anticomidae** | | |
| 2 | *Anticoma* sp. | 0.29 | 0.04 |
| | **Family Oncholaimidae** | | |
| 3 | *Viscosia* sp. 1 | 3.16 | 0.97 |
| 4 | *Viscosia* sp. 2 | 6.66 | 0.93 |
| 5 | *Viscosia* sp. 3 | 0.19 | 0.04 |
| 6 | Oncholaimidae gen. sp. | 0.05 | – |
| | **Family Enchelidiidae** | | |
| 7 | *Symplocostoma* sp. | 0.10 | 0.04 |
| 8 | *Polygastrophora* sp. | 1.05 | 0.04 |
| | **Family Ironidae** | | |
| 9 | *Thalassironus* sp. | 0.34 | 0.43 |
| | **Family Oxystominidae** | | |
| 10 | *Oxystomina* sp. 1 | 0.24 | 0.23 |
| 11 | *Oxystomina* sp. 2 | 1.34 | 0.12 |
| 12 | *Oxystomina* sp. 3 | 0.05 | 0.04 |
| 13 | *Lithinium* sp. | 0.05 | – |
| 14 | *Thalassoalaimus* sp. 1 | 0.19 | 0.66 |
| 15 | *Thalassoalaimus* sp. 2 | 0.05 | 0.16 |
| 16 | *Thalassoalaimus* sp. 3 | – | 0.04 |
| 17 | *Halalaimus* sp. 1 | 2.59 | 1.48 |
| 18 | *Halalaimus* sp. 2 | 1.15 | 0.74 |
| 19 | *Halalaimus* sp. 3 | 0.43 | 0.97 |
| 20 | *Halalaimus* sp. 4 | 1.49 | 1.13 |
| 21 | *Halalaimus* sp. 5 | 0.10 | 0.16 |
| 22 | *Halalaimus* sp. 6 | 0.10 | – |
| 23 | *Halalaimus* sp. 7 | – | 0.08 |
| 24 | *Halalaimus* sp. 8 | 0.86 | 1.13 |
| 25 | *Wieseria* sp. | 0.29 | 2.95 |
| 26 | Oxystominidae gen. sp. | 0.81 | 0.93 |
| | **Family Tripyloididae** | | |
| 27 | *Tripyloides* sp. | 0.43 | 0.04 |
| 28 | *Bathylaimus* sp. | 1.49 | 0.23 |
| | **Family Trefusiidae** | | |

| | Classification and identity | Siphoning | Flotation with $H_2O$ |
|---|---|---|---|
| 29 | *Trefusia* sp. | – | 1.44 |
| | **ORDER TRIPLONCHIDA** | | |
| | **Family Pandolaimidae** | | |
| 30 | *Pandolaimus* sp. | 0.48 | 1.63 |
| | **Family Rhabdodemaniidae** | | |
| 31 | *Rhabdodemania* sp. | 0.72 | 0.19 |
| | **ORDER DESMOSCOLECIDA** | | |
| | **Family Desmoscolecidae** | | |
| 32 | *Desmoscolex* sp. | 0.67 | 1.71 |
| 33 | *Tricoma (Quadricoma)* sp. 1 | 2.83 | 16.11 |
| 34 | *Tricoma (Quadricoma)* sp. 2 | 5.56 | 1.28 |
| 35 | *Tricoma (Quadricoma)* sp. 3 | 0.19 | 0.39 |
| | **Family Cyartonematidae** | | |
| 36 | *Cyartonema* sp. | – | 1.05 |
| | **ORDER CHROMADORIDA** | | |
| | **Family Chromadoridae** | | |
| 37 | *Chromadorita* sp. 1 | 0.58 | 0.47 |
| 38 | *Chromadorita* sp. 2 | 0.14 | – |
| 39 | *Neochromadora* sp. 1 | 0.05 | – |
| 40 | *Neochromadora* sp. 2 | 0.38 | 0.04 |
| 41 | *Actinonema* sp. | 2.25 | 0.74 |
| 42 | *Trochamus* sp. | 1.49 | 0.78 |
| 43 | *Chromadoridae* gen. spp. | 0.14 | 0.74 |
| 44 | *Acantholaimus* sp. | – | 0.08 |
| | **Family Cyatholaimidae** | | |
| 45 | *Paracanthonchus* cf. *longicaudatus* | 0.05 | 2.06 |
| 46 | *Marylinnia* sp | 6.28 | 6.29 |
| 47 | *Paralongicyatholaimus* sp. | 0.10 | 1.01 |
| 48 | *Craspodema* sp. | 0.19 | – |
| | **Family Selachinematidae** | | |
| 49 | *Halichoanolaimus* sp. | 0.10 | – |
| 50 | *Cheironchus* sp. | – | 0.12 |
| 51 | *Synonchiella* sp. | – | 0.19 |
| 52 | *Choniolaimus* sp. 1 | 0.24 | 1.55 |
| 53 | *Choniolaimus* sp. 2 | – | 0.04 |
| | **ORDER DESMODORIDA** | | |
| | **Family Desmodoridae** | | |
| 54 | *Desmodora* cf. *communis* | 0.05 | – |
| 55 | *Desmodora pontica* | 0.62 | 0.19 |
| 56 | *Desmodorella tenuispiculum* | 3.60 | 3.18 |
| 57 | *Chromaspirina* sp. | – | 0.12 |

| | Classification and identity | Siphoning | Flotation with $H_2O$ |
|---|---|---|---|
| 58 | *Spirinia* sp. | 0.10 | 0.39 |
| | **Family Microlaimidae** | | |
| 59 | *Microlaimus* sp. 1 | 0.24 | 0.85 |
| 60 | *Microlaimus* sp. 2 | – | 2.02 |
| 61 | *Microlaimus* sp. 3 | – | 0.74 |
| 62 | *Bolbolaimus* sp | – | 0.04 |
| | **Family Monoposthiidae** | | |
| 63 | Monoposthiidae gen. sp. | 0.05 | – |
| | **Family Richtersiidae** | | |
| 64 | *Richtersia* sp. | 2.01 | 5.90 |
| | **ORDER MONHYSTERIDA** | | |
| | **Family Xyalidae** | | |
| 65 | *Daptonema* sp. | 1.58 | 0.08 |
| 66 | *Metadesmolaimus* sp. | 3.07 | 4.62 |
| 67 | *Gnomoxyala* sp. | – | 0.19 |
| 68 | Xyalidae gen. sp. | 0.14 | 0.97 |
| | **Family Sphaerolaimidae** | | |
| 69 | *Sphaerolaimus* sp. 1 | 1.77 | 1.82 |
| 70 | *Sphaerolaimus* sp. 2 | 0.58 | 0.12 |
| 71 | *Parasphaerolaimus* sp. | 0.38 | 0.47 |
| | **Family Siphonolaimidae** | | |
| 72 | *Siphonolaimus* sp. | – | 0.08 |
| | **Family Linhomoeidae** | | |
| 73 | *Disconema* sp. | 0.10 | – |
| 74 | *Eleutherolaimus* sp. | 0.58 | 1.32 |
| 75 | *Linchomoeidae* gen. sp. | 2.11 | 0.16 |
| 76 | *Monhysteroides* sp. | – | 0.97 |
| 77 | *Terschellingia* sp. 1 | – | 1.71 |
| 78 | *Terschellingia* sp. 2 | 0.05 | 0.08 |
| | **ORDER ARAEOLAIMIDA** | | |
| | **Family Comesomatidae** | | |
| 79 | *Sabatieria* spp. | 18.31 | 8.46 |
| 80 | *Setosabatieria* sp. | 5.61 | 2.80 |
| 81 | *Dorylaimopsis* sp. | 0.86 | 3.69 |
| 82 | *Laimella* sp. | – | 0.04 |
| | **Family Axonolaimidae** | | |
| 83 | *Axonolaimus* sp. | 5.42 | 0.85 |
| 84 | *Odontophora* sp. | 0.05 | 0.08 |
| | **Family Diplopeltidae** | | |
| 85 | *Araeolaimus spinosus* | 0.96 | 0.08 |
| 86 | *Campylaimus rimatus* | 0.14 | 0.04 |

| | Classification and identity | Siphoning | Flotation with $H_2O$ |
|---|---|---|---|
| 87 | *Campylaimus amphidialis* | – | 0.31 |
| 88 | *Campylaimus tkatchevi* | 0.05 | 0.39 |
| 89 | *Campylaimus orientalis* | – | 0.31 |
| 90 | *Campylaimus* sp. | 0.05 | 0.19 |
| 91 | *Pararaeolaimus* sp. | 0.19 | 0.04 |
| 92 | *Diplopeltula* sp. | 0.10 | 0.04 |
| | **ORDER PLECTIDA** | | |
| | **Family Leptolaimidae** | | |
| 93 | *Leptolaimus danicus* | 0.05 | 0.43 |
| 94 | *Leptolaimus septempapillatus* | – | 0.19 |
| 95 | *Leptolaimus venustus* | – | 0.04 |
| 96 | *Leptolaimus* sp. | – | 0.08 |
| 97 | *Antomicron lorenzeni* | – | 0.08 |
| 98 | *Antomicron quindecimpapillatus* | – | 0.04 |
| 99 | *Leptolaimoides* sp. | 0.05 | – |
| | **Family Camacolaimidae** | | |
| 100 | *Alaimella* sp. | 0.34 | 0.19 |
| 101 | Camacolaimidae gen. sp. | 0.05 | 0.12 |
| 102 | *Deontolaimus catalinae* | – | 0.04 |
| 103 | *Deontolaimus* sp. | – | 0.08 |
| | **Family Rhadinematidae** | | |
| 104 | *Rhadinema timmi* | 0.05 | 0.04 |
| | **Family Ceramonematidae** | | |
| 105 | *Pselionema* sp. | 0.62 | 0.82 |
| 106 | *Dasynemoides* sp. | – | 0.04 |
| | **Family Diplopeltoididae** | | |
| 107 | *Diplopeltoides ornatus* | 0.05 | 0.04 |
| 108 | *Diplopeltoides linkei* | – | 0.04 |
| 109 | *Diplopeltoides nudus* | – | 0.08 |
| 110 | *Diplopeltoides asetosus* | – | 0.04 |
| 111 | *Diplopeltoides* sp. n. | – | 1.67 |
| | **Family Aegialoalaimidae** | | |
| 112 | *Aegialoalaimus* sp. | 0.29 | 0.16 |
| | **Family Paramicrolaimidae** | | |
| 113 | *Paramicrolaimus* sp. | – | 0.04 |
| | | | |
| 114 | Nematoda indet. | 0.38 | 0.89 |

**Supplementary Table 4.** Results of alignment-based taxonomy assignment using BLASTN 2.5.0+ against the nucleotide collection of the NCBI database.

| OTU ID | Best hit | identity % | cover % | Family identification |
|---|---|---|---|---|
| HE1.SSU848264 | *Paracyatholaimus intermedius* | 92 | 100 | Cyatholaimidae |
| HE1.SSU850987 | *Eubostrichus topiarius* | 89 | 100 | unassigned |
| HE1.SSU856624 | *Thalassoalaimus* sp. | 94 | 88 | unassigned |
| HE1.SSU856738 | *Calomicrolaimus parahonestus* | 92 | 100 | Microlaimidae |
| HE1.SSU858060 | *Dichromadora* sp. | 94 | 98 | unassigned |
| HE1.SSU867071 | *Synonchiella* sp. | 95 | 93 | unassigned |
| HE2.SSU637072 | *Campydora* sp. | 90 | 99 | unassigned |
| HE2.SSU637135 | *Neochromadora* sp. | 90 | 99 | unassigned |
| HE2.SSU644966 | *Paracanthonchus* sp. | 91 | 100 | Cyatholaimidae |
| HE2.SSU654005 | *Rhabdodemania* sp. | 93 | 84 | unassigned |
| HE2.SSU655107 | *Theristus* sp. | 88 | 99 | unassigned |
| HE2.SSU659506 | *Neochromadora* sp. | 97 | 100 | Chromadoridae |
| HE3.SSU110275 | *Enoplus communis* sp. | 98 | 100 | Enoplidae |
| HE3.SSU117415 | *Paracyatholaimus* sp. | 91 | 100 | Cyatholaimidae |
| HE3.SSU118424 | *Oxystomina* sp. | 89 | 98 | unassigned |
| HE3.SSU124287 | *Enoploides brunettii* | 98 | 100 | Thoracostomopsidae |
| HE3.SSU124998 | *Astomonema* sp. | 85 | 100 | unassigned |
| HE4.SSU913283 | *Anaplectus* sp. | 89 | 100 | unassigned |
| HE5.SSU181724 | *Neochromadora* sp. | 87 | 100 | unassigned |
| HE5.SSU188855 | *Symplocostoma* sp. | 99 | 98 | unassigned |
| HE6.SSU355777 | *Achromadora terricola* | 93 | 100 | Achromadoridae |
| HE6.SSU358048 | *Neochromadora* sp. | 92 | 100 | Chromadoridae |
| HE6.SSU360897 | *Desmodora ovigera* | 93 | 92 | unassigned |
| HE6.SSU361449 | *Syringolaimus* sp. | 94 | 100 | Ironidae |
| HE6.SSU365256 | *Desmodora ovigera* | 97 | 76 | unassigned |
| HE6.SSU368318 | *Prochaetosoma* sp. | 93 | 92 | unassigned |
| HE6.SSU370544 | *Daptonema normandicum* | 97 | 99 | unassigned |
| HE6.SSU378839 | *Calomicrolaimus parahonestus* | 82 | 99 | unassigned |
| HE6.SSU383414 | *Sabatieria* sp. | 94 | 100 | Comesomatidae |
| HE6.SSU383888 | *Neochromadora* sp. | 90 | 100 | Chromadoridae |
| HE7.SSU232624 | *Leptolaimus* sp. | 91 | 98 | unassigned |
| HE7.SSU256492 | *Neochromadora* sp. | 95 | 100 | Chromadoridae |
| HE8.SSU829972 | *Anticoma* sp. | 98 | 88 | unassigned |
| HE8.SSU843570 | *Neochromadora* sp. | 91 | 100 | Chromadoridae |
| HE9.SSU305678 | *Theristus* sp. | 88 | 100 | unassigned |
| HF1.SSU759758 | *Camacolaimus* sp. | 98 | 98 | unassigned |
| HF1.SSU763392 | *Praeacanthonchus punctatus* | 94 | 100 | Cyatholaimidae |

| OTU ID | Best hit | identity % | cover % | Family identification |
|---|---|---|---|---|
| HF1.SSU764346 | *Paracanthonchus* sp. | 93 | 100 | Cyatholaimidae |
| HF1.SSU774294 | *Mermis nigrescens* | 95 | 100 | Mermithidae |
| HF1.SSU779114 | *Odontophora rectangula* | 99 | 98 | unassigned |
| HF1.SSU780927 | *Desmolaimus* sp. | 95 | 76 | unassigned |
| HF2.SSU192072 | *Chromadorina* sp. | 100 | 98 | unassigned |
| HF2.SSU204352 | *Anaplectus* sp. | 87 | 100 | unassigned |
| HF2.SSU205129 | *Neochromadora* | 89 | 100 | unassigned |
| HF2.SSU208147 | *Synonchiella* or *Halichoanolaimus* | 98 | 98 | unassigned |
| HF2.SSU210357 | *Calyptronema* sp. | 92 | 100 | Enchelidiidae |
| HF3.SSU989895 | *Setostephanolaimus* sp. | 92 | 98 | unassigned |
| HF3.SSU990962 | *Neochromadora* sp. | 90 | 100 | Chromadoridae |
| HF4.SSU606153 | *Neochromadora* sp. | 99 | 100 | Chromadoridae |
| HF4.SSU614317 | *Sabatieria* sp. | 92 | 100 | Comesomatidae |
| HF4.SSU619471 | *Molgolaimus demani* | 90 | 100 | Microlaimidae |
| HF4.SSU620879 | *Punctodora ratzeburgensis* | 92 | 100 | Chromadoridae |
| HF4.SSU622464 | *Plectus aquatilis* | 95 | 100 | Plectidae |
| HF4.SSU624085 | *Prochaetosoma* sp. | 90 | 93 | unassigned |
| HF4.SSU625424 | *Sabatieria* sp. | 87 | 100 | unassigned |
| HF4.SSU628562 | *Stilbonema majum* | 82 | 100 | unassigned |
| HF4.SSU631524 | *Leptolaimus* sp. | 97 | 92 | unassigned |
| HF4.SSU632264 | *Diplopeltula* sp. | 95 | 98 | unassigned |
| HF4.SSU635045 | *Chromadorina* sp. | 96 | 98 | unassigned |
| HF5.SSU991188 | *Oncholaimidae indet.* | 93 | 88 | unassigned |
| HF5.SSU995414 | *Rhabdolaimus aquaticus* | 90 | 100 | Rhabdolaimidae |
| HF6.SSU329881 | *Desmodora ovigera* | 99 | 93 | unassigned |
| HF6.SSU338435 | *Neochromadora* sp. | 89 | 91 | unassigned |
| HF6.SSU338739 | *Astomonema* sp. | 89 | 100 | unassigned |
| HF7.SSU385021 | *Neochromadora* sp. | 90 | 100 | Chromadoridae |
| HF7.SSU390110 | *Desmolaimus* sp. | 97 | 98 | unassigned |
| HF7.SSU398053 | *Camacolaimus* sp. | 94 | 77 | unassigned |
| HF7.SSU407024 | *Achromadora ruricola* | 95 | 95 | unassigned |
| HF7.SSU407761 | *Desmolaimus* sp. | 94 | 98 | unassigned |
| HF7.SSU409331 | *Pomponema* sp. | 94 | 76 | unassigned |
| HF8.SSU795426 | *Cyatholaimus* sp. | 90 | 100 | Cyatholaimidae |
| HF9.SSU14048 | *Calomicrolaimus* sp. | 93 | 76 | unassigned |
| HF9.SSU14296 | *Pomponema* sp. | 85 | 90 | unassigned |
| HF9.SSU17250 | *Enoploides brunettii* | 91 | 100 | Thoracostomopsidae |
| HF9.SSU17844 | *Pomponema* sp. | 94 | 89 | unassigned |
| HF9.SSU18227 | *Neochromadora* sp. | 91 | 100 | Chromadoridae |
| HF9.SSU19963 | *Desmoscolex* sp. | 87 | 100 | unassigned |
| HF9.SSU20251 | *Calomicrolaimus parahonestus* | 98 | 100 | Microlaimidae |

| OTU ID | Best hit | identity % | cover % | Family identification |
|---|---|---|---|---|
| HF9.SSU22538 | *Mermis nigrescens* | 90 | 100 | Mermithidae |
| TF1.SSU676746 | *Haliplectus* or *Prodesmodora* | 90 | 100 | unassigned |
| TF1.SSU677162 | *Anaplectus* sp. | 89 | 100 | unassigned |
| TF1.SSU681557 | *Pseudocella* sp. | 87 | 100 | unassigned |
| TF1.SSU688192 | *Terschellingia longicaudata* | 90 | 100 | Linhomoeidae |
| TF1.SSU692690 | *Synonchiella* or *Halichoanolaimus* | 91 | 100 | Selachinematidae |
| TF1.SSU694267 | *Desmoscolex* sp. | 89 | 98 | unassigned |
| TF1.SSU694751 | *Neochromadora* sp. | 90 | 100 | Chromadoridae |
| TF1.SSU698227 | *Teratocephalus lirellus* | 91 | 100 | Teratocephalidae |
| TF1.SSU700188 | *Terschellingia longicaudata* | 96 | 100 | Linhomoeidae |
| TF1.SSU703579 | *Astomonema* sp. | 90 | 100 | Siphonolaimidae |
| TF1.SSU710679 | *Cyatholaimus* sp. | 99 | 100 | Cyatholaimidae |
| TF1.SSU734804 | *Astomonema* sp. | 90 | 100 | Siphonolaimidae |
| TF3.SSU956521 | *Sabatieria* sp. | 90 | 100 | Comesomatidae |
| TF3.SSU960449 | *Desmoscolex* sp. | 88 | 98 | unassigned |
| TF3.SSU966338 | *Theristus* sp. | 91 | 99 | unassigned |
| TF4.SSU144249 | *Paracanthonchus* sp. | 97 | 99 | unassigned |
| TF4.SSU150234 | *Metachromadora* sp. | 91 | 100 | Desmodoridae |
| TF5.SSU410031 | *Praeacanthonchus* sp. | 88 | 100 | unassigned |
| TF5.SSU419519 | *Desmoscolex* sp. | 92 | 98 | unassigned |
| TF5.SSU430294 | *Theristus* sp. | 97 | 99 | unassigned |
| TF5.SSU437076 | *Setosabatieria hilarula* | 92 | 100 | Comesomatidae |
| TF5.SSU444034 | *Tripylella* sp. | 93 | 98 | unassigned |
| TF5.SSU446087 | *Tarvaia* sp. | 91 | 97 | unassigned |
| TF5.SSU453472 | *Tripylina* sp. | 89 | 100 | unassigned |
| TF5.SSU457543 | *Oxystomina* sp. | 92 | 98 | unassigned |
| TF5.SSU459305 | *Viscosia viscosa* | 92 | 100 | Oncholaimidae |
| TF5.SSU466315 | *Daptonema* sp. | 96 | 99 | unassigned |
| TF6.SSU33463 | *Oxystomina* sp. | 96 | 98 | unassigned |
| TF6.SSU33935 | *Desmolaimus* sp. | 93 | 98 | unassigned |
| TF6.SSU36442 | *Desmoscolex* sp. | 89 | 98 | unassigned |
| TF6.SSU37421 | *Prochaetosoma* sp. | 93 | 92 | unassigned |
| TF6.SSU41803 | *Daptonema* sp. | 88 | 98 | unassigned |
| TF6.SSU47996 | *Viscosia* sp. | 92 | 100 | Oncholaimidae |
| TF6.SSU48167 | *Sabatieria* sp. | 99 | 100 | Comesomatidae |
| TF6.SSU53456 | *Viscosia* sp. | 99 | 98 | unassigned |
| TF6.SSU54250 | *Calomicrolaimus parahonestus* | 93 | 100 | Microlaimidae |
| TF6.SSU58877 | *Tarvaia* or *Desmoscolex* | 91 | 97 | unassigned |
| TF6.SSU74955 | *Paracanthonchus* sp. | 93 | 100 | Cyatholaimidae |
| TF6.SSU82210 | *Stilbonema* sp. | 85 | 100 | unassigned |
| TF6.SSU84268 | *Spectatus spectatus* | 82 | 90 | unassigned |

| OTU ID | Best hit | identity % | cover % | Family identification |
|---|---|---|---|---|
| TF6.SSU98667 | *Campydora* sp. | 86 | 92 | unassigned |
| TS1.SSU270885 | *Tarvaia* sp. | 88 | 97 | unassigned |
| TS1.SSU284163 | *Desmoscolex* sp. | 87 | 98 | unassigned |
| TS2.SSU821962 | *Bathylaimus* sp. | 91 | 100 | Tripyloididae |
| TS2.SSU823349 | *Bathylaimus* sp. | 90 | 88 | unassigned |
| TS3.SSU475561 | *Astomonema* sp. | 91 | 100 | Siphonolaimidae |
| TS3.SSU489684 | *Desmoscolex* sp. | 95 | 97 | unassigned |
| TS3.SSU503133 | *Tripyloides* sp. | 94 | 100 | Tripyloididae |
| TS3.SSU508400 | *Desmolaimus* sp. | 90 | 87 | unassigned |
| TS4.SSU543236 | *Thalassoalaimus* sp. | 90 | 88 | unassigned |
| TS4.SSU544032 | *Desmoscolex* sp. | 92 | 98 | unassigned |
| TS5.SSU874117 | *Halalaimus* sp. | 91 | 98 | unassigned |
| TS5.SSU875407 | *Sabatieria celtica* | 93 | 100 | Comesomatidae |
| TS5.SSU881546 | *Theristus* sp. | 90 | 99 | unassigned |
| TS5.SSU900338 | *Leptolaimus* sp. | 94 | 98 | unassigned |
| TS5.SSU901243 | *Dolicholaimus* sp. | 96 | 88 | unassigned |
| TS6.SSU559765 | *Tripyloides* sp. | 95 | 89 | unassigned |
| TS6.SSU570763 | *Axonolaimus* sp. | 88 | 98 | unassigned |
| TS6.SSU587229 | *Viscosia viscosa* | 99 | 100 | Oncholaimidae |
| HE6.SSU372021 | *Eumonhystera filiformis* | 91 | 99 | unassigned |

**Supplementary Table 5.** Results of alignment-based taxonomy assignment using LCAClassifier of CREST against built-in reference database.

| OTU ID | Best hit | Family identification |
|---|---|---|
| HE1.SSU848264 | Chromadorea; Chromadorida | unassigned |
| HE1.SSU850987 | Chromadorea | unassigned |
| HE1.SSU856624 | Enoplea; Enoplida; Oxystominoidea | unassigned |
| HE1.SSU856738 | Chromadorea; Desmodorida; Richtersioidea; Microlaimidae | Microlaimidae |
| HE1.SSU858060 | Chromadorea; Chromadorida | unassigned |
| HE1.SSU867071 | Chromadorea | unassigned |
| HE2.SSU637072 | Enoplea; Enoplida; Enoploidea | unassigned |
| HE2.SSU637135 | Chromadorea; Chromadorida; Chromadoridae | Chromadoridae |
| HE2.SSU644966 | Chromadorea; Chromadorida | unassigned |
| HE2.SSU654005 | Enoplea | unassigned |
| HE2.SSU655107 | Chromadorea; Unknown Chromadorea | unassigned |
| HE2.SSU659506 | Chromadorea; Chromadorida | unassigned |
| HE3.SSU110275 | Enoplea; Enoplida; Enoploidea; Enoplidae; Enoplus | Enoplidae |
| HE3.SSU117415 | Chromadorea; Chromadorida | unassigned |
| HE3.SSU118424 | Arthropoda; Crustacea; Maxillopoda; Copepoda | unassigned |
| HE3.SSU124287 | Enoplea; Enoplida; Enoploidea; Thoracostomopsidae; Enoploides | Thoracostomopsidae |
| HE3.SSU124998 | Chromadorea | unassigned |
| HE4.SSU913283 | Chromadorea | unassigned |
| HE5.SSU181724 | Chromadorea; Chromadorida | unassigned |
| HE5.SSU188855 | Enoplea; Enoplida; Oncholaimoidea | unassigned |
| HE6.SSU355777 | Chromadorea; Chromadorida | unassigned |
| HE6.SSU358048 | Chromadorea; Chromadorida | unassigned |
| HE6.SSU360897 | Chromadorea; Desmodorida | unassigned |
| HE6.SSU361449 | Enoplea; Enoplida; Ironoidea | unassigned |
| HE6.SSU365256 | Chromadorea; Desmodorida | unassigned |
| HE6.SSU368318 | Chromadorea; Desmodorida; Desmodoridae | Desmodoridae |
| HE6.SSU370544 | Chromadorea; Monhysterida; Xyalidae; Daptonema | Xyalidae |
| HE6.SSU378839 | Chromadorea | unassigned |
| HE6.SSU383414 | Chromadorea; Monhysterida | unassigned |
| HE6.SSU383888 | Chromadorea | unassigned |
| HE7.SSU232624 | Chromadorea; Araeolaimida; Leptolaimoidea | unassigned |
| HE7.SSU256492 | Chromadorea; Chromadorida | unassigned |
| HE8.SSU829972 | Enoplea | unassigned |
| HE8.SSU843570 | Chromadorea; Chromadorida | unassigned |
| HE9.SSU305678 | Chromadorea | unassigned |
| HF1.SSU759758 | Chromadorea; Araeolaimida; Leptolaimoidea; Leptolaimidae | Leptolaimidae |
| HF1.SSU763392 | Chromadorea; Chromadorida | unassigned |

| OTU ID | Best hit | Family identification |
|---|---|---|
| HF1.SSU764346 | Chromadorea; Chromadorida | unassigned |
| HF1.SSU774294 | Enoplea; Mononchida; Mononchina; Anatonchoidea | unassigned |
| HF1.SSU779114 | Chromadorea; Araeolaimida; Axonolaimoidea; Axonolaimidae; *Odontophora rectangula* | Axonolaimidae |
| HF1.SSU780927 | Chromadorea; Araeolaimida | unassigned |
| HF2.SSU192072 | Chromadorea; Chromadorida; Chromadoridae; *Chromadora nudicapitata* | Chromadoridae |
| HF2.SSU204352 | Chromadorea; Araeolaimida; Leptolaimoidea | unassigned |
| HF2.SSU205129 | Chromadorea; Chromadorida | unassigned |
| HF2.SSU208147 | Nematoda | unassigned |
| HF2.SSU210357 | Enoplea; Enoplida; Oncholaimoidea | unassigned |
| HF3.SSU989895 | Chromadorea; Araeolaimida; Plectoidea | unassigned |
| HF3.SSU990962 | Chromadorea | unassigned |
| HF4.SSU606153 | Chromadorea; Chromadorida; Chromadoridae | Chromadoridae |
| HF4.SSU614317 | Chromadorea; Monhysterida; Comesomatidae; *Sabatieria* sp. | Comesomatidae |
| HF4.SSU619471 | Chromadorea; Desmodorida; Richtersioidea | unassigned |
| HF4.SSU620879 | Chromadorea; Chromadorida | unassigned |
| HF4.SSU622464 | Nematoda | unassigned |
| HF4.SSU624085 | Chromadorea; Desmodorida | unassigned |
| HF4.SSU625424 | Chromadorea | unassigned |
| HF4.SSU628562 | Nematoda | unassigned |
| HF4.SSU631524 | Chromadorea; Araeolaimida; Leptolaimoidea; Leptolaimidae; *Leptolaimus* sp. | Leptolaimidae |
| HF4.SSU632264 | Chromadorea | unassigned |
| HF4.SSU635045 | Chromadorea; Chromadorida | unassigned |
| HF5.SSU991188 | Enoplea; Enoplida; Oncholaimoidea | unassigned |
| HF5.SSU995414 | Enoplea; Enoplida; Ironoidea; Ironidae | Ironidae |
| HF6.SSU329881 | Chromadorea; Desmodorida; Desmodoridae; Desmodorinae | Desmodoridae |
| HF6.SSU338435 | Chromadorea | unassigned |
| HF6.SSU338739 | Chromadorea | unassigned |
| HF7.SSU385021 | Chromadorea | unassigned |
| HF7.SSU390110 | Chromadorea; Araeolaimida | unassigned |
| HF7.SSU398053 | Chromadorea; Araeolaimida; Leptolaimoidea | unassigned |
| HF7.SSU407024 | Chromadorea; Chromadorida | unassigned |
| HF7.SSU407761 | Chromadorea; Araeolaimida | unassigned |
| HF7.SSU409331 | Enoplea | unassigned |
| HF8.SSU795426 | Chromadorea | unassigned |
| HF9.SSU14048 | Chromadorea; Desmodorida; Richtersioidea | unassigned |
| HF9.SSU14296 | Nematoda | unassigned |
| HF9.SSU17250 | Enoplea; Enoplida; Enoploidea; Thoracostomopsidae | Thoracostomopsidae |
| HF9.SSU17844 | Chromadorea | unassigned |
| HF9.SSU18227 | Chromadorea; Chromadorida | unassigned |
| HF9.SSU19963 | Chromadorea | unassigned |

| OTU ID | Best hit | Family identification |
|---|---|---|
| HF9.SSU20251 | Chromadorea; Desmodorida; Richtersioidea; Microlaimidae; *Calomicrolaimus* sp. | Microlaimidae |
| HF9.SSU22538 | Enoplea; Mononchida; Mononchina; Anatonchoidea | unassigned |
| TF1.SSU676746 | Chromadorea | unassigned |
| TF1.SSU677162 | Chromadorea; Araeolaimida | unassigned |
| TF1.SSU681557 | Enoplea | unassigned |
| TF1.SSU688192 | Chromadorea; Monhysterida | unassigned |
| TF1.SSU692690 | Chromadorea | unassigned |
| TF1.SSU694267 | Chromadorea | unassigned |
| TF1.SSU694751 | Chromadorea | unassigned |
| TF1.SSU698227 | Chromadorea; Rhabditida; Teratocephaloidea | unassigned |
| TF1.SSU700188 | Chromadorea; Monhysterida; Monhysterida incertae sedis | unassigned |
| TF1.SSU703579 | Chromadorea | unassigned |
| TF1.SSU710679 | Chromadorea; Chromadorida; Cyatholaimidae; *Paracanthonchus* sp. | Cyatholaimidae |
| TF1.SSU734804 | Chromadorea | unassigned |
| TF3.SSU956521 | Chromadorea; Monhysterida | unassigned |
| TF3.SSU960449 | Chromadorea | unassigned |
| TF3.SSU966338 | Chromadorea | unassigned |
| TF4.SSU144249 | Chromadorea; Chromadorida | unassigned |
| TF4.SSU150234 | Chromadorea; Desmodorida | unassigned |
| TF5.SSU410031 | Nematoda | unassigned |
| TF5.SSU419519 | Chromadorea | unassigned |
| TF5.SSU430294 | Chromadorea; Monhysterida | unassigned |
| TF5.SSU437076 | Chromadorea; Monhysterida; Comesomatidae; *Setosabatieria hilarula* | Comesomatidae |
| TF5.SSU444034 | Enoplea; Enoplida; Tripyloidea | unassigned |
| TF5.SSU446087 | Nematoda | unassigned |
| TF5.SSU453472 | Enoplea | unassigned |
| TF5.SSU457543 | Chromadorea | unassigned |
| TF5.SSU459305 | Enoplea; Enoplida; Oncholaimoidea; Oncholaimidae; *Oncholaimus* sp. | Oncholaimidae |
| TF5.SSU466315 | Chromadorea; Monhysterida | unassigned |
| TF6.SSU33463 | Enoplea | unassigned |
| TF6.SSU33935 | Chromadorea; Araeolaimida | unassigned |
| TF6.SSU36442 | Chromadorea | unassigned |
| TF6.SSU37421 | Chromadorea; Desmodorida; Desmodoridae; Spiriniinae | Desmodoridae |
| TF6.SSU41803 | Chromadorea; Monhysterida; Xyalidae | Xyalidae |
| TF6.SSU47996 | Enoplea; Enoplida; Oncholaimoidea; Enchelidiidae; *Pareurystomina* sp. | Enchelidiidae |
| TF6.SSU48167 | Chromadorea; Monhysterida; Comesomatidae; *Sabatieria* sp. | Comesomatidae |
| TF6.SSU53456 | Enoplea; Enoplida; Oncholaimoidea; Oncholaimidae; *Viscosia* sp. | Oncholaimidae |
| TF6.SSU54250 | Chromadorea; Desmodorida; Richtersioidea | unassigned |
| TF6.SSU58877 | Chromadorea | unassigned |
| TF6.SSU74955 | Chromadorea | unassigned |

| OTU ID | Best hit | Family identification |
|---|---|---|
| TF6.SSU82210 | Chromadorea | unassigned |
| TF6.SSU84268 | Chromadorea | unassigned |
| TF6.SSU98667 | Nematoda | unassigned |
| TS1.SSU270885 | Chromadorea | unassigned |
| TS1.SSU284163 | Annelida; Polychaeta; Scolecida | unassigned |
| TS2.SSU821962 | Enoplea; Enoplida; Tripyloidoidea | unassigned |
| TS2.SSU823349 | Enoplea; Enoplida; Tripyloidoidea | unassigned |
| TS3.SSU475561 | Chromadorea | unassigned |
| TS3.SSU489684 | Chromadorea | unassigned |
| TS3.SSU503133 | Enoplea; Enoplida; Tripyloidoidea | unassigned |
| TS3.SSU508400 | Chromadorea | unassigned |
| TS4.SSU543236 | Enoplea | unassigned |
| TS4.SSU544032 | Chromadorea | unassigned |
| TS5.SSU874117 | Enoplea | unassigned |
| TS5.SSU875407 | Chromadorea; Monhysterida; Comesomatidae; *Setosabatieria hilarula* | Comesomatidae |
| TS5.SSU881546 | Chromadorea | unassigned |
| TS5.SSU900338 | Chromadorea; Araeolaimida; Leptolaimoidea | unassigned |
| TS5.SSU901243 | Enoplea | unassigned |
| TS6.SSU559765 | Enoplea; Enoplida; Tripyloidoidea | unassigned |
| TS6.SSU570763 | Chromadorea; Araeolaimida | unassigned |
| TS6.SSU587229 | Enoplea; Enoplida; Oncholaimoidea; Oncholaimidae | Oncholaimidae |
| HE6.SSU372021 | Chromadorea; Monhysterida | unassigned |

**Supplementary Table 6.** Results of tree-based taxonomy assignment using complete and partitioned query dataset.

| OTU ID | Family identification (complete) | bootstrap % | Family identification (partitioned) | bootstrap % |
|---|---|---|---|---|
| HE1.SSU848264 | unassigned | – | unassigned | – |
| HE1.SSU850987 | unassigned | – | unassigned | – |
| HE1.SSU856624 | Oxystominidae | 90 | Oxystominidae | 93 |
| HE1.SSU856738 | unassigned | – | unassigned | – |
| HE1.SSU858060 | unassigned | – | Chromadoridae | 92 |
| HE1.SSU867071 | Selachinematidae | 86 | unassigned | – |
| HE2.SSU637072 | unassigned | – | unassigned | – |
| HE2.SSU637135 | unassigned | – | unassigned | – |
| HE2.SSU644966 | unassigned | – | unassigned | – |
| HE2.SSU654005 | Rhabdodemaniidae | 100 | Rhabdodemaniidae | 100 |
| HE2.SSU655107 | Xyalidae | 70 | Xyalidae | 76 |
| HE2.SSU659506 | unassigned | – | Chromadoridae | 91 |
| HE3.SSU110275 | Enoplidae | 100 | Enoplidae | 100 |
| HE3.SSU117415 | unassigned | – | unassigned | – |
| HE3.SSU118424 | unassigned | – | Oxystominidae | 72 |
| HE3.SSU124287 | Thoracostomopsidae | 84 | Thoracostomopsidae | 100 |
| HE3.SSU124998 | unassigned | – | unassigned | – |
| HE4.SSU913283 | unassigned | – | unassigned | – |
| HE5.SSU181724 | unassigned | – | unassigned | – |
| HE5.SSU188855 | Enchelidiidae | 100 | Enchelidiidae | 100 |
| HE6.SSU355777 | unassigned | – | unassigned | – |
| HE6.SSU358048 | unassigned | – | unassigned | – |
| HE6.SSU360897 | Desmodoridae | 77 | unassigned | – |
| HE6.SSU361449 | Ironidae | 90 | Ironidae | 93 |
| HE6.SSU365256 | unassigned | – | unassigned | – |
| HE6.SSU368318 | unassigned | – | unassigned | – |
| HE6.SSU370544 | Xyalidae | 70 | Xyalidae | 99 |
| HE6.SSU378839 | Microlaimidae | 99 | Microlaimidae | 98 |
| HE6.SSU383414 | unassigned | – | Comesomatidae | 90 |
| HE6.SSU383888 | unassigned | – | Chromadoridae | 86 |
| HE7.SSU232624 | Leptolaimidae | 82 | Leptolaimidae | 92 |
| HE7.SSU256492 | unassigned | – | unassigned | – |
| HE8.SSU829972 | Anticomidae | 100 | Anticomidae | 100 |
| HE8.SSU843570 | unassigned | – | unassigned | – |
| HE9.SSU305678 | Xyalidae | 70 | unassigned | – |
| HF1.SSU759758 | Camacolaimidae | 99 | Camacolaimidae | 100 |
| HF1.SSU763392 | unassigned | – | unassigned | – |

| OTU ID | Family identification (complete) | bootstrap % | Family identification (partitioned) | bootstrap % |
|---|---|---|---|---|
| HF1.SSU764346 | unassigned | – | unassigned | – |
| HF1.SSU774294 | unassigned | – | unassigned | – |
| HF1.SSU779114 | Axonolaimidae | 82 | Axonolaimidae | 75 |
| HF1.SSU780927 | unassigned | – | unassigned | – |
| HF2.SSU192072 | Chromadoridae | 91 | Chromadoridae | 100 |
| HF2.SSU204352 | Leptolaimidae | 82 | unassigned | – |
| HF2.SSU205129 | unassigned | – | unassigned | – |
| HF2.SSU208147 | Selachinematidae | 86 | Selachinematidae | 88 |
| HF2.SSU210357 | Enchelidiidae | 85 | Enchelidiidae | 85 |
| HF3.SSU989895 | unassigned | – | Camacolaimidae | 77 |
| HF3.SSU990962 | unassigned | – | unassigned | – |
| HF4.SSU606153 | unassigned | – | unassigned | – |
| HF4.SSU614317 | unassigned | – | unassigned | – |
| HF4.SSU619471 | Microlaimidae | 99 | Microlaimidae | 100 |
| HF4.SSU620879 | unassigned | – | Chromadoridae | 100 |
| HF4.SSU622464 | unassigned | – | unassigned | – |
| HF4.SSU624085 | unassigned | – | unassigned | – |
| HF4.SSU625424 | unassigned | – | Comesomatidae | 95 |
| HF4.SSU628562 | unassigned | – | unassigned | – |
| HF4.SSU631524 | Leptolaimidae | 94 | Leptolaimidae | 100 |
| HF4.SSU632264 | Diplopeltidae | 100 | Diplopeltidae | 100 |
| HF4.SSU635045 | Chromadoridae | 91 | Chromadoridae | 100 |
| HF5.SSU991188 | unassigned | – | Oncholaimidae | 100 |
| HF5.SSU995414 | unassigned | – | unassigned | – |
| HF6.SSU329881 | Desmodoridae | 77 | Desmodoridae | 85 |
| HF6.SSU338435 | unassigned | – | Chromadoridae | 97 |
| HF6.SSU338739 | unassigned | – | unassigned | – |
| HF7.SSU385021 | unassigned | – | Chromadoridae | 97 |
| HF7.SSU390110 | unassigned | – | unassigned | – |
| HF7.SSU398053 | unassigned | – | unassigned | – |
| HF7.SSU407024 | unassigned | – | unassigned | – |
| HF7.SSU407761 | unassigned | – | unassigned | – |
| HF7.SSU409331 | unassigned | – | unassigned | – |
| HF8.SSU795426 | unassigned | – | unassigned | – |
| HF9.SSU14048 | unassigned | – | unassigned | – |
| HF9.SSU14296 | unassigned | – | unassigned | – |
| HF9.SSU17250 | unassigned | – | Thoracostomopsidae | 89 |
| HF9.SSU17844 | unassigned | – | unassigned | – |
| HF9.SSU18227 | unassigned | – | unassigned | – |
| HF9.SSU19963 | unassigned | – | unassigned | – |
| HF9.SSU20251 | Microlaimidae | 99 | Microlaimidae | 97 |

| OTU ID | Family identification (complete) | bootstrap % | Family identification (partitioned) | bootstrap % |
|---|---|---|---|---|
| HF9.SSU22538 | unassigned | – | unassigned | – |
| TF1.SSU676746 | Ceramonematidae | 79 | Ceramonematidae | 80 |
| TF1.SSU677162 | unassigned | – | unassigned | – |
| TF1.SSU681557 | Oxystominidae | 90 | Oxystominidae | 89 |
| TF1.SSU688192 | Linhomoeidae | 96 | Linhomoeidae | 89 |
| TF1.SSU692690 | Selachinematidae | 86 | Selachinematidae | 95 |
| TF1.SSU694267 | unassigned | – | unassigned | – |
| TF1.SSU694751 | unassigned | – | Chromadoridae | 98 |
| TF1.SSU698227 | unassigned | – | unassigned | – |
| TF1.SSU700188 | Cyartonematidae | 94 | Cyartonematidae | 89 |
| TF1.SSU703579 | unassigned | – | unassigned | – |
| TF1.SSU710679 | unassigned | – | Cyatholaimidae | 95 |
| TF1.SSU734804 | unassigned | – | unassigned | – |
| TF3.SSU956521 | unassigned | – | unassigned | – |
| TF3.SSU960449 | unassigned | – | unassigned | – |
| TF3.SSU966338 | Xyalidae | 70 | Xyalidae | 78 |
| TF4.SSU144249 | unassigned | – | Cyatholaimidae | 74 |
| TF4.SSU150234 | unassigned | – | unassigned | – |
| TF5.SSU410031 | unassigned | – | unassigned | – |
| TF5.SSU419519 | unassigned | – | unassigned | – |
| TF5.SSU430294 | Xyalidae | 70 | Xyalidae | 78 |
| TF5.SSU437076 | Comesomatidae | 93 | unassigned | – |
| TF5.SSU444034 | Rhabdodemaniidae | 100 | Rhabdodemaniidae | 98 |
| TF5.SSU446087 | unassigned | – | Tarvaiidae | 91 |
| TF5.SSU453472 | Oxystominidae | 85 | Oxystominidae | 90 |
| TF5.SSU457543 | Oxystominidae | 85 | Oxystominidae | 94 |
| TF5.SSU459305 | Oncholaimidae | 75 | Oncholaimidae | 74 |
| TF5.SSU466315 | Xyalidae | 70 | Xyalidae | 100 |
| TF6.SSU33463 | Oxystominidae | 99 | Oxystominidae | 99 |
| TF6.SSU33935 | unassigned | – | unassigned | – |
| TF6.SSU36442 | unassigned | – | unassigned | – |
| TF6.SSU37421 | unassigned | – | unassigned | – |
| TF6.SSU41803 | unassigned | – | Xyalidae | 93 |
| TF6.SSU47996 | Enchelidiidae | 75 | Enchelidiidae | 91 |
| TF6.SSU48167 | unassigned | – | Comesomatidae | 90 |
| TF6.SSU53456 | Oncholaimidae | 71 | Oncholaimidae | 100 |
| TF6.SSU54250 | Microlaimidae | 99 | Microlaimidae | 100 |
| TF6.SSU58877 | unassigned | – | Tarvaiidae | 91 |
| TF6.SSU74955 | unassigned | – | unassigned | – |
| TF6.SSU82210 | unassigned | – | unassigned | – |
| TF6.SSU84268 | unassigned | – | unassigned | – |

| OTU ID | Family identification (complete) | bootstrap % | Family identification (partitioned) | bootstrap % |
|---|---|---|---|---|
| TF6.SSU98667 | unassigned | – | unassigned | – |
| TS1.SSU270885 | unassigned | – | unassigned | – |
| TS1.SSU284163 | unassigned | – | unassigned | – |
| TS2.SSU821962 | Tripyloididae | 100 | Tripyloididae | 100 |
| TS2.SSU823349 | Tripyloididae | 100 | Tripyloididae | 100 |
| TS3.SSU475561 | unassigned | – | unassigned | – |
| TS3.SSU489684 | Desmoscolecidae | 84 | unassigned | – |
| TS3.SSU503133 | Tripyloididae | 100 | Tripyloididae | 100 |
| TS3.SSU508400 | unassigned | – | unassigned | – |
| TS4.SSU543236 | Oxystominidae | 90 | Oxystominidae | 93 |
| TS4.SSU544032 | unassigned | – | unassigned | – |
| TS5.SSU874117 | Oxystominidae | 100 | Oxystominidae | 100 |
| TS5.SSU875407 | Comesomatidae | 93 | unassigned | – |
| TS5.SSU881546 | Xyalidae | 70 | unassigned | – |
| TS5.SSU900338 | Leptolaimidae | 94 | Leptolaimidae | 99 |
| TS5.SSU901243 | Ironidae | 81 | Ironidae | 100 |
| TS6.SSU559765 | Tripyloididae | 100 | Tripyloididae | 100 |
| TS6.SSU570763 | unassigned | – | unassigned | – |
| TS6.SSU587229 | Oncholaimidae | 75 | Oncholaimidae | 100 |
| HE6.SSU372021 | Monhysteridae | 82 | Monhysteridae | 92 |

**Supplementary Table 7.** Results of phylogeny-based taxonomy assignment using mothur and PaPaRa-based alignments.

| OTU ID | Family identification (mothur) | bootstrap % | Family identification (PaPaRa) | bootstrap % |
|---|---|---|---|---|
| HE1.SSU848264 | Cyatholaimidae | 100 | Cyatholaimidae | 100 |
| HE1.SSU850987 | unassigned | – | unassigned | – |
| HE1.SSU856624 | Oxystominidae | 100 | Oxystominidae | 100 |
| HE1.SSU856738 | Microlaimidae | 100 | Microlaimidae | 100 |
| HE1.SSU858060 | Chromadoridae | 100 | Chromadoridae | 100 |
| HE1.SSU867071 | Selachinematidae | 100 | Selachinematidae | 100 |
| HE2.SSU637072 | unassigned | – | unassigned | – |
| HE2.SSU637135 | Chromadoridae | 100 | Chromadoridae | 100 |
| HE2.SSU644966 | Cyatholaimidae | 100 | Cyatholaimidae | 100 |
| HE2.SSU654005 | Rhabdodemaniidae | NA | Rhabdodemaniidae | NA |
| HE2.SSU655107 | Xyalidae | 97 | Xyalidae | 97 |
| HE2.SSU659506 | Chromadoridae | 100 | Chromadoridae | 100 |
| HE3.SSU110275 | Enoplidae | 100 | Enoplidae | 100 |
| HE3.SSU117415 | Cyatholaimidae | 100 | Cyatholaimidae | 100 |
| HE3.SSU118424 | Oxystominidae | 100 | Oxystominidae | 100 |
| HE3.SSU124287 | Thoracostomopsidae | 100 | Thoracostomopsidae | 100 |
| HE3.SSU124998 | unassigned | – | unassigned | – |
| HE4.SSU913283 | unassigned | – | unassigned | – |
| HE5.SSU181724 | unassigned | – | unassigned | – |
| HE5.SSU188855 | Enchelidiidae | 94 | Enchelidiidae | 94 |
| HE6.SSU355777 | unassigned | – | unassigned | – |
| HE6.SSU358048 | Chromadoridae | 100 | Chromadoridae | 100 |
| HE6.SSU360897 | unassigned | – | unassigned | – |
| HE6.SSU361449 | Ironidae | 100 | Ironidae | 100 |
| HE6.SSU365256 | Desmodoridae | 72 | Desmodoridae | 72 |
| HE6.SSU368318 | unassigned | – | unassigned | – |
| HE6.SSU370544 | Xyalidae | 97 | Xyalidae | 97 |
| HE6.SSU378839 | Microlaimidae | 100 | Microlaimidae | 100 |
| HE6.SSU383414 | Comesomatidae | 99 | Comesomatidae | 99 |
| HE6.SSU383888 | Chromadoridae | 100 | Chromadoridae | 100 |
| HE7.SSU232624 | Leptolaimidae | 99 | Leptolaimidae | 99 |
| HE7.SSU256492 | Chromadoridae | 100 | Chromadoridae | 100 |
| HE8.SSU829972 | Anticomidae | 100 | Anticomidae | 100 |
| HE8.SSU843570 | Chromadoridae | 100 | Chromadoridae | 100 |
| HE9.SSU305678 | Xyalidae | 97 | Xyalidae | 97 |
| HF1.SSU759758 | Camacolaimidae | 100 | Camacolaimidae | 100 |
| HF1.SSU763392 | Cyatholaimidae | 100 | Cyatholaimidae | 100 |

| OTU ID | Family identification (mothur) | bootstrap % | Family identification (PaPaRa) | bootstrap % |
|---|---|---|---|---|
| HF1.SSU764346 | Cyatholaimidae | 100 | Cyatholaimidae | 100 |
| HF1.SSU774294 | Mermithidae | 100 | Mermithidae | 100 |
| HF1.SSU779114 | Axonolaimidae | 100 | Axonolaimidae | 100 |
| HF1.SSU780927 | unassigned | – | unassigned | – |
| HF2.SSU192072 | Chromadoridae | 100 | Chromadoridae | 100 |
| HF2.SSU204352 | unassigned | – | unassigned | – |
| HF2.SSU205129 | Chromadoridae | 100 | Chromadoridae | 100 |
| HF2.SSU208147 | Selachinematidae | 100 | Selachinematidae | 100 |
| HF2.SSU210357 | Enchelidiidae | 94 | Enchelidiidae | 94 |
| HF3.SSU989895 | Camacolaimidae | 100 | Camacolaimidae | 100 |
| HF3.SSU990962 | Chromadoridae | 100 | Chromadoridae | 100 |
| HF4.SSU606153 | Chromadoridae | 100 | Chromadoridae | 100 |
| HF4.SSU614317 | unassigned | – | unassigned | – |
| HF4.SSU619471 | Microlaimidae | 100 | Microlaimidae | 100 |
| HF4.SSU620879 | Chromadoridae | 100 | Chromadoridae | 100 |
| HF4.SSU622464 | unassigned | – | unassigned | – |
| HF4.SSU624085 | unassigned | – | unassigned | – |
| HF4.SSU625424 | Comesomatidae | 99 | Comesomatidae | 99 |
| HF4.SSU628562 | Desmodoridae | 72 | Desmodoridae | 72 |
| HF4.SSU631524 | Leptolaimidae | 99 | Leptolaimidae | 99 |
| HF4.SSU632264 | Diplopeltidae | 100 | Diplopeltidae | 100 |
| HF4.SSU635045 | Chromadoridae | 100 | Chromadoridae | 100 |
| HF5.SSU991188 | Oncholaimidae | 100 | Oncholaimidae | 100 |
| HF5.SSU995414 | Ironidae | 100 | Ironidae | 100 |
| HF6.SSU329881 | Desmodoridae | 72 | Desmodoridae | 72 |
| HF6.SSU338435 | Chromadoridae | 100 | Chromadoridae | 100 |
| HF6.SSU338739 | unassigned | – | unassigned | – |
| HF7.SSU385021 | Chromadoridae | 100 | Chromadoridae | 100 |
| HF7.SSU390110 | unassigned | – | unassigned | – |
| HF7.SSU398053 | Camacolaimidae | 100 | Camacolaimidae | 100 |
| HF7.SSU407024 | Achromadoridae | 97 | Achromadoridae | 97 |
| HF7.SSU407761 | Linhomoeidae | 99 | Linhomoeidae | 99 |
| HF7.SSU409331 | unassigned | – | unassigned | – |
| HF8.SSU795426 | unassigned | – | unassigned | – |
| HF9.SSU14048 | Microlaimidae | 100 | Microlaimidae | 100 |
| HF9.SSU14296 | Selachinematidae | 100 | Selachinematidae | 100 |
| HF9.SSU17250 | Thoracostomopsidae | 100 | Thoracostomopsidae | 100 |
| HF9.SSU17844 | unassigned | – | unassigned | – |
| HF9.SSU18227 | Chromadoridae | 100 | Chromadoridae | 100 |
| HF9.SSU19963 | unassigned | – | unassigned | – |
| HF9.SSU20251 | Microlaimidae | 100 | Microlaimidae | 100 |

| OTU ID | Family identification (mothur) | bootstrap % | Family identification (PaPaRa) | bootstrap % |
|---|---|---|---|---|
| HF9.SSU22538 | unassigned | – | unassigned | – |
| TF1.SSU676746 | Ceramonematidae | 100 | Ceramonematidae | 100 |
| TF1.SSU677162 | unassigned | – | unassigned | – |
| TF1.SSU681557 | Oxystominidae | 100 | Oxystominidae | 100 |
| TF1.SSU688192 | unassigned | – | unassigned | – |
| TF1.SSU692690 | Selachinematidae | 100 | Selachinematidae | 100 |
| TF1.SSU694267 | Desmoscolecidae | 99 | Desmoscolecidae | 99 |
| TF1.SSU694751 | Chromadoridae | 100 | Chromadoridae | 100 |
| TF1.SSU698227 | Benthimermithidae | NA | Benthimermithidae | NA |
| TF1.SSU700188 | Cyartonematidae | NA | Cyartonematidae | NA |
| TF1.SSU703579 | unassigned | – | unassigned | – |
| TF1.SSU710679 | Cyatholaimidae | 100 | Cyatholaimidae | 100 |
| TF1.SSU734804 | Siphonolaimidae | 99 | Siphonolaimidae | 99 |
| TF3.SSU956521 | unassigned | – | unassigned | – |
| TF3.SSU960449 | Desmoscolecidae | 99 | Desmoscolecidae | 99 |
| TF3.SSU966338 | Xyalidae | 97 | Xyalidae | 97 |
| TF4.SSU144249 | Cyatholaimidae | 100 | Cyatholaimidae | 100 |
| TF4.SSU150234 | Desmodoridae | 99 | Desmodoridae | 99 |
| TF5.SSU410031 | unassigned | – | unassigned | – |
| TF5.SSU419519 | Desmoscolecidae | 99 | Desmoscolecidae | 99 |
| TF5.SSU430294 | Xyalidae | 97 | Xyalidae | 97 |
| TF5.SSU437076 | Comesomatidae | 99 | Comesomatidae | 99 |
| TF5.SSU444034 | Rhabdodemaniidae | NA | Rhabdodemaniidae | NA |
| TF5.SSU446087 | Tarvaiidae | NA | Tarvaiidae | NA |
| TF5.SSU453472 | Oxystominidae | 100 | Oxystominidae | 100 |
| TF5.SSU457543 | Oxystominidae | 100 | Oxystominidae | 100 |
| TF5.SSU459305 | Oncholaimidae | 100 | Oncholaimidae | 100 |
| TF5.SSU466315 | Xyalidae | 100 | Xyalidae | 100 |
| TF6.SSU33463 | Oxystominidae | 100 | Oxystominidae | 100 |
| TF6.SSU33935 | unassigned | – | unassigned | – |
| TF6.SSU36442 | Desmoscolecidae | 99 | Desmoscolecidae | 99 |
| TF6.SSU37421 | unassigned | – | unassigned | – |
| TF6.SSU41803 | Xyalidae | 97 | Xyalidae | 97 |
| TF6.SSU47996 | Enchelidiidae | 90 | Enchelidiidae | 90 |
| TF6.SSU48167 | Comesomatidae | 99 | Comesomatidae | 99 |
| TF6.SSU53456 | Oncholaimidae | 81 | Oncholaimidae | 81 |
| TF6.SSU54250 | Microlaimidae | 100 | Microlaimidae | 100 |
| TF6.SSU58877 | Tarvaiidae | 100 | Tarvaiidae | 100 |
| TF6.SSU74955 | Cyatholaimidae | 100 | Cyatholaimidae | 100 |
| TF6.SSU82210 | unassigned | – | unassigned | – |
| TF6.SSU84268 | unassigned | – | unassigned | – |

| OTU ID | Family identification (mothur) | bootstrap % | Family identification (PaPaRa) | bootstrap % |
|---|---|---|---|---|
| TF6.SSU98667 | unassigned | – | unassigned | – |
| TS1.SSU270885 | Desmoscolecidae | 99 | Desmoscolecidae | 99 |
| TS1.SSU284163 | Desmoscolecidae | 99 | Desmoscolecidae | 99 |
| TS2.SSU821962 | Tripyloididae | 100 | Tripyloididae | 100 |
| TS2.SSU823349 | Tripyloididae | 100 | Tripyloididae | 100 |
| TS3.SSU475561 | unassigned | – | unassigned | – |
| TS3.SSU489684 | Desmoscolecidae | 99 | Desmoscolecidae | 99 |
| TS3.SSU503133 | Tripyloididae | 100 | Tripyloididae | 100 |
| TS3.SSU508400 | unassigned | – | unassigned | – |
| TS4.SSU543236 | Oxystominidae | 100 | Oxystominidae | 100 |
| TS4.SSU544032 | Desmoscolecidae | 99 | Desmoscolecidae | 99 |
| TS5.SSU874117 | Oxystominidae | 100 | Oxystominidae | 100 |
| TS5.SSU875407 | Comesomatidae | 99 | Comesomatidae | 99 |
| TS5.SSU881546 | Xyalidae | 97 | Xyalidae | 97 |
| TS5.SSU900338 | Leptolaimidae | 99 | Leptolaimidae | 99 |
| TS5.SSU901243 | Ironidae | 100 | Ironidae | 100 |
| TS6.SSU559765 | Tripyloididae | 100 | Tripyloididae | 100 |
| TS6.SSU570763 | unassigned | – | unassigned | – |
| TS6.SSU587229 | Oncholaimidae | 100 | Oncholaimidae | 100 |
| HE6.SSU372021 | Monhysteridae | 100 | Monhysteridae | 100 |

**Supplementary Table 8.** Comparison of the results produced by different taxonomy assignment approaches and distribution (presence vs. absence) of different OTUs among study sites (HE – Hållö, flotation with $MgCl_2$; HF – Hållö, flotation with $H_2O$; TS – Telekabeln, syphoning; TF – Telekabeln, flotation with $H_2O$).

| | Distribution | | | | BLASTN | CREST | TREE-C | TREE-P | EPA (BOTH) |
|---|---|---|---|---|---|---|---|---|---|
| OTU ID | HE | HF | TS | TF | 52/139 | 26/139 | 56/139 | 67/139 | 105/139 |
| HE1.SSU848264 | * | * | – | – | Cyatholaimidae | | | | Cyatholaimidae |
| HE1.SSU850987 | * | * | – | – | | | | | |
| HE1.SSU856624 | * | – | – | * | | | Oxystominidae | Oxystominidae | Oxystominidae |
| HE1.SSU856738 | * | * | – | – | Microlaimidae | Microlaimidae | | | Microlaimidae |
| HE1.SSU858060 | * | – | – | – | | | | Chromadoridae | Chromadoridae |
| HE1.SSU867071 | * | – | – | – | | | Selachinematidae | | Selachinematidae |
| HE2.SSU637072 | * | * | – | – | | | | | |
| HE2.SSU637135 | * | * | – | – | | Chromadoridae | | | Chromadoridae |
| HE2.SSU644966 | * | – | – | – | Cyatholaimidae | | | | Cyatholaimidae |
| HE2.SSU654005 | * | * | – | – | | | Rhabdodemaniidae | Rhabdodemaniidae | Rhabdodemaniidae |
| HE2.SSU655107 | * | * | – | – | | | Xyalidae | Xyalidae | Xyalidae |
| HE2.SSU659506 | * | * | * | * | Chromadoridae | | | Chromadoridae | Chromadoridae |
| HE3.SSU110275 | * | * | – | – | Enoplidae | Enoplidae | Enoplidae | Enoplidae | Enoplidae |
| HE3.SSU117415 | * | – | – | * | Cyatholaimidae | | | | Cyatholaimidae |
| HE3.SSU118424 | * | – | – | – | | | | Oxystominidae | Oxystominidae |
| HE3.SSU124287 | * | * | – | – | Thoracostomopsidae | Thoracostomopsidae | Thoracostomopsidae | Thoracostomopsidae | Thoracostomopsidae |
| HE3.SSU124998 | * | – | – | – | | | | | |
| HE4.SSU913283 | * | – | – | – | | | | | |
| HE5.SSU181724 | * | * | – | – | | | | | |
| HE5.SSU188855 | * | * | * | – | | | Enchelidiidae | Enchelidiidae | Enchelidiidae |
| HE6.SSU355777 | * | * | – | – | Achromadoridae | | | | |

| | Distribution | | | | BLASTN | CREST | TREE-C | TREE-P | EPA (BOTH) |
|---|---|---|---|---|---|---|---|---|---|
| OTU ID | HE | HF | TS | TF | 92/139 | 26/139 | 54/139 | 67/139 | 105/139 |
| HE6.SSU358048 | * | – | – | – | Chromadoridae | | | | Chromadoridae |
| HE6.SSU360897 | * | – | – | – | | | Desmodoridae | | |
| HE6.SSU361449 | * | * | – | – | Ironidae | | Ironidae | Ironidae | Ironidae |
| HE6.SSU365256 | * | – | – | – | | | | | Desmodoridae |
| HE6.SSU368318 | * | * | – | – | | Desmodoridae | | | |
| HE6.SSU370544 | * | – | – | – | | Xyalidae | Xyalidae | Xyalidae | Xyalidae |
| HE6.SSU378839 | * | – | – | – | | | Microlaimidae | Microlaimidae | Microlaimidae |
| HE6.SSU383414 | * | – | – | – | Comesomatidae | | | Comesomatidae | Comesomatidae |
| HE6.SSU383888 | * | * | – | * | Chromadoridae | | | Chromadoridae | Chromadoridae |
| HE7.SSU232624 | * | * | – | * | | | Leptolaimidae | Leptolaimidae | Leptolaimidae |
| HE7.SSU256492 | * | * | * | * | Chromadoridae | | | | Chromadoridae |
| HE8.SSU829972 | * | * | * | * | | | Anticomidae | Anticomidae | Anticomidae |
| HE8.SSU843570 | * | * | – | – | Chromadoridae | | | | Chromadoridae |
| HE9.SSU305678 | * | * | – | * | | | Xyalidae | | Xyalidae |
| HF1.SSU759758 | – | * | – | – | | Leptolaimidae | Camacolaimidae | Camacolaimidae | Camacolaimidae |
| HF1.SSU763392 | * | * | – | – | Cyatholaimidae | | | | Cyatholaimidae |
| HF1.SSU764346 | * | * | * | * | Cyatholaimidae | | | | Cyatholaimidae |
| HF1.SSU774294 | * | * | – | – | Mermithidae | | | | Mermithidae |
| HF1.SSU779114 | * | * | * | * | | Axonolaimidae | Axonolaimidae | Axonolaimidae | Axonolaimidae |
| HF1.SSU780927 | – | * | * | – | | | | | |
| HF2.SSU192072 | – | * | – | – | | Chromadoridae | Chromadoridae | Chromadoridae | Chromadoridae |
| HF2.SSU204352 | – | * | – | – | | | Leptolaimidae | | |
| HF2.SSU205129 | * | * | – | – | | | | | Chromadoridae |
| HF2.SSU208147 | * | * | – | * | | | Selachinematidae | Selachinematidae | Selachinematidae |

| | Distribution | | | | BLASTN | CREST | TREE-C | TREE-P | EPA (BOTH) |
|---|---|---|---|---|---|---|---|---|---|
| OTU ID | HE | HF | TS | TF | 92/139 | 26/139 | 54/139 | 67/139 | 105/139 |
| HF2.SSU210357 | – | * | – | – | Enchelidiidae | | Enchelidiidae | Enchelidiidae | Enchelidiidae |
| HF3.SSU989895 | * | * | – | – | | | | Camacolaimidae | Camacolaimidae |
| HF3.SSU990962 | – | * | – | – | Chromadoridae | | | | Chromadoridae |
| HF4.SSU606153 | * | * | * | * | Chromadoridae | Chromadoridae | | | Chromadoridae |
| HF4.SSU614317 | * | * | – | * | Comesomatidae | Comesomatidae | | | |
| HF4.SSU619471 | * | * | – | – | Microlaimidae | | Microlaimidae | Microlaimidae | Microlaimidae |
| HF4.SSU620879 | – | * | – | – | Chromadoridae | | | Chromadoridae | Chromadoridae |
| HF4.SSU622464 | – | * | – | – | Plectidae | | | | |
| HF4.SSU624085 | * | * | – | – | | | | | |
| HF4.SSU625424 | * | * | – | – | | | | Comesomatidae | Comesomatidae |
| HF4.SSU628562 | – | * | – | – | | | | | Desmodoridae |
| HF4.SSU631524 | – | * | – | – | | Leptolaimidae | Leptolaimidae | Leptolaimidae | Leptolaimidae |
| HF4.SSU632264 | – | * | – | – | | | Diplopeltidae | Diplopeltidae | Diplopeltidae |
| HF4.SSU635045 | * | * | – | – | | | Chromadoridae | Chromadoridae | Chromadoridae |
| HF5.SSU991188 | * | * | – | – | | | | Oncholaimidae | Oncholaimidae |
| HF5.SSU995414 | – | * | – | – | Rhabdolaimidae | Ironidae | | | Ironidae |
| HF6.SSU329881 | * | * | * | * | | Desmodoridae | Desmodoridae | Desmodoridae | Desmodoridae |
| HF6.SSU338435 | – | * | – | – | | | | Chromadoridae | Chromadoridae |
| HF6.SSU338739 | * | * | – | – | | | | | |
| HF7.SSU385021 | – | * | – | – | Chromadoridae | | | Chromadoridae | Chromadoridae |
| HF7.SSU390110 | * | * | * | * | | | | | |
| HF7.SSU398053 | – | * | – | – | | | | | Camacolaimidae |
| HF7.SSU407024 | * | * | – | * | | | | | Achromadoridae |
| HF7.SSU407761 | * | * | – | – | | | | | Linhomoeidae |

| | Distribution | | | | BLASTN | CREST | TREE-C | TREE-P | EPA (BOTH) |
|---|---|---|---|---|---|---|---|---|---|
| OTU ID | HE | HF | TS | TF | 92/139 | 26/139 | 54/139 | 67/139 | 105/139 |
| HF7.SSU409331 | – | * | – | – | | | | | |
| HF8.SSU795426 | – | * | – | – | Cyatholaimidae | | | | |
| HF9.SSU14048 | – | * | – | – | | | | | Microlaimidae |
| HF9.SSU14296 | – | * | – | – | | | | | Selachinematidae |
| HF9.SSU17250 | – | * | – | – | Thoracostomopsidae | Thoracostomopsidae | | Thoracostomopsidae | Thoracostomopsidae |
| HF9.SSU17844 | * | * | – | – | | | | | |
| HF9.SSU18227 | – | * | – | – | Chromadoridae | | | | Chromadoridae |
| HF9.SSU19963 | – | * | – | – | | | | | |
| HF9.SSU20251 | * | * | * | * | Microlaimidae | Microlaimidae | Microlaimidae | Microlaimidae | Microlaimidae |
| HF9.SSU22538 | – | * | – | – | Mermithidae | | | | |
| TF1.SSU676746 | – | – | – | * | | | Ceramonematidae | Ceramonematidae | Ceramonematidae |
| TF1.SSU677162 | – | – | – | * | | | | | |
| TF1.SSU681557 | – | – | * | * | | | Oxystominidae | Oxystominidae | Oxystominidae |
| TF1.SSU688192 | – | – | – | * | Linhomoeidae | | Linhomoeidae | Linhomoeidae | |
| TF1.SSU692690 | – | – | – | * | Selachinematidae | | Selachinematidae | Selachinematidae | Selachinematidae |
| TF1.SSU694267 | – | – | – | * | | | | | Desmoscolecidae |
| TF1.SSU694751 | – | – | – | * | Chromadoridae | | | Chromadoridae | Chromadoridae |
| TF1.SSU698227 | – | – | – | * | Teratocephalidae | | | | Benthimermithidae |
| TF1.SSU700188 | – | – | – | * | Linhomoeidae | | Cyartonematidae | Cyartonematidae | Cyartonematidae |
| TF1.SSU703579 | – | – | – | * | Siphonolaimidae | | | | |
| TF1.SSU710679 | * | * | * | * | Cyatholaimidae | Cyatholaimidae | | Cyatholaimidae | Cyatholaimidae |
| TF1.SSU734804 | – | – | – | * | Siphonolaimidae | | | | Siphonolaimidae |
| TF3.SSU956521 | – | – | – | * | Comesomatidae | | | | |
| TF3.SSU960449 | – | – | – | * | | | | | Desmoscolecidae |

| | Distribution | | | | BLASTN | CREST | TREE-C | TREE-P | EPA (BOTH) |
|---|---|---|---|---|---|---|---|---|---|
| OTU ID | HE | HF | TS | TF | 92/139 | 26/139 | 54/139 | 67/139 | 105/139 |
| TF3.SSU966338 | * | * | * | * | | | Xyalidae | Xyalidae | Xyalidae |
| TF4.SSU144249 | * | * | * | * | | | | Cyatholaimidae | Cyatholaimidae |
| TF4.SSU150234 | – | – | * | * | Desmodoridae | | | | Desmodoridae |
| TF5.SSU410031 | – | – | – | * | | | | | |
| TF5.SSU419519 | * | * | * | * | | | | | Desmoscolecidae |
| TF5.SSU430294 | * | * | * | * | | | Xyalidae | Xyalidae | Xyalidae |
| TF5.SSU437076 | – | – | * | * | Comesomatidae | Comesomatidae | Comesomatidae | | Comesomatidae |
| TF5.SSU444034 | * | * | * | * | | | Rhabdodemaniidae | Rhabdodemaniidae | Rhabdodemaniidae |
| TF5.SSU446087 | – | – | – | * | | | | Tarvaiidae | Tarvaiidae |
| TF5.SSU453472 | – | – | – | * | | | Oxystominidae | Oxystominidae | Oxystominidae |
| TF5.SSU457543 | – | – | * | * | | | Oxystominidae | Oxystominidae | Oxystominidae |
| TF5.SSU459305 | – | – | – | * | Oncholaimidae | Oncholaimidae | Oncholaimidae | Oncholaimidae | Oncholaimidae |
| TF5.SSU466315 | – | – | * | * | | | Xyalidae | Xyalidae | Xyalidae |
| TF6.SSU33463 | * | * | * | * | | | Oxystominidae | Oxystominidae | Oxystominidae |
| TF6.SSU33935 | * | – | – | * | | | | | |
| TF6.SSU36442 | – | – | – | * | | | | | Desmoscolecidae |
| TF6.SSU37421 | – | – | * | * | | Desmodoridae | | | |
| TF6.SSU41803 | – | – | – | * | | Xyalidae | | Xyalidae | Xyalidae |
| TF6.SSU47996 | – | – | * | * | Oncholaimidae | Enchelidiidae | Enchelidiidae | Enchelidiidae | Enchelidiidae |
| TF6.SSU48167 | * | * | * | * | Comesomatidae | Comesomatidae | | Comesomatidae | Comesomatidae |
| TF6.SSU53456 | * | * | * | * | | Oncholaimidae | Oncholaimidae | Oncholaimidae | Oncholaimidae |
| TF6.SSU54250 | – | – | * | * | Microlaimidae | | Microlaimidae | Microlaimidae | Microlaimidae |
| TF6.SSU58877 | – | – | – | * | | | | Tarvaiidae | Tarvaiidae |
| TF6.SSU74955 | – | – | – | * | Cyatholaimidae | | | | Cyatholaimidae |

| | Distribution | | | | BLASTN | CREST | TREE-C | TREE-P | EPA (BOTH) |
|---|---|---|---|---|---|---|---|---|---|
| OTU ID | HE | HF | TS | TF | 92/139 | 26/139 | 54/139 | 67/139 | 105/139 |
| TF6.SSU82210 | – | – | * | * | | | | | |
| TF6.SSU84268 | – | – | – | * | | | | | |
| TF6.SSU98667 | – | – | – | * | | | | | |
| TS1.SSU270885 | – | – | * | – | | | | | Desmoscolecidae |
| TS1.SSU284163 | – | – | * | – | | | | | Desmoscolecidae |
| TS2.SSU821962 | * | * | * | * | Tripyloididae | | Tripyloididae | Tripyloididae | Tripyloididae |
| TS2.SSU823349 | * | – | * | * | | | Tripyloididae | Tripyloididae | Tripyloididae |
| TS3.SSU475561 | – | – | * | – | Siphonolaimidae | | | | |
| TS3.SSU489684 | * | * | * | * | | | Desmoscolecidae | | Desmoscolecidae |
| TS3.SSU503133 | – | – | * | – | Tripyloididae | | Tripyloididae | Tripyloididae | Tripyloididae |
| TS3.SSU508400 | – | – | * | * | | | | | |
| TS4.SSU543236 | – | – | * | – | | | Oxystominidae | Oxystominidae | Oxystominidae |
| TS4.SSU544032 | – | * | * | * | | | | | Desmoscolecidae |
| TS5.SSU874117 | – | – | * | * | | | Oxystominidae | Oxystominidae | Oxystominidae |
| TS5.SSU875407 | – | – | * | * | Comesomatidae | Comesomatidae | Comesomatidae | | Comesomatidae |
| TS5.SSU881546 | – | – | * | – | | | Xyalidae | | Xyalidae |
| TS5.SSU900338 | – | – | * | – | | | Leptolaimidae | Leptolaimidae | Leptolaimidae |
| TS5.SSU901243 | – | – | * | * | | | Ironidae | Ironidae | Ironidae |
| TS6.SSU559765 | – | – | * | – | | | Tripyloididae | Tripyloididae | Tripyloididae |
| TS6.SSU570763 | – | – | * | – | | | | | |
| TS6.SSU587229 | * | * | * | * | Oncholaimidae | Oncholaimidae | Oncholaimidae | Oncholaimidae | Oncholaimidae |
| HE6.SSU372021 | * | – | – | – | | | Monhysteridae | Monhysteridae | Monhysteridae |

**Supplementary Table 9.** Number of taxa (species/OTUs) per family recovered in morphology-based identification and best scoring taxonomy assignment approach (EPA) grouped per each site and extraction method.

| | Hållö | | | | Telekabeln | | | |
|---|---|---|---|---|---|---|---|---|
| | Flotation (MgCl2) | | Flotation ($H_2O$) | | Siphoning | | Flotation ($H_20$) | |
| | morphology based | barcoding | morphology based | barcoding | morphology based | barcoding | morphology based | barcoding |
| **ORDER ENOPLIDA** | | | | | | | | |
| Enoplidae | 1 | 1 | 1 | 1 | 0 | 0 | 0 | 0 |
| Thoracostomopsidae | 2 | 1 | 2 | 2 | 1 | 0 | 1 | 0 |
| Phanodermatidae | 0 | 0 | 1 | 0 | 0 | 0 | 0 | 0 |
| Anticomidae | 1 | 1 | 1 | 1 | 1 | 1 | 1 | 1 |
| Oncholaimidae | 3 | 3 | 3 | 3 | 4 | 2 | 3 | 3 |
| Enchelidiidae | 2 | 1 | 2 | 2 | 2 | 2 | 2 | 1 |
| Leptosomatidae | 0 | 0 | 1 | 0 | 0 | 0 | 0 | 0 |
| Ironidae | 0 | 1 | 0 | 2 | 1 | 1 | 1 | 1 |
| Oxystominidae | 4 | 3 | 5 | 1 | 15 | 5 | 15 | 6 |
| Tripyloididae | 0 | 2 | 1 | 1 | 2 | 4 | 2 | 2 |
| Xennellidae | 1 | NA | 1 | NA | 0 | NA | 0 | NA |
| Trefusiidae | 0 | 0 | 0 | 0 | 0 | 0 | 1 | 0 |
| **ORDER MERMITHIDA** | | | | | | | | |
| Mermithidae | 0 | 1 | 0 | 1 | 0 | 0 | 0 | 0 |
| **ORDER TRIPLONCHIDA** | | | | | | | | |
| Pandolaimidae | 0 | NA | 0 | NA | 1 | NA | 1 | NA |
| Rhabdodemaniidae | 1 | 2 | 1 | 2 | 1 | 1 | 1 | 1 |
| **ORDER DESMOSCOLECIDA** | | | | | | | | |
| Desmoscolecidae | 3 | 2 | 3 | 3 | 4 | 5 | 4 | 6 |
| Meyliidae | 0 | NA | 2 | NA | 0 | NA | 0 | NA |
| Cyartonematidae | 0 | 0 | 0 | 0 | 0 | 0 | 1 | 1 |
| **ORDER CHROMADORIDA** | | | | | | | | |
| Achromadoridae | 0 | 1 | 0 | 1 | 0 | 0 | 0 | 1 |
| Chromadoridae | 14 | 10 | 13 | 14 | 7 | 3 | 6 | 5 |
| Cyatholaimidae | 8 | 7 | 9 | 5 | 4 | 3 | 3 | 5 |
| Selachinematidae | 4 | 2 | 4 | 2 | 2 | 0 | 4 | 2 |
| **ORDER DESMODORIDA** | | | | | | | | |
| Desmodoridae | 10 | 2 | 11 | 2 | 4 | 2 | 4 | 2 |
| Epsilonematidae | 2 | 0 | 2 | 0 | 0 | 0 | 0 | 0 |

| | Hållö | | | | Telekabeln | | | |
|---|---|---|---|---|---|---|---|---|
| | Flotation (MgCl2) | | Flotation ($H_2O$) | | Siphoning | | Flotation ($H_20$) | |
| | morphology based | barcoding | morphology based | barcoding | morphology based | barcoding | morphology based | barcoding |
| Draconematidae | 0 | 0 | 1 | 0 | 0 | 0 | 0 | 0 |
| Microlaimidae | 6 | 4 | 7 | 4 | 1 | 2 | 4 | 2 |
| Monoposthiidae | 1 | 0 | 1 | 0 | 1 | 0 | 0 | 0 |
| Richtersiidae | 0 | NA | 0 | NA | 1 | NA | 1 | NA |
| **ORDER MONHYSTERIDA** | | | | | | | | |
| Xyalidae | 7 | 5 | 9 | 4 | 3 | 4 | 4 | 5 |
| Sphaerolaimidae | 0 | 0 | 0 | 0 | 3 | 0 | 3 | 0 |
| Monhysteridae | 1 | 1 | 1 | 0 | 0 | 0 | 0 | 0 |
| Siphonolaimidae | 1 | 0 | 1 | 0 | 0 | 0 | 1 | 1 |
| Linhomoeidae | 2 | 1 | 2 | 1 | 5 | 0 | 6 | 0 |
| **ORDER ARAEOLAIMIDA** | | | | | | | | |
| Comesomatidae | 2 | 3 | 1 | 2 | 3 | 3 | 4 | 3 |
| Axonolaimidae | 3 | 1 | 2 | 1 | 2 | 1 | 2 | 1 |
| Diplopeltidae | 2 | 0 | 2 | 1 | 6 | 0 | 6 | 0 |
| **ORDER PLECTIDA** | | | | | | | | |
| Leptolaimidae | 4 | 1 | 4 | 2 | 2 | 1 | 6 | 1 |
| Camacolaimidae | 2 | 1 | 4 | 3 | 2 | 0 | 4 | 0 |
| Rhadinematidae | 0 | NA | 0 | NA | 1 | NA | 1 | NA |
| Ceramonematidae | 0 | 0 | 0 | 0 | 1 | 0 | 2 | 1 |
| Diplopeltoididae | 1 | NA | 1 | NA | 1 | NA | 4 | NA |
| Tarvaiidae | 0 | 0 | 1 | 0 | 0 | 0 | 0 | 2 |
| Tubolaimoididae | 0 | NA | 1 | NA | 0 | NA | 0 | NA |
| Paramicrolaimidae | 0 | NA | 0 | NA | 1 | NA | 1 | NA |
| Aegialoalaimidae | 0 | NA | 0 | NA | 0 | NA | 1 | NA |
| **ORDER BENTHIMERMITHIDA** | | | | | | | | |
| Benthimermithidae | 0 | 0 | 0 | 0 | 0 | 0 | 0 | 1 |
| | | | | | | | | |
| Unidentified taxa | 1 | 14 | 11 | 17 | 8 | 7 | 21 | 13 |

**Supplementary Data 1.** Nematode OTU sequences used in this study.

```
>HE1_SSU848264
ATGCATGTCTAAGCAAGAGCCTAGAAATGGTGAAGCCGCGAACAGCTCATTACAACAGCCATAGTTTATTGAATCGTCTCTCATATACT
TGGATACCTTTGGTAATTCTAGAGCTAATACACGCACCAAACTCTGAGCTGAGGCAAGGAGTGCATTTATTAGAACAAAACCAATGGGC
TTCGGCCCTGTTTTGGTGAATCTGAATAACTCAGTTGATCGCACAGTCTCGCACTGGCGACGTATCTTTCAAGTGTCTGCCTTATCAAC
TGTCGATGGTAGTTTATATGACTACCATGGTTGTAACGGGTAACGGAGAATAAGGGTTCGACTCCGGAGAGGGAGCCTGAGAAATGGCT
ACCACA
>HE1_SSU850987
ATGCGTGTCTAAGCACAAACTGATTTAAGGTGAAGCCGCGAATGGCTCATTACAACAGCCATAGTTTCTTGGATCTTACTTCTACTTGG
ATAACTGTAGTAATTCTAGAGCTAATACATGCGTCAAGCTTCAACCTCATGGAAGAAACACATTTATTAGATCAAAACCAATCGGGCTT
TGCTTGTTTGTTTGATGAATCTGAATAACTCTGCCAATTGCATGGTCTTTGCACCGGTGACATATCTTTCAAGTGTCTGCCTTATCAAC
TGTTGATGGTAGTTTATATGACTACCATGGTTGTAACGGGTAACGGAGAATCAGGGTTCGACTCCGGAGAGGGAGCCCGAGAAACGGCT
ACCACA
>HE1_SSU856624
ATGCATGTCTCAGCACACGCCAATGTATGGTAAAGCCGCGAATGGCTCATTACAACAGCCACCGTTTATTAGATCATCCTTCTTACTTG
GATAACTGTGGAAAAGCTAGAGCTAATACATGCTACAAGCTCTGACCTTACGGAAAGAGCGCATTTATTAGAACAAAACCAATCGGACT
TTGTCCGTCGTTTGGTGAATCTGAATAACTAAGCAGATCGCACGGTCTTAGAACCGGCGACATATCGTTCAAATGTCTGCCTTATCAAC
TTTCGATGGTAGTTTATGCGCCTACCATGGTTGTAACGGGTAACGAAGAATCAGGGTTTGATTTCGGAGAGGGAGCCTGAGAAACGGCT
ACCACA
>HE1_SSU856738
ATGCATGTCTAAGTACAAGCCTCTTTAAGGTGAAACCGCGAATGGCTCATTAAATCACACCTAATATACTGGATAGTATACAGTTACTT
GGATAACTGTGGTAATTCTAGAGCTAATACACGCATTCAAGTTCCGACGCAAGAAGGAACGCATTTATTAGAACTAAACCAATCGGGCT
TGCCCGTTGTTTGGTGAATCTGAATAACTCCGCAGATCGCATGGTCTAGCACCGGCGACATATCTTTCAAGTGTCTGCCTTATCAACTT
TCGATGGTAGTTTATGTGACTACCATGGTTGTTACGGGTAACGGAGAATTAGGGTTCGACTCCGGAGAGGGAGCCTGAGAAACGGCTAC
CACA
>HE1_SSU858060
ATGCATGTCTAAGCATAAACCGAATATGGTAAAGCCGCGAATGGCTCATTACAACAGCCATAGTTTATTGGATCTTAGAGTCCTACTTG
GATACCTGTGGTAATTCTAGAGCTAATACACGCAATCAAGCCCCAACCTTACGGGCGGGGCGCATTTATTAGAACAAAACCAATTGGCT
TCGGCCATAGATTGGTGAATCTGAATAACTACGCTGATCGCACGGGCTTGTCCCGGCGACGTATCCTTCCAAGGTGTGCCCTATCAACT
GTCGACTGTGGCATAGACGCCCACAGTGGTTTTGACGGGTAACGGGGAATCAGGGTTCGATTCCGGAGAGGGAGCCTGAGAAACGGCTA
CCACA
>HE1_SSU867071
ATGCATGTCTAAGTACAGGCCTCACTAAGGTGAAACCGCGAATGGCTCATTAAATCACACATGATTTATTCGATCATAAAATCCTACTT
GGATAACTGTGGTAATTCTAGAGCTAATACACGCAACTAAACTCCAACCTTGCGGAAGGAGTGCATTTATTAGAACAAAACCAATCGGC
TTCGGCCGTTACTTGGTGAATCTGAATAACTCAGTTGATCGCACGGTCTTTGTACTGGCGACATATCTTTCAAGTGTCTGCCTTATCAA
CTTTCGATGGTAGTTTATACGACTACCATGGTTGTTACGGGTAACGGAGAATTAGGGTTCGACTCCGGAGAGGGAGCCTGAGAAATGGC
TACCACA
>HE2_SSU637072
ATGCATGTCCAAGTACAAGCCTCATTAAGGTGAAACCGCGAATGGCTCATTAAATCACACCTAATGCACTGGACGTTGCCAGTTACTTG
GATAACTGTGGTAATTCTAGAGCTAATACACGCATCAAAGCTTCGACCTTGCGGAAGGAGCGCATTTATTAGAACAAAACCAATCGGAC
TTCGGTCCGTCATTTGGTGAATCTGAATAACTTTGCTGATCGCACGGTCCTCGTACCGGCGACGTATCTTTCAAATGTCTGCCTTATCA
ACTTTCGATGGTAGTTTATGCGCCTACCATGGTTGTAACGGGTAACGGAGAATCAGGGTTTGATTCCGGAGAGGGAGCCTGAGAAACGG
CTACCACT
>HE2_SSU637135
ATGCATGTCCAAGTACAAGCCTCATTAAGGTGAAACCGCGAATGGCTCATTAAATCACACCTAATGTCCTGGATAGTGTCAGTTACTTG
GATAACTGCGGTAATTCTGGAGCTAATACACGCAATCAAGCCCCAACCTGACGGACGGGGCGCATTTATTAGAACAAAACCAATTGGCT
TCGGCCATAAATTGGTGAATCTGAATAACTACGCTGATCGCATGGTCTCGTACCGGCGACGTATCCTTCAAGTGTCTGCCTTATCAACT
TTCGATGGTAGTCTATAAGCCTACCATGGTTGTAACGGGTAACGGAGAATAAGGGTTCGACTCCGGAGAGGGAGCCTGAGAAACGGCTA
CCACT
>HE2_SSU644966
ATGCATGTCTAAACAAAAGCCCCAATATGGTGAAGCCGCGAACAGCTCATTACAACAGCCATAGTTTATTGGATCTTATCTTATACTTG
GATACCTGTACTAATTGTAGAGCTAATACATGCAAAAAAGCCACACTCTCTGGGTGTGGTGCATTTATTAGAACAAAACCAATCGGCTT
CGGCCGTTGTTTGGTGAATCTGAATAACTCAGCTGATCGCACAGTCTAGCACTGGCGATGTATCTTTCAAGTGTCTGCCTTATCAACTT
TCGATGGTAGTTTACATGACTACCATGGTTGTAACGGGTAACGGAGAATAAGGGTTCGACTCCGGAGAGGGAGCCTGAGAAATGGCTAC
CACA
>HE2_SSU654005
ATGCATTGTCTAAGCAGAAGCCGATTAATGGTAAAGCCGCGAATGGCTCATTACAACAGCTATTGTTTATTGGACACTATCATCCTACT
TGGATAACTGTGGCAATTCTAGAGCTAATACACGCAGCAAAACGGGGACCTTACGGAACCTGTGCATTTATTAGAACAAAACCAATCGG
GTTTTACCCGTCGTTTGGTGAATCTAGGTAACTCTGCTAATCGCATGGTCTAAGAACCGGCGATATATCTTTCAAATGTCTGCCTTATC
AACTTTCGATGGTAGTTTATGCGCCTACCATGGTTGTAACGGGTAACGGAGAATCAGGGTTCGACTCCGGAGAGGGAGCCTGAGAAACG
GCTACCACA
>HE2_SSU655107
ATGCATGTCTAAGCACAAACTTTATGAGTGAAGCCGCGAAAAGCTCATTACAACAGCCGTCGTTTCTTGGATCTCCGAATTTACTTGGA
TAACTGTGGTAATTCTAGAGCTAATACATGCAATCGAGCCCTGAACGTAAGTGATGGGCGCACTTATTAGTAACAAAACCGATCGGTGT
TATGCACCGTACGTTTGGTGAATCTGAATAACTGAGCAGATCGCTTTGGTCTTTGCACCGGCGACGTATCTTTCAAATGTCTGCCCTAT
CAACTTTCGATGGTACGTGATATGCCTACCATGGTTGTAACGGGTAACGGGGAATCAGGGTTCGATTCCGGAGAGGGAGCATGAGAAAC
GGCTACCACA
>HE2_SSU659506
```

```
ATGCATGTCTAAGCATAAGCCGAATATGGTAAAGCCGCGAATGGCTCATTACAACAGCCATAGTTTATTGGATCTGAATATCCTACTTG
GATACCTGTGGTAATTCTAGAGCTAATACACGCAAAAAAGCCCTGACTTTACGGAAAGGGCGCATTTATTATAACAAGACCAATTGGCT
TCGGCCATCCATTGGTGACTCTGAATAACTACGCAGATCGCACGGTCTCGTACCGGCGACATATCCTTCAAGTGTCTGCCTTATCAACT
GTCGATGGTAGTTTACATGACTACCATGGTTGTAACGGGTAACGGAGAATTAGGGTTCGACTCCGGAGAGGGAGCCTGAGAAACGGCTA
CCACA
>HE3_SSU110275
ATGCATGTCTAAGTACATACTGATAAATAGTGAAGCTGTGAATGGCTCATTACAACAGCCGTAGTTTATTTGATTTATAGAGTTACATG
GATACCTGTGGTAACCTAAGAGCTAATACGCGCAATTAAGTCCAGACCTTAGGGGACGGACGCGGTTATTAGACCAAAACCAATCGGGC
TTGTCCCGGTATTTTGGTGAATCTGAATAACTCTGCAGATCGCACGGTCCTCGCACCGGCGACATGTCATTCAAATGTCTGCCTTATCA
ACTTTCGATGGTAGTTTATGCGCCTACCATGGTTGTAACGGGTAACGGAGAATTAGGGTTCGACTCCGGAGAGGGAGCCTGAGAAACGG
CTACCACA
>HE3_SSU117415
ATGCATGTCTAAGCATAAGCCGATTAATGGTGAAGCCGCGAATAGCTCATTACAACAGCCATAGTTTATTGGATCTTCTCTCATACTTG
GATACCTGTGGTAATTCTAGAGCTAATACACGCAACAAAACCCTGACTTCGGAAGGGGTGCATTTATTAGAACAAAACCAATTGGCTTC
GGCCATTCATTGGTGACTCTGGATAACTTTGGGCTGATCGCACGGTCTAGCACCGGCGACGTATCTTTCAAATGTCTGCCCTATCAAAT
GTCGAAGGGAGGTGATATGCCTCCCTTGTTTTTAACGGGTAACGGGGAATCAGGGTTCGATTCCGGAGAGGGAGCATGAGAAACGGCTA
CCACA
>HE3_SSU118424
ATGCATGTGTAAGCATGAATCACTTAATGGTGAAGCCGCGAATGGCTCATTATAACAGCCATAGTTTATTAGATTTTTTTTTACTTGGA
TAACTGTGGTAATTCTAGAGCTAATACACGCAGCAAAGCTTCGACCTTACGGAAGGAGCGCATTTATTAGATCAAAACCAATCAGCTCC
GGCTGTATCCTTGACGAATCTGAATAACTTTCCGCTGATCGCATGGCTTCACGCCGGCGACGTATCTTTCTAAGGTGTGCCCTATCAAC
TGTCGACTGTGGCATAGACGCCCACAGTGGTTTTGACGGGTAACGGGGAATCAGGGTTCGATTCCGGAGAGGGAGCATGAGAAACGGCT
ACCACA
>HE3_SSU124287
ATGCATGTCTCAGTACATACTGATTAATAGTGAAACCGCAAATGGCTCATTACAACAGCTATAGTTTATTAGATCTTACTCTACATGGA
TACCTGTGGTAACCTAAGAGCTAATACATGCAACAAAGCCCTATTGCAGGGCGCATTTATTAGGACAAAACCAATCGGGCTTGCCCGCT
TATTGGTGGATCTGAATAACTTTGCTAATCGCACAGTCCTCGTACTGGCGATGTATCTTTCAAATATCTGCCTTATCAACTGTCGATGG
TAGCTTACACGACTACCATGGTTGTTACGGGTAACGGAGAATTAGGGTTCGACTCCGGAGAGGGAGCCTGAGAAACGGCTACCACA
>HE3_SSU124998
ATGCATGTCCAAGTACAAGCCTCATTAAGGTGAAACCGCGAATGGCTCATTAAATCACACCTAATGCACTGGACGTTGCCAGTTACTTG
GATAACTGTGGTAATTCTAGAGCTAATACATGCTGCAAGCTCTGACAAGTTCTTGGACTTGAAGGAGTGCATTTATTAGTACAAAACCA
ATCGGGCTTCTGCCTGCAAATTGGTGAATCTGAATAACTCAGCTGATCGCACAGTCTTGAACTGGCGACGTATCTTTCAAGTGTCTGCC
GTATCAACTGTTGATGGTAGTTTATGTGACTACCATGGTTGTAACGGGTAACGGAGAATAAGGGTTCGACTCCGGAGAGGGAGCCTGAG
AAACGGCTACCACA
>HE4_SSU913283
ATGCATGTCTAAGCACAAACCGAATATGGTAAAGCCGCGCATTGCTCATTACAACAGCCATTGTTTACTGGATCTTGAAAAGTTACTTG
GATAACTGTGGTAATTCTAGAGCTAATACACGCACAAAACTCTAGCCGTCTGGCAAGAGTGCATTTATTAGAACAAAACCAATCAGGCT
TCGGTCTGCTGTTTGGTGAATCTGAATAACTGAGCTGATCGTACCGGTCTTCGCACCGGCGACGTATCTTTCAAGTGTCTGCCTTATCA
ACTTTCGATGGTAGCTTATGTGGCTACCATGGTTGTAACGGGTAACGGAGAATCAGGGTTCGACTCCGGAGAGGGAGCCTGAGAAATGG
CTACCACA
>HE5_SSU181724
ATGCATGTCCAAGTACAAGCCTCATTAAGGTGAAACCGCGAATGGCTCATTAAATCACACCTAATGCACTGGACGTTGCCAGTTACTTG
GATAACTGCGGTAATTCTAGAGCTAATACACGCAATTCAGCCCTGACCTCACGGGAGGGGAGCATTTATTAGAACAAAACCAATTGGCC
TCGGCCATCTGTTGGTGAATCTGAATAACTACGCAGATCGCATGGTCTCGTACCGGCGACGTATCCTTCAAGGGTCTGCCTTATCAACT
TTCGATGGTAGTGTATTTGCCTACCATGGTTGTAACGGGTAACGGAGAATTAGGGTTTGACTCCGGAGAGGGAGCCTGAGAAACGGCTA
CCACA
>HE5_SSU188855
ATGCATGTCTAAGCACAAGCTATTAATTGTGAAGCCGCGAATGGCTCATTATAACAGCCATTGTTTACAGGATATATTATTACTACATG
GATAACTGTGGTAATTCTACAGCTAATACACGCATCAAAACCCCGACTTCGTGAAGGGGTGCGTTTGTTACTTCAAATCAATCGGACTT
CGGTCCGCACTCAAGTGAGATTGAACAATTCAGCTGATCGAACGGTCTAAGAACCGACGACATATCCTTCAAACGTCTGCCTTATCAAC
TTTCGACGTGTGTCTATGCGACAAACGTGGTCGTGACGGGTAACGGAGAATCAGGGTTTGATTCCGGAGAGGGAGCCTGAGAAACGGCT
ACCACA
>HE6_SSU355777
ATGCATGTCTAAGCATAAGCCAAAAAATGGTGAAGCCGCGAATAGCTCATTACAACAGCCATAGTTTATTAGATCTACCAATCCTACTT
GGATAACTTTAGTAATTCTAGAGCTAATACACGCAATCAAGCTCCAACCTCTGGGCGGAGCGCATTTATTAGAACAAAACCAATCGGGT
CCGCCCGTCATTTGGTGAATCTGAATAACTCAGCCGATCGCATGGTCCTCGTACCGGCGACGTATCTTTCAAGTGTCTGCCTTATCAAC
TTTCGATGGTAGTTTACACGCCTACCATGGTTGTAACGGGTAACGGAGAATAAGGGTTCGACTCCGGAGAGGGAGCCTGAGAAACGGCT
ACCACA
>HE6_SSU358048
ATGCATGTGTAAGTACAGACTGTACAACGGTGAAACTGCGAATGGCTCATTAAATCAGTTATGGTTCCTTAGATATTAAAAATCTACAT
GGATAACTGTGGTAATTCTAGAGCTAATACACGCAATCAAGCCCCAACCTGACGGGCGGGGCGCATTTATTAGAACAAAACCAATTGGC
TTCGGCCATTCATTGGTGAATCTGAATAACTACGCTGATCGCACGGTCTCGCACCGGCGACGTATCCTTCAAGTGTCTGCCTTATCAAC
TTTCGATGGTAGTTTACATGACTACCATGGTTGTAACGGGTAACGGAGAATAAGGGTTCGACTCCGGAGAGGGAGCCTGAGAAACGGCT
ACCACA
>HE6_SSU360897
ATGCATGTCTAAGCATGAACCGAATATGGTGAAGCCGCGAATGGCTCATTACAACAGCCGTTGTTTCTTGGAGCTTGATTTACTTGGAT
AACTGTGGTAATTCTAGAGCTAATACATGCAACCAAGCTCTGACCTTTGGAAGGAGCGCGTTTATTAGACCAAGACCAATCAGACTTTG
TCTGGAATCTGGTGACTCTGAATAACTTTGCTGATCACACAGTCCTCTCACTGGTGACATATCTTTCAAGTGTCTGCCCTATCAACTGT
CGACTGTGGCATAGACGCCCACAGTGGTTTTGACGGGTAACGGGGAATCAGGGTTCGATTCCGGAGAGGGAGCATGAGAAACGGCTACC
TCA
>HE6_SSU361449
```

ATGCATGTGTAAGCACAAGCCTTATATGGTGAAGCCGCGAATGGCTCATTACAACAGCCATAGTTTATTAGATCTTATCCTATTACATG
GATAACTGTGGTAATTCTAGAGCTAATACAAGCATTCAAGCTCAGACCTTACGGAATGAGCGCATTTATTAGAACAAAACCAATCGGAC
TTCGGTCCGCTCTTTGGTGAATCTGAATACCTTAGCAGATCGCACGGTCTTTGAACCGGCGACATATCTTTCAAATGTCTGCCTTATCA
ACTTTCGTTGGTAGTTTATGCGCCTACCATGGTTGTAACGGGTAACGGAGAATAAGGGTTCGACTCCGGAGAGGGAGCCTGAGAAACGG
CTACCACA
>HE6_SSU365256
ATGCATGTCCAAGTACAAGCTCGTCCCGAGCGAAACTGCGGATGGCTCATTAAATCAGTTATGGTTCATTGGATCGAGTACCCCCCGAC
ATGGATAACTGTGGTAATTCTAGAGCTAATACATGCAACCAAGGTCTGACCTTTGGAAGGAGCGCGTTTATTAGACCAAGACCAATCAG
ACTTTGTCTGGAATCTGGTGACTGACTAACTTTGCTGATCACACAGTCCTCGCACTGGTGACATATCTTTCAAGTGTCTGCCCTATCAA
CTTTCGATGGTAGTTTATGTGCCTACCATGGTGGTAACAGGTAACAGAGAATAAGGGTTCGACTCCGGAGAGGGAGCCTGAGAAATGGT
TACCACA
>HE6_SSU368318
ATGCATGTGTAAGTACAGACTGTACAACGGTGAAACTGCGAATGGCTCATTAGATCAGTTATGGTTCCTTAGATCGTACAATCCTACTT
GGATAACTGTGGTAATTCTAGAGCTAATACATGCAACCGAGCTCCGACCTCAGGGAAGGAGCGCATTTATTAGACCAAGACCAATCAGG
CTCTGCCTGTCATCTGGTGACTCTGAATAACTTTGCTGATCACATGGTCCTAGTACCGGTGACATATCTTTCAAGTGTCTGCCCTATCA
ACTTTCGATGGTAGTTTATGTGCCTACCATGGTTGTAACGGGTAACGGAGAATAAGGGTTCGACTCCGGAGAGGGAGCCTGAGAAATGG
CTACCACA
>HE6_SSU370544
ATGCATGTCTAAGCATAAACCGAACTAAAGTGAAGCCGCGAATAGCTCATTACAACAGCCGTTGTTTCTTGGATCTCCGCAATACTTGG
ATAACTGAGGTAATTCTTGAGCTAATACACGCAATCGAGCTCCGACCTTCGGGACGAGCGCATTTATTAGAACAAAACCAATCGGTGCT
TGCACTGTGGTTTGGTGAATCTGAATAACTGAGCAGATCGCTTCGGTCTCGTACCGGCGATGTATCCTTCAAGTGTCTGCCTTATCAAC
TTTCGATGGTAGTTTATGTGCCTACCATGGTTGTAACGGGTAACGGGGAATCAGGGTTCGATTCCTGAGAGGGAGCATGAGAAACGGCT
GCCACA
>HE6_SSU378839
ATGCACGTTTAAATATAGGCCGCTTTAAGGTGAAATCGCGAATAGCTCATTACAACAGCCATTGTTTCTTGGATCTTGCTTTCCTACTT
GGATAACTGTGGTAATTTAGGAGCTAATACATGCAACAAAAAACCGATGCAAGAGGAACACATGTATTAGAGTTAAACCAGTCGGTGAA
TCTGAATAACTCAGCAGAGCACATGGGCTAGTCCTGGTGCCATATCTTTCAACTGTTTGTCTTACCCACTTTCTAAGGTTATTTGTGTG
CCTACCATGGTTGTAACGGGTAACGGAGACTAAGGGTTCGACTCCGGAGAGGGAGCCTGAGAAACGGCTACCACT
>HE6_SSU383414
ATGCATGTCTAAGCAGAAGCCGCACAACGGTAAAGCCGCGAATGGCTCATTACAACAGCCGTCGTTTCTTGGATCTCTAATTTTACTTG
GATAACTGTGGTAATTCTAGAGCTAATACACGCACTAAAGCTCCGACCTTACGGGACGAGCGCATTTATTAGAACAAAACTAATCGCGT
TTCGGCCCGTTCGTTGGTGACTCTGAATAACTAAGCCGATCGCACGGTCTCGTACCGGCGACGTATCTTTCAAGTGTCTGCCCTATCAA
ATGTCGAAGGGAGGTGATATGCCTCCCTTGTTTTTAACGGGTAACGGGGAATCAGGGTTCGATTCCGGAGAGGGAGCATGAGAAACGGC
TACCACA
>HE6_SSU383888
ATGCATGTCTAAGAATAAACCGAAAATGGTAACTCTGTGTACGGCTCATTATATCAGCTCAAATTTATTGGATCATATCATCCTACTTG
GATACCTGTGGTAATTCTAGAGCTAATACACGCAATTCAGCCCTGACCTCACGGGAGGGGAGCATTTATTAGAACAAAACCAATTGGCC
TCGGCCATCTGTTGGTGAATCTGAATAACTACGCAGATCGCATGGTCTCGTACCGGCGACGTATCCTTCAAGGTCTGCCTTATCAACTT
TCGATGGTAGTGTATTTGCCTACCATGGTTGTAACGGGTAACGGAGAATTAGGGTTTGACTCCGGAGAGGGAGCCTGAGAAACGGCTAC
CACA
>HE7_SSU232624
ATGCATGTCTAAGCACAAGCCGATATATGGCAAACCCGCGAATGGCTCATTACAACAGCCACTGTTCACTTGATCTGTACCATATCTAC
TTGGATAACTGTGGTAATTCTAGAGCTAATACACGCACCCATGCTCCGTCCGTGAGGAACAAGCGCATTTATTAGAACAAAACCAATCG
ACTTCGGTCGTTCGTTTGTGACTCTGAATAACTTTGCTGATCGCACAGTCATTGTACTGGCGACGCATCTTTCAAGTGTCTGCCTTATC
AACTTTCGACAGTAGTTTCTGTGCCTACCGTAGTTGCAACGGGTAACGGAGAATAAGGGTTCGACTCCGGAGAGGGAGCCTGAGAAACG
GCTACCACA
>HE7_SSU256492
ATGCATGTCTAAGCATAAACCGAATATGGTAAAGCCGCGAATGGCTCATTACAACAGCCATAGTTTATTGGATCTTACTATCCTACTTA
GATAACTGTGGTAATTCTAGAGCTAATACACGCACTCAAGCCCCAACCTGACGGTAGGGGCGCATTTATTAGAACAAGACCAATTGGCC
TCGGCCATCTATTGGTGAATCTGAATAACTACGCTGATCGCACACTCTCGCAGTGGCGACGTATCCTTCAAGTGTCTGCCTTATCAACT
TTCGATGGTAGTTTATATGACTACCATGGTTGTAACGGGTAACGGAGAATAAGGGTTCGACTCCGGAGAGGGAGCCTGAGAAACGGCTA
CCACA
>HE8_SSU829972
ATGCATGTGTAAGCACAAGCCAATGAATGGTAAAGCTGCGAATGGCTCATTACAACAGCCACTGTTCATTTGATCTTAATCCATTACTT
GGATACCTGTTCTAATTGAAGAGCTAATACATGCAACTAAGTCCCAACCGCAAGGGCGGGATGCACTTATTAGACCAAAACCAATCGGG
CTTGCCCGAGGTTTGGTGACTCTGAATAATTTCGCTGACCGCACGGTCTCGCACCGGCGGCGCATCTTTCAAATGTCTGCCTTATCAAC
TTTCGATGGTAGTTTATGCGCCTACCATGGTTGTAACGGGTAACGGAGAATAAGGGTTCGACTCCGGAGAGGGAGCCTGAGAAACGGCT
ACCACA
>HE8_SSU843570
ATGCATGTCCAAGTACAAGCCTCATTAAGGTGAAACCGCGAATGGCTCATTAAATCACACCTAATGCACTGGACGTTGCCAGTTACTTG
GATACCTGTGGTAATTCTAGAGCTAATACACGCAATCAAGCCCTGACCTTACGGGAAGGGCGCATTTATTAGAACAAGACCAATTGGCT
TCGGCCATTTATTGTTGAATCTGAATAACTACGCAGATCGCACGGTCTCGCACCGGCGACATATCCTTCAAGTGTCTGCCTTATCAACT
GTCGATGGTAGTTTACATGACTACCATGGTTTTAACGGGTAACGGAGAATTAGGGTTCGACTCCGGAGAGGGAGCCTGAGAAACGGCTA
CCACA
>HE9_SSU305678
ATGCATGTCTAAGCACAAACTTTATGAGTGAAGCCGCGAAAAGCTCATTACAACAGCCGTCGTTTCTTGGATATCCGAATTTACTTGGA
TAACTGTGGTAATTCTAGAGCTAATACATGCAATCGAGCCCTGAACGTAAGTGATGGGCGCATTTATTAGTAACAAAACCGATCGGTGT
TATGCACCGTACGTTTGGTGAATCTGAATAATTGAGCAGATCGCTTTGGTCTTGTACCGGCGACGTATCTTTCAAGTGTCTGTTTTATC
AACTTTAGATGTTAGTTTATATGACTAACATGGTTGTCACGGATAACGGAGAATAAGGGTTCGACTCCGGAGAGGGAGCCTGAGAAACG
GCTACCACT
>HF1_SSU759758

ATGCAAGTGTCAGCTCAAGCCATATTATGGTTAAGCCGCGGAAAGCTCATTACAACAGCCATTGTTCACTTGATCTTGACTATCCTACT
TGGATAACTGTGGTAATTCTAGAGCTAATACGTGCAACAATGCTCAGGTAGTCCTTCGGGGCGACGAGCGCAGTTATTAGAACAAAACC
AATCGGGCTTCGGTCCGTCGGTTTGGTGGATCTGAATAACTACAGCTGATCGCACAGTCTTCGTACTGGCGACGAATCTTTCAAGTGTC
TGCCTTATCAGCTGTCGATGGTAGTCTACGTGGCTACCATGGCTGTAACGGGTAACGGAGAATAAGGGTTCGACTCCGGAGAGGGAGCC
TGAGAAACGGCTACCACA
>HF1_SSU763392
ATGCATGTCTAAGTACAAGCCGAGTTAAGGTGAAACCGCGAATGGCTCATTAAATCACACCTAATATACTGGATAGTGTCAGCTACTTG
GATACCTGTGGTAATTCTAGAGCTAATACACGCACGAAAGCCCTGACTTCGGGAGGGGCGCATTTATTAGAACAAAACCAATCGGGCTT
GCCCGTCATTTGGTGAATCTGAATAACTCAGTTGATCGCACAGTCCTCGCACTGGCGACGTATCTTTCAAGTGTCTGCCTTATCAACTT
TCGATGGTAGTTTACATGACTACCATGGTTGTAACGGGTAACGGAGAATAAGGGTTCGACTCCGGAGAGGGAGCCTGAGAAATGGCTAC
CACA
>HF1_SSU764346
ATGCATGTCTAAGCAAAAGCCTCAAAATGGTGAAGCCGCGAATAGCTCATTACAACAGCCATAGTTTATTGGATCTTCTCTCATACTTG
GATACCTGTGGTAATTCTAGAGCTAATACACGCAAACAAACTCTGAGCTCTGGCGAGGAGTGCATTTATTAGAACAAAACCAATGGACC
CCGGTCCTTGTTTGGTGAATCTGAATAACTCAGTTGATCGCACAGTCTAGCACTGGCGACGTATCTTTCAAGTGTCTGCCTTATCAACT
GTCGATGGTAGTTTATATGACTACCATGGTTGTAACGGGTAACGGAGAATAAGGGTTCGACTCCGGAGAGGGAGCCTGAGAAATGGCTA
CCACA
>HF1_SSU774294
ATGCATGTCTAAGCACACGCCTTAAAATGGTAAAGCCGCGAATGGCTCGGTATAACAGCTACGGTTTATTAGATATTAGTTGTTTACTT
GGATAACTGTGGTAATTCTAGAGCTAATACATGCACTTTAGCTCGGACCTCACGGAAAGAGCGCATTTATTAGATCAAAACCAATCGGG
CCTCGGTCCGTGTTTTGGTGACTCTGAATAACTCAGTTGATCGCACAGTCTTGTACTGGCGACGTATCTTTCAAATGTCTGCCTTATCA
ACTTTCGATGGTAGGTTATACGCCTACCATGGTTATAACGGGTAACGGAGAATCAGGGTTCGACTCCGGAGAGGGAGCCTGAGAAATGG
CTACCACA
>HF1_SSU779114
ATGCATGTCTATGCATAAGCCTAAATAAGGTGAAGTCGCGAATTGCTCATTACAACAGCCATTGTTTACTGGATCTTAATATCCTACTT
GGATAACTGTGGTAATTCTAGAGCTAATACACGCACCAAGCTCTGACCGCAAGGGATGAGCGCATTTATTAGAACAAAACCAATCGGGT
TCGTCCCGTCTTTGGTGGATCTGAATAACTCAGCTGATCGCATGGTCTCGCACCGGCGACGTATCTTCCAAGTGTCTGCCTTATCAACT
TTTGATGGTAGTTTATGCGACTACCATGGTTGTAACGGGTAACGGAGAATAAGGGTTCGACTCCGGAGAGGGAGCCTGAGAAACGGCTA
CCACA
>HF1_SSU780927
ATGCATGTGTAAGTACAAACCTGTACATGGTGAAACTACGAATGGCTCATTAAATCAGTTGTGGTTCCTTAGATCGTTTTACAGTTTGG
ATAACTGTAGTAATTCTAGAGCTAATACACGCAACAAGCTCTGACCTCTCGGGGAAAGAGTGCATTTATTAGAACAAAACCAATCGGGC
TTCGGTCTGTCAATTGGTGAATCTGAATAACTCAGCTGATCGCACGGTCTTGTACCGGTGACGCATCTTTCAAGTGTCTGCCTTATCAA
CTGTTGATGGTAGTTTATGTGACTACCATGGTTGTAACGGGTAACGGAGAATAAGGGTTCGACTCCGGAGAGGGAGCCTGAGAAACGGC
TACCACA
>HF2_SSU192072
ATGCATGTGTAAGAATAAACCGAATATGGTAAATCCGCGAATGGCTCATTATTCAGCCACAAATCATTGGATCTAATCAGTTACTTGGA
TAACTGTTCAAAAGGAAGAGCTAAGACATGCCTCGAAGGCCAAGCGCAAGCTTGGTCGCACTTCTTAGAAAAGACCAATTGGCCTCGGC
CATCCATTGGTGAATCTTCCGAAGAAAGCAGATCGCACGGTCTAGTACCGGCGACATATCCTTCATGTGTCTGCCTTATCAACTGTCGA
TGGTAGTTTATTGGACTACCATGGTTGTAACGGGTAACGGAGAATTAGGGTTCGACTCCGGAGAGGGAGCCTGAGATACGGCTACCACA
>HF2_SSU204352
ATGCATGTCTAAGTACAGGCCTCACTAAGGTGAAACCGCGAATGGCTCATTAAATCACACCTAATATACTGGATAGTATCAGTTACTTG
GATAACTGCGGTAATTCTAGAGCTAATACACGCACCCATGCTCCGTCCGTGAGGAACGAGTGCATTTATTAGAACAAAACCAATCGACT
TCGGTCGTTCGTTTGTGACTCTGAATAACTTTGCTGATCGCACAGTCATTGTACTGGCGACGCATCTTTCAAGTGTCTGCCTTATCAAC
TTTCGACGGTAGTTTCTGTGCCTACCGTGGTTGCAACGGGTAACGGAGAATAAGGGTTCGACTCCGGAGAGGGAGCCTGAGAAACGGCT
ACCACA
>HF2_SSU205129
ATGCATGTCTAAGTACAGGCCTCACTAAGGTGAAACCGCGAATGGCTCATTAAATCACACCTAATATACTGGATAGTATCAGTTACTTG
GATAACTGCGGTAATTCTAGAGCTAATACACGCACTCAAGCCCCAACCTGACGGTAGGGGCGCATTTATTAGAACAAGACCAATTGGCC
TCGGCCATCTATTGGTGAATCTGAATAACTACGCTGATCGCACACTCTCGCAGTGGCGACGTATCCTTCAAGTGTCTGCCTTATCAACT
TTCGATGGTAGTTTATATGACTACCATGGTTGTAACGGGTAACGGAGAATAAGGGTTCGACTCCGGAGAGGGAGCCTGAGAAACGGCTA
CCACA
>HF2_SSU208147
ATGCATGTCTATGCACGAGCCGAAAATGGTGAAGCCGCGAATGGCTCATTACAACAGCCTTGGTTTATTGGATCTATTATATCCACTTG
GATAACTGTGGTAATTCTAGAGCTAATACACGCAACTAAACTCCAACCTTGCGGAAGGAGTGCATTTATTAGAACAAAACCAATCGGCT
TCGGCCGTTACTTGGTGAATCTGAATAACTCAGTTGATCGCACGGTCTTTGTACTGGCGACATATCTTTCAAGTGTCTGCCTTATCAAC
TTTCGATGGTAGTTTAGACGACTACCATGGTTGTTACGGGTAACGGAGAATTAGGGTTCGACTCCGGAGAGGGAGCCTGAGAAATGGCT
ACCACA
>HF2_SSU210357
ATGCATGTCTAAGCACAAACTATTTAATTGTGAAGCCGCGAATGGCTCATTATAACAGCCATTGTTTACTGGATATATTTTTACTACAT
GGATAACTGTGGTAATTCTACAGCTAATACACGCAGCAGAACCCCGACTTAATGAAGGGGTGCGTTTGTTACTTCAAACCAATCAGGCT
TCGGTCTGAATTCAAGTGATATTGAACAATTTAGCTGATCGAACGGTCTATGAACCGACGACATATCCTTCAAACGTCTGCCTTATCAA
CTTTCGATGGTAGCCTACACGTCTACCATGGTTGTAACGGGTAACGGAGAATCAGGGTTTGATTCCGGAGAGGGAGCCTGAGAAATGGC
TACCACA
>HF3_SSU989895
ATGCAAGTGTCAGCTCAAGCCTATGTATGGTTAAGCCGCGAATGGCTCATTACAACAGCCACTGTTTACTTGATCTTGATAATCCTACT
TGGATAACTGTGGTAATTCTAGAGCTAATACATGCAACTATGCTCCGACCTTACGGGACGAGCGCAACTATTAGAACCAAACCAATCGG
GTTTCGGCCCGTTCGGTGGTGAATCTGAATAACTGTTTGCTGATCGCACGGTCTTTGCACCGGCGACGCATCTTTCAAGTGTCTGCCTT
ATCAACTTTCGATGGTAGTTTATGTGCCTACCATGGTTGTAACGGGTAACGGAGAATAAGGGTTCGACTCCGGAGAGGGAGCCTGAGAA
ACGGCTACCACA
>HF3_SSU990962

ATGCATGTCTATGCATAAACCGAATATGGTAAAGCCGCGCATGGCTCATTACAACAGCCATAGTTTATAGGATCTTACTATCCTACTTT
GATAACTGTGGTAATTCTAGAGCTAATACACGCACTCAAGCCCCAACCTGACGGTAGGGGCGCATTTATTAGAACAAGACCAATTGGCT
TCGGCCATCTATTGGTGAATCTGAATAACTACGCTGATCGCACACTCTCGAAGTGGCGACGTATCCTTCTAAGGTGTGCCCTATCAACT
GTCGACTGTGGCATAGACGCCCACAGTGGTTTTGACGGGTAACGGGGAATCAGGGTTCGATTCCGGAGAGGGAGCCAGAGAAACGGCTA
CCACA
>HF4_SSU606153
ATGCATGTCTAAGCATAAGCCGAATATGGTAAAGCCGCGAATGGCTCATTACAACAGCCATAGTTTATTGGATCTTAGAGTCCTACTTG
GATACCTGTGGTAATTCTAGAGCTAATACACGCATTCAAGCCCCAACCTGACGGCGGGGCGCATTTATTAGAACAAAACCAATTGGCT
TCGGCCATTCATTGGTGAATCTGAATAACTACGCTGATCGCACGGTCTCGCACCGGCGACGTATCCTTCAAGTGTCTGCCTTATCAACT
TTCGATGGTAGTTTACATGACTACCATGGTTGTAACGGGTAACGGAGAATAAGGGTTCGACTCCGGAGAGGGAGCCTGAGAAACGGCTA
CCACA
>HF4_SSU614317
ATGCATGTCTAAGTACATACCTTCACACGGTGAAACTGCGAATGGCTCATTAAATCAGTTATGGTTCCTTAGATCGATACACTCCTACT
TGGATAACTGTGGCAATTCTAGAGCTAATACACGCACTAAAGCTCCGACCTTACGGGACGAGCGCATTTATTAGAACAAAACCAATCGG
GTTTCGGCCCGTTCGTTGGTGACTCTGAATAACTAAGCCGATCGCACGGTCTCGTACCGGCGACGTATCTTTCAAGTGTCTGCCTTATC
AACTTTCGATGGTAGTTTATGTGCCTACCATGGTTGTAACGGGTAACGGAGAATAAGGGTTCGACTCCGGAGAGGGAGCCTGAGAAACG
GCTACCACA
>HF4_SSU619471
ATGCACGTTCTAATATGAGCATTAAAAATGTGAAATCGCGAATAGCTCATTACAACAGCCATTGTTTCTTGGATCTTATATTCCTACTT
GGATAACTGTGGTAATTCTAGAGCTAATACATGCAATGAAGTTCCAACGCAAGAGGAATGCATTTATTAGAGCTAAACCAATCAGGGGC
AACCATGTTTGTTTGGTGAATCTGAATAACTTAGCAGAGCACATGGGCTAGTCCTGGTGCCATATCTTTCAAGTGTCTGCCTTATCAAC
TTTCGATGGTAGTTTATGTGCCTACCATGGTTGTAACGGGTAACGGAGAATAAGGGTTCGACTCCGGAGAGGGAGCCTGAGAAACGGCT
ACCACA
>HF4_SSU620879
ATGCATGTGTAAGTACAGACTGTACAACGGTGAAACTGCGAATGGCTCATTAGATCAGTTATGGTTCCTTAGATCGTACAATCCTACTT
GGATAACTGTGGCAATTCTAGAGCTAATACACGCACTCAATCCCTGACTTCGGAAAGGGAGCATTTATTAGAACAAGACCAATTGGCTT
CGGCCATCTATTGGTGAATCTGAATAACTACGCAGATCGCACAGGCTCGTCCTGGCGACATATCCTTCAAGTGTCTGCCTTATCAACTG
TCGATGGTAGTTTATTGGACTACCATGGTTGTAACGGGTAACGGAGAATTAGGGTTCGACTCCGGAGAGGGAGCCTGAGAAACGGCTAC
CACA
>HF4_SSU622464
ATGCATGTCTATGCACAAGCCGAAAATGGTGAAGCCGCGAATGGCTCATTACAACAGCCACTGTTTACTTGATCTTGATTATCCTACTT
GGATAACTGTGGTAATTCTAGAGCTAATACATGCCAAGATGCTCCGACCTTACGGGACGAGCGCACTTATTAGACCAAGACCAATCGGG
CTTCGGCTCGTAGTCTGGTGACTCTGAATAACTCTGCCGATCGCACGGTCTTTGTACCGGCGACGCATCTTTCAAGTGTCTGCCTTATC
AACTTTCGATGGTAGTTTCTGTGCCTACCATGGTTGTAACGGGTAACGGAGAATAAGGGTTCGACTCCGGAGAGGGAGCCTGAGAAACG
GCTACCACA
>HF4_SSU624085
ATGCATGTCTAAGTACAAGCCTCATTAAGGTGAAACCGCGAATGGCTCATTAAATCACACCTAATATACTGGATAGTATCAGTTACTTG
GATAACTGCGGTAATTCTGGAGCTAATACATGCGTTCAAGCCCCAAACTTGCGTGCGGGGCGCTTTTATTAGACCAAGACCAATCAGGC
ATTGCCTGGAATCTGGTGACTCTGAATAACTTTGCTGATCACATGGTCCTAGCACCGGTGACATATCTTTCAAGTGTCTGCCCTATCAA
CTTTCGATGGTAGTTTATGTGCCTACCATGGTGGTAACGGGTAACGGAGAATAAGGGTTCGACTCCGGAGAGGGAGCCTGAGAAATGGC
TACCACA
>HF4_SSU625424
ATGCATGTCTATGCAGAAGCCGAACTATCGCAAAGCTGCAAATGGCTCATTACAACAACTTTTGTTTCTTGGATCTCTCTGATCTACTT
GGATACCTGTGGTAATTCTAGAGCTAATACACGCACCAAATCTCTGACCTTTGGGGATGAGTGCATTTATTAGAACAAAATCAATCGGG
CTCTGCTCGTATGATGGTGACTCTGAATAACTACGCTGATCGCATGGTCTCATACTGGCGACGTATCTTCCAAGTATCTGCCTTATCAA
CTGTTGATGGTAGTTTATTTGCCTACCATGGTTGTAATGGGTAATGGAGAATAAGGGTTCAACTCCAGAGAGGGAGCCCGGGAAACGGC
TACCACA
>HF4_SSU628562
ATGCATGTCTAAGCATGAGCCGTACTATGGTGAAGCCGCGAACAGCTCATTACAATAGCCGTTGTTTCTTGGATCTCCAAACACTACAT
GGATAACTGTGGTAATTCTAGAGCTAATACATGCGTACATGCGTCAACTGGTGCAAGCCGGGAGACGTGCATTTATTGGATCAGAACCA
TCCGGCCTTCGGGCCGTACTCTGGTGAATCTAAATAACTGAGCGGAGCACACGCTCTTGCAGCGGTGCCAGTTCGTTCAAGTGTCTACC
CTATCAACTTGCGATGGTAAATTACAAGCTTACCATGGTGGTAACGGGTAACGGAGAATCAGGGTTCGATACCGGAGAGGGAGCCTGAG
AAATTGCTACCACA
>HF4_SSU631524
ATGCATGTCTAAGTACACTTCCTTGTATGGAGAAACTGCGAATGGCTCATTACAACAGCCACTGTTCACTTGATCTGTATTATCCTACT
TGGATAACTGTGGTAATTCTAGAGCTAATACACGCACCCATGCTCCGACCTGAGGGGACGAGCGCATTTATTAGAACAAAACCAATCGG
GCCTCGGCCTGTTTCGTTTGTGACTCTGAATAACTCTGCTGATCGTACGGTCCCGTACCGACGACGCATCTTTCAAGTGTCTGCCTTAT
CAACTTTCGATGGTAAGTTCTGTGCTTACCATGGTTGTAACGGGTAACGGAGAATAAGGGTTCGACTCCGGAGAGGGAGCCTGAGAAAC
GGCTACCACA
>HF4_SSU632264
ATGCATGTCTATGCACAAGCCGATTCGGCGAAGCCGCGAATGGCTCATTACAACAGCTGTGGTTTCTTGGATCTTTCAATCCTACTTGG
ATAACTGTGGCAATTCTAGAGCTAATACACGCACTGAAACCGCGGTCCTTGGGCTGCGGTGCATTTATTAGAACAAAGCCAACCGGGCC
TTGGCCTGACTGCTTGGCGAATCTGAATAACCTGGCTGATCGCACGGTCTCGCACCGGCGACGCATCTTTCAAGTGTCTGCCTTATCAA
CTGTCGATGGTAGGTTACGTGCCTACCATGGTTGTAACGGGTAACGGAGAATAAGGGTTCGACTCCGGAGAGGGAGCCTGAGAAATGGC
TACCACA
>HF4_SSU635045
ATGCATGTGTAAGAATAAACCGAATATGGTAAATCCGCGAATGGCTCATTATTCAGCCTCAATTTATTGGATCTAATCAGTTACTTGGA
TAACTGTTCAAAATGAAGAGCTAAGACATGCCTCGAAAATCCAGCGCAAGCCGGATTGCACTTCTTAGAAAAGACCAATTGGCTTCGGC
CATCCATTGGTGAATCTTCTGAAATTCGCAGATCGCACGGTCTAGTACCGGCGACATACCCTTCAAATGTCTTCCTTATCAACTGTCGA
TGGTAGTTTATTGGACTACCATGGTTGTAACGGGTAACGGAGAATTAGGGTTCGACTCCGGAGAGGGAGCCTGAGATACGGCTACCACA
>HF5_SSU991188

```
ATGCATGTCTAAGCACAACTATTTTATTTGTGAAGCCGCGAATGGCTCATTACAACAGCCATAGTTCACTGGATATATTCCTTTTACAT
GGATAACTGAGGTAATTCTTCAGCTAATACACGCTTCAAAACCCCGACTTTTTGGAGGGGTGCGTTTGTTACTTCAAATCAATCGGGTT
TCGGCCCGTTTATAAGTGATATTGAACAATTTAGCTGATCGCACGGTCTGAGCACCGGCGACATATCCTTCAAATGTCTGCCTTATCAA
CTTTCGATGGTAGATTATGCGCCTACCATGGTTGTTACGGGTAACGGAGAATCAGGGTTTGATTCCGGAGAGGGCGCCTGAGAGACGGC
GGCCACA
>HF5_SSU995414
ATGCATGTCTAAGTACAAGCCGAGTTAAGGTGAAACCGCGAATGGCTCATTAAATCACACCTAATATACTGGATAGTGTCAGTTACTTG
GATAACTGTGGTAATTCTAGAGCTAATACAAGCATTCAAGCTCAGACCTTACGGAATGAGCGCATTTATTAGAACAAAACCAATCGGAC
TTCGGTCCGCTCTTTGGTGAATCTGAATACCTTAGCAGATCGCACGGTCTTTGAACCGGCGACATATCTTTCAAATGTCTGCCTTATCA
ACTTTCGTTGGTAGTTTATGCGCCTACCATGGTTGTAACGGGTAACGGAGAATAAGGGTTCGACTCCGGAGAGGGAGCCTGAGAAACGG
CTACCACA
>HF6_SSU329881
ATGCATGTCTAAGCATGAACCGAATATGGTGAAGCCGCGAATGGCTCATTACAACAGCCGTTGTTTCTTGGATCTTGATTTACTTGGAT
AACTGTGGTAATTCTAGAGCTAATACATGCAACCAAGCTCTGACCTTGGAAGGAGCGCGTTTATTAGACCAAGACCAATCAGACTTTGT
CTGGAATCTGGTGACTCTGAATAACTTTGCTGATCACACAGTCCTCGCACTGGTGACATATCTTTCAAGTGTCTGCCCTATCAACTTTC
GATGGTAGTTTATGTGCCTACCATGGTGGTAACGGGTAACGGAGAATAAGGGTTCGACTCCGGAGAGGGAGCCTGAGAAATGGCTACCA
CA
>HF6_SSU338435
ATGCATGTCTACGTACAGAGATTTTTCTCGAAACCGCGAACGGCTCATTACAACAGCCATAGTTTATTGGATCTACAAATCCTACTTGG
ATACCTGTGGTAATTCTAGAGCTAATACACGCAACAGATCCCTGACCTTGCGGAAGGGGAGCATTTATTAGAACAAAACCAATTGGCTT
CGGCCATTAGTTGGTGAATCTGAATAACTACGCAGAGCATACGAGCTCGTCTCGATGCCATATCCTTCAAGTGTCTGCCCTATCAACTG
TCGATGGTAGGTGATATGCCTACCATGGTTGCAACGGGTAACGGGGAATCAGGGTTCGATTCCGGAGAGGGAGCCTGAGAAACGGCTAC
CACG
>HF6_SSU338739
ATGCATGTCTAAGCACAAGCTGATTTAAAGTGAAGCCGCGAATAGCTCATTACAACAGTCATAGTTTACTTGATTTTGTTTTACTTGGA
TAACTGTGGTAATTCTAGAGCTAATACATGCTGCAAGCTCCAACAAGTTCTTTGGACTTGAAGGAGTGCATTTATTAGAACAAAACCAA
TCGGGTCTTTGACCTGTCCATTGGTGAATCTGAATAACTCAGCTGATCGCACAGTCCTGAACTGGCGACGTATCTTTCAAGTGTCTGCC
TTATCAACTGTTGATGGTAGTTTATGTGACTACCATGGTTGTAACGGGTAACGGAGAATAAGGGTTCGACTCCGGAGAGGGAGCCTGAG
AAACGGCTACCACA
>HF7_SSU385021
ATGCATGTCTAAGAATAGGGATTATTCCCAAATCCGCGAATGGCTCATTACAACAGCCTTAGTTTATTGGATCTACAAATCCTACATGG
ATACCTGTGGTAATTCTAGAGCTAATACACGCAAGAAAGCCCCGACCTTACGGGAGGGGTGCATTTATTAGAACAAGACCAATTGGCTT
CGGCCATCCATTGGTGAATCTGAATAACCTAGCAGAGCATACGAGCTCGTCTCGATGCCATATCCTTCAAGTGTCTGCCCTATCAACTG
TTGACGGTAGGTTACATGCCTACCGTGGTTGTAACGGGTAACGGAGAATTAGGGTTCGACTCCGGAGAGGGAGCCTGAGAAACGGCTAC
CACA
>HF7_SSU390110
ATGCATGTCTAAGCACAAGCTTAAACAAAGTGAAGCCGCGAATAGCTCATTACAACAGCCATTGTTTACTTGATCTTGAAATCCTACTT
GGATAACTGTGGTAATTCTAGAGCTAATACACGCAACAAGCTCTGACCCCTCGGGGAAAGAGTGCATTTATTAGAACAAAACCAATCGG
ACTTCGGTCTGTCAATTGGTGAATCTGAATAACTCAGCTGATCGCACGGTCTTGTACCGGTGACGCATCTTTCAAGTGTCTGCCTTATC
AACTGTTGATGGTAGTTTATGTGACTACCATGGTTGTAACGGGTAACGGAGAATAAGGGTTCGACTCCGGAGAGGGAGCCTGAGAAACG
GCTACCACA
>HF7_SSU398053
ATGCATGTCTAAGTACAAGCCGAGTTAAGGTGAAACCGCGAATGGCTCATTAAATCACACCTAATATACTGGATAGTGTCAGTTACTTG
GATAACTGCGGTAATTCTAGAGCTAATACATGCAACAATGCTCAGGTAGCCCTTCGGGGTGACGAGCGCAATTATTAGAACAAAACCAA
TCGGGCTTCGGTCCGTTCGGTTTGGTGGATCTGAATAACTATAGCTGATCGCACAGTCTTTGCACTGGCGACGCATCTTTCAAGTGTCT
GTCCTATCAGCTGTCGATGGTATTCTATACGATTACCATGGCTGTAACGGGTAACGGAGAATAAGGGTTCGACTCCGGAGAGGGAGCCT
GAGAAACGGCTACCACA
>HF7_SSU407024
ATGCATGTCTAAGCAAAATGGTGAAGCCGCGAATAGCTCATTACAACAGCCATAGTTTATTGGATCTTGACTTCCTACTTGGAATACCT
GTGGTAATTCTAGAGCTAATACATGCAAATAAGCTCCAACCTCTGGGCGGAGCGCATTTATTAGAACAAAACCAATCGGCTTCGGCCGT
AGTTTGGTGAATCTGAATAACTCAGCTGATCGCATGGTCCTCGCACTGGCGACGTATCTTTCAAGTGTCTGCCTTATCAACTTTCGATG
GTAGTTTACACGCCTACCATGGTTGTAACGGGTAACGGAGAATAAGGGTTCGACTCCGGAGAGGGAGCCTGAGAAACGGCTACCACA
>HF7_SSU407761
ATGCATGTCTAAGCACAAGCTTAAATAAAGTGAAGCCGCGAATAGCTCATTACAACAGCCATTGTTCACTTGATCTTGAAATCCTACTT
GGATAACTGTGGTAATTCTAGAGCTAATACACGCAAAAAGCTCTGACCTTACAGGGAAGAGTGCATTTATTAGAACAAAACCAATCGGG
CTTCGGCCTGTCAATTGGTGAATCTGAATAACTCAGCCGATTGCACGGTCTTGAACCGGCAACATATCTTTCAAGTGTCTGCCTTATCA
ACTGTTGATGGTAGTTTTTGTGACTACCATGGTTGTAACGGGTAACGGAGAATAAGGGTTCGACTCCGGAGAGGCAGCCTGAGAAACGG
CTACCACA
>HF7_SSU409331
ATGCATGTCTAAGTACAAGCCCCATTAAGGTGAAACCGCGAATGGCTCATTAAATCACACCTAATATAATGGATAGTGTCCGTTACTTT
GATAACTGTGGTAATTCTAGAGCTAATACACGCACATAAGCTCTAACCGTAAGGAAAGAGCGCATTTATTAGAACAAAAGCATCCGGCA
TTTAGCCGTTACTTGGTGAATCTGAATAACTCTGCTAATCGCACAGTCAGAGTACTGGAGATGCATCTTTCAAGTGTCTGCCTTATCAA
CTTTCGATGGTAGTTTATATGCCTACCATGGTTATAACGGGTAACGGAGAATAAGGGTTCGACTCCGGAGAGGGAGCCTGAGAAATGGC
TACCACA
>HF8_SSU795426
ATGCATGTCTAAGCAGAAGCCGAAAATGGTGAAGCCGCGAATAGCTCATTACAACAGCCTTAGTTTCTTGGATCTTCAGTTCCTACTTG
GATAACTGTGGTAATTCTAGAGCTAATACACGCTAACAAGTCCCGGCCTCTGGCTGGGCTGCATTTATTAGACCAAAACCAATCGGACT
CTGGTCCGTGTATGGTGAATCTGAATAACTTTGCTGATCGCACGGTCCTCGTACCGGCGACGCGTCTTTCAAGTGTCTGGCTTATCAAC
TTTCGATGGTAGTTTATACGACTACCATGGTTGTAACGGGTAACGGAGAATAAGGGTTCGACTCCGGAGAGGGAGCCCGAGAAATGGCT
ACCACA
>HF9_SSU14048
```

ATGCATGTCCAAGTACAAGCCTCATTAAGGTGAAACCGCGAATGGCTCATTAAATCACACCTAATGCACTGGACGTTGCCAGTTACTTG
GATAACTGCGGTAATTCTGGAGCTAATACACGCATACAAGTTCCGACGTAAGAAGGAATGCATTTATTAGAACTAAACCAATCGGGCTT
GCCCGTTGTTTGGTGAATCTGAATAACTCCGCAGATCGCATGGTCTAGCACCGGCGACATATCTTTCAAGTGTCTGCCTTATCAAATGT
CGAAGGGACGTGATATGCCTCCCTTGTTTGTAACGGGTAACGGGGAATCAGGGTTCGATTCCGGAGAGGGAGCATGAGAAACGGCTACC
ACA
>HF9_SSU14296
ATGCACGTGCAAGCACAAGCCAAATTATGGTGAAGCCGCGAATGGCTCATTACAACAGCTATTGTTTATTGGATTTGATGTTCACTTGG
ATAACTGTGGTAATTCTAGAGCTAATACATGAAACCTAGCTTTTGCCGCAAGGTTAGAGCGCAATTATTAGACTAAAACCAACCAGCTT
TAGCTGATACCTGGTGAATCTGAATAAACCTGCTTATTGCGTAATCAAAGCACTGGCGAAGCATCTTTCAAGTGTCTGCCTTATCAACT
GTTGAAGGTAGTTTACATGCCTACCATGGTTATAACGGGTAACGGAGAATAAGGGTTTTACTCCGGAGAGGGAGCCTGAGAAATGGCTA
CCACA
>HF9_SSU17250
ATGCATGTCTAAGTACAGGCCTCACTAAGGTGAAACCGCGAATGGCTCATTAAATCACACCTAATATACTGGATAGTATCAGTTACTTG
GATAACTGCGGTAATTCTGGAGCTAATACATGCAACAAAGCCCTATTGCAGGGCGCATTTATTAGGACAAAACCAATCGGGCTTGCCCG
CTTATTGGTGGATCTGAATAACTTTGCTAATCGCACAGTCCTCGTACTGGCGATGTATCTTTCAAATATCTGCCTTATCAACTGTCGAT
GGTAGCTTACACGACTACCATGGTTGTTACGGGTAACGGAGAATTAGGGTTCGACTCCGGAGAGGGAGCCTGAGAAACGGCTACCACA
>HF9_SSU17844
ATGCACGTCTAAGCACAAGCCAAATTGAAGGTAAAGCCGCGAATGGCTCATTACAACAGCCATTGTTTATTGGATCTGTTTAACTTGGA
TAACTGTGGTAATTCTAGAGCTAATACACGCACATAAGCTCTAACCGTAAGGAAAGAGCGCATTTATTAGAACAAAACCATCCGGCATT
TAGCCGTTACTTGGTGAATCTGAATAACTCTGCTAATCGCACAGTCAGAGTACTGGCGATGCATCTTTCAAGTGTCTGCCTTATCAACT
TTCGATGGTAGTTTATATGCCTACCATGGTTATAACGGGTAACGGAGAATAAGGGTTCGACTCCGGAGAGGGAGCCTGAGAAATGGCTA
CCACA
>HF9_SSU18227
ATGCATGTCTAAGTACAAGCCTCATTAAGGTGAAACCGCGAATGGCTCATTAAATCACACCTAATGTACTGGACAGTCTCAGTTACTTG
GATACCTGTGGTAATTCTAGAGCTAATACACGCAAAAAAGCCCTGACTTTACGGAAAGGGCGCATTTATTAGAACAAGACCAATTGGCT
TCGGCCATCCATTGGTGACTCTGAATAACTACGCAGATCGCACGGTCTCGTACCGGCGACATATCCTTCAAGTGTCTGCCTTATCAACT
TTCGATGGTAGTTTACATGACTACCATGGTTGTAACGGGTAACGGAGAATAAGGGTTCGATTCCGGAGAGGGAGCCTGAGAAACGGCTA
CCACA
>HF9_SSU19963
ATGCATGTCTAAGCACAAGCCTAATACGGTGAAGCCGCGAATGGCTCATTACAACAGCCTTTGTCTATTTGATGTTGAAAATCTACTTG
GATAACTGTGGAAAAGCCAGAGCTAATACATGCTTACAAACTCCGACCTTGCGGAAGGGGTGCATTTATTAGTGCAAAACCAATCGGGC
GCTTCGCGTCCCGTCGTATGGTGAATCTGAATAACTAAGCCGATCGTACGGTCCTCGCACCGGCGACGAATCATTCGAGGTTCTGCCAT
ATCAACTTTGACGGTAGTTTACGTGACTACCATGGTTATAACGGGTAACGGAGAATAAGGGTTCGACTCCGGAGAAGCAGCCTGAGAAA
CGGCTACTACA
>HF9_SSU20251
ATGCACGTTTAAATACAAGCCTTAAAATGGTGAAATCGCGAATAGCTCATTACAACAGCCATTGTTTCTTGGATCTTACTTTCTACTTG
GATAACTGTGGTAATTCTAGAGCTAATACACGCATACAAGTTCCGACGCAAGAAGGAATGCATTTATTAGAACTAAACCAATCGGGCTT
GCCCGTTGTTTGGTGAATCTGAATAACTCCGCAGATCGCATGGTCTAGCACCGGCGACATATCTTTCAAGTGTCTGCCTTATCAACTTT
CGATGGTAGTTTATGTGACTACCATGGTTGTAACGGGTAACGGAGAATTAGGGTTCGACTCCGGAGAGGGAGCCTGAGAAACGGCTACC
ACA
>HF9_SSU22538
ATGCATGTGTAAGTACAGACTGTACAACGGTGAAACTGCGAATGGCTCATTAGATCAGTTATGGTTCCTTAGATCGTACAATCCTACTT
GGATAACTGTGGTAATTCTAGAGCTAATACATGCACTTTAGCTCGGACCTCACGGAAAGAGCGCATTTATTAGATCAAAACCAATCGGG
CCTCGGTCCGTGTTTTGGTGACTCTGAATAACTCAGTTGATCGCACAGTCTTGTACTGGCGACGTATCTTTCAAATGTCTGCCTTATCA
ACTTTCGATGGTAGGTTATACGCCTACCATGGTTATAACGGGTAACGGAGAATCAGGGTTCGACTCCGGAGAGGGAGCCTGAGAAATGG
CTACCACA
>TF1_SSU676746
ATGCATGTCTAAGCACAAGCCAAAAATGGTGAAGCCGCGAATGGCTCATTACAACAGCCGTAGTTTCTTGGATCTTTCTAATCTACTTG
GATAACTGTGGTAATTCTAGAGCTAATACACGCACAAAAGCTCTATCCGAATGGAAGAGCGCATTTATTAGAACTAAACCGACCGGGTG
TTCGCACCTGTCGTTTGGCGAATCTGAATAACTGAGCTGATCGCACGGTCCTAGTACCGGCGACACATCTTTCAAGTGTCTGCCTTATC
AACTGTCGATGGTAGTTTATGTGCCTACCATGGTTGTAACGGGTAACGGAGAATCAGGGTTCGACTCCGGAGAGGGAGCCTGAGAAATG
GCTACCACA
>TF1_SSU677162
ATGCATGTCTAAGTACAGCTCTCGTATAGTGAAACCGCGAATGGCTCATTAAATCAGTTACTATTTCTTAGATCTTACTTTTGTTACTT
GGATAACTGTGGTAATTCTAGAGCTAATACACGCACCAAAGCTCTGACCTTACGGGAAGAGTGCATTTATTAGAACAAAACCAATCGGG
CTCTGCCTGCTGTTTGGTGAATCTGAATAACTGAGCTGATCGCACTGGTCTAGTACCGGCGACATATCTTTCAAGTGTCTGCCTTATCA
ACTTTCGATGGTAGCTTATGTGGCTACCATGGTTGTAACGGGTAACGGAGAATAAGGGTTCGACTCCGGAGAGGGAGCCTGAGAAACGG
CTACCACA
>TF1_SSU681557
ATGCATGTCTCAGCACATGCCTATGAACGGCTAAGCCGCGAATGGCTCATTACAACAGCTATGTTTATTGGGTCTAATCAGTTACTTGG
ATAACTGTGGAAAATCCAGAGCTAATACGTGCACTAACTCCGACCGTAAGGAAGGAGGCATTTTATTAGAACAAAACCAATCGGACTTG
TCCGTGGTTTGTTGACTCTGAATAACGGCAGCGATCGTACGGTCTTTGAACCGACGACATATCTTTCAAATGTCTGCCTTATCAACTTT
CGATGGTAGTTTATGCGCCTACCACGGTTGTAACGGGTAACGGAGAATCAGGGTTTGATTCCGGAGAGGGAGCCTGAGAAATGGCTACC
ACA
>TF1_SSU688192
ATGCATGTCTAAGTGCAAGCTGAAATAAAGTGAAACCGCGAAGGGCTCATTACAACAGCCGTTGTTTCCTGGAGACTCAAAATACTTGG
ATAACTGTGGTAATTCTAGAGCTAATACATGCAGCCAAGCTCTGACCGTAAGGGAAGAGTGCATTTATTAGAACAAAACCAATCGGACT
TGTCCGTAGTTTGGTGAATCTGAATAACCTAGCTGATCGCACAGTCTTTGAACTGGCGACGCATCTTTCAAGTGTCTGCCTTATCAACT
TTCGATGGTAGTTTATGTGACTACCATGGTTATAACGGGTAATGGAGAATCAGGGTTCGACTCCGGAGAGGGAGCCTGAGAAATGGCTA
CCACA
>TF1_SSU692690

ATGCATGTCTAAGCATAAACCGAATATGGTAAAGCCGCGAATGGCTCATTACAACAGCCATAGTTTATTGGATCAGTATTTTCCACTTG
GATAACTGTGGTAATTCTAGAGCTAATACATGCGATGAAACTCTGACCTTGGTGGAAAGAGTGCATTTATTAGAACAAAACCAATCGGC
TTTTGTCGTTACTTGGTGAATCTGAATAACTCGGTTGATCGCACAGTCTTTGTACTGGCGACATATCTTTCAAGTGTCTGCCTTATCAA
CTTTCGATGGTAGTTTATACGACTACCATGGTTTTGACGGTAACGGAGGATTAGGGTTTGACTCCGGAGAGGGAGCCTGAGATATGGC
TACCACA
>TF1_SSU694267
ATGCATGTCTAAGCTCAAGCCGAAAATGGTGAAGCCGCGAATGGCTCATTACAACAGCCTTTGTTTATTTGATCTTGAAAATTACTTGG
ATAACTGTGGTAATTCTAGAGCTAATACACGCGAAAAAATTCGGAACTTCTGTTCCGAGTGCATTTATTAGTACAAGACCAACCGGCGC
AAGCCGTAAATTGTCGAATCTGAATAACTGAGCCGATCGCATGGCCTCGTGCCGGCGACGTATCTTTCAAGTGTCTGCCCTATCAACTT
TTGATGGTAGTTTATGTGACTACCATGGTGATCACGGGTAACGGAGAATAAGGGTTCGACTCCGGAGAAGCAGCCTGAGAAACGGCTAC
TACA
>TF1_SSU694751
ATGCATGTGTAAGTATAAGCTTTTAGAACGGTGAAACCGCGAATGGCTCATTAGATCAGTTAATATTTATTAGATCGTAGAAAGTTACT
TGGATAACTGTGGTAATTCTAGAGCTAATACACGCAATCAAGCCCTGACCTGACGGGATGGGCGCATTTATTAGAATAAGACCAATTGG
CTTCGGCCATTTATTGGTGACTCTGAATAACTACGCAGATCGCATGGTCTCGTACCGGCGACATATCCTTCAAGTGTCTGCCTTATCAA
CTTTCGATGGTAGTTTACACGACTACCATGGTTGTAACGGGTAACGGAGAATAAGGGTTCGACTCCGGAGAGGGAGCCTGAGAAACGGC
TACCACA
>TF1_SSU698227
ATGCATGTCTAAGCACAAACCGATTTATGGTAAAGCCGCGAATTGCTCATTACAACAGCCATTGTTTACTGGATCTTGATAAGTTACTT
GGATAACTGTGGTAATTCTAGAGCTAATACACGCACCAAAGCTCTGACTTTATGGAAGAGCGCATTTATTAGATCAAAACCAATCAGGC
TTTGCCTGCTGTTTGGTGAATCTGAATAACTGAGCTGATCGCATTGGTCTAGTACCGGCGACATATCTTTCAAGTGTCTGCCTTATCAA
CTGTCGATGGTAGCTTATGTGGCTACCATGGTTATAACGGGTAACGGAGAATAAGGGTTCGACTCCGGAGAGGGAGCCTGAGAAACGGC
TACCACA
>TF1_SSU700188
ATGCATGTCTCAGTACAAGCTGCATTAAAGTGAAACCGCGAATGGCTCATTACAACAGCCGTTGTTCCCTAGAGACTATCAATACTTGG
ATAACTGTGGCAATTCTAGAGCTAATACACGCAGAGAAACCCCGAACTTGCGTGAGGGGTGCATTTATTAGAACAAAACCAATCGGGGC
AACCCGTAAGTTGGTGAATCTGAATAACCCAGCTGATCGCACGGTCTTTGCACTGGCGACGCATCTTTCAAGTGTCTGCCTTATCAACT
TTCGATGGTAGTTTATGTGACTACCATGGTTATAACGGGTAACGGAGAATAAGGGTTCGACTCCGGAGAGGGAGCCTGAGAAACGGCTA
CCACA
>TF1_SSU703579
ATGCATGTCTCAGTATGAGCTGAAAAAAGTGAAACCGCGAATAGCTCATTACAACAGCCATTGTTCACTGGATCTTTGTATCCTACGTG
GATAACTGTGGCAATTCTAGAGCTAATACATGCAAAAAAGCTCCGACCACTCGTTGGGAGGAGCGCATTTATTAGAACAAAACCAATCG
GGCTTCGGCCTGTCTTTGGTGACTCTGAATAACTCAGTTGATCGCACGGTCTCGTACCGGCGACTCATCTTTCAAGTGTCTGCCTTATC
AACTGTTGATGGTAGTTTACGTGACTACCATGGTTGCAACGGGTAACGGAGAATAAGGGTTCGACTCCGGAGAGGGAGCCTGAGAAACA
GCTACCACA
>TF1_SSU710679
ATGCATGTCTAAGCATAGGCCGATTAATGGTGAAGCCGCGAATAGCTCATTACAACAGCCATAGTTTATTGGATCTTCTCTCATACTTG
GATACCTGTGGTAATTCTAGAGCTAATACACGCAAGAAAACCCTGACTTCGGAAGGGGTGCATTTATTAGAACAAAACCAATCGGACTT
GTCCGTAGTTTGGTGAATCTGAATAACTCAGCTGATCGCACAGTCCTCGCACTGGCGACGTATCTTTCAAGTGTCTGCCTTATCAACTT
TCGATGGTAGTTTATATGACTACCATGGTTGTAACGGGTAACGGAGAATAAGGGTTCGACTCCGGAGAGGGAGCCTGAGAAACGGCTAC
CACA
>TF1_SSU734804
ATGCATGTCTGAGCACAAGCTCAAGAAAAGTGAAGCCGCGAATAGCTCATTACAACAGCCACTTTTCACTTGATCTTGATATCCTACTT
GGATAACTGTGGCAATTCTAGAGCTAATACATGCATTGAAGCTCTGACCAGCTTGCTGGGAAGAGCGCATTTATTAGAACAAAACCAAT
TGGACTTCGGTTCGTAATTGGTGACTCTGAATAACTCAGATGATCGCACGGTCTTGTACCGGTGACAGATCATTCAAGTGTCTGCCTTA
TCAACTGTTGATGGTAGTTTATATGACTACCATGGTTGCAACGGGTAACGGAGAATAAGGGTTCGTCTCCGGAGAGGGAGCCTGAGAAA
CGGCTACCACA
>TF3_SSU956521
ATGCGTGTCTAGGTACAAGCCTAAAAACGGTAAAGCCGCGAATGGCTCATTACAACAGCCATAGTTTATTGGATCTTGACTATCTTACT
TGGATAACTGTAGTAATTCTAGAGCTAATACACGCACCAAAGCCCAGACCTTACGGAACGGGCGCATTTATTAGACCAAAACCAATCGG
GCTTCGGCCCGTCTTTGGTGACTCTGAATAACTACGCTGAGTGCACGGTCCTCGAACCGGCACCGTATCTTTCAAGTGTCTGCCTTATC
AACTGTCGATGGTAGTTTACGTGACTACCATGGTTGTAATGGGTAACGGAGAATAAGGGTTCGACTCCGGAGAGGGAGCCTGAGAAATG
GCTACCACA
>TF3_SSU960449
ATGCATGTGTAAGTACAAGCTTTTAGAACGGTGAAACCGCGAATGGCTCATTAGATCAGTTAATATTTATTAGATCGTAGAAAGTTACT
TGGATAACTGTGGTAATTCTAGAGCTAATACACGCAAAAGAACTTGGAACGTAGGTTCCGGGTGCATTTATTAGTACAAGACCATCAGG
GCTCGTCCCTTCCAATGGTGAATCTGAATAACTGAGCCGATCGCATGGTCTCGCACCGGCGACGTATCTTTCAAGTGTCTGCCCTATCA
ACTTTCGATGGTAGTTTATATGACTACCATGGTTATAACGGGTAACGGAGAATAAGGGTTCGACTCCGGAGAAGCAGCCTGAGAAACGG
CTACTACA
>TF3_SSU966338
ATGCATGTCTAAGCACAAACGAAATTAACGTGAAGCCGCGAAAAGCTCATTACAACAGCCGTCGTTTCTTGGATCTCCGAACTTACTTG
GATAACTGTGGTAATTCTAGAGCTAATACATGCAATCGAGTCCTGAGCGTAAGCGATGGGCGCATTTATTAGTAACAAGACCAATCGGT
GCTTGCACCGTGGTTTGGTGAATCTGAATAACTGAGCAGATCGCTTCGGTCTTTGTACCGGCGACGTATCTTTCAAGTGTCTGTTTTAT
CAACTTTCGATGTTAGTTTATGTGACTAACATGGTTGTCACGGATAACGGAGAATAAGGGTTCGACTCCGGAGAGGGAGCATGAGAAAC
GGCTACCACA
>TF4_SSU144249
GTGCATGTCTAAGCATGAGCCATCAAATGGTGAAGCCGCGAACAGCTCATTACAACAGCCATAGTTCATTGGACTTCTCTCAATACTTG
GATATCTGTAATAATTTTAGAGCTAATACACGCAAGCAAACTCCAATCTCACGAGCGGAGTGCATTTATTAGAACAAAACCAATCGGGC
TTGCCCGTGCGTTTGGTGAATCTGAATAACTCAGCTGATCGCACAGTCTAGCACTGGCGACATATCTTTCAAGTGTCTGCCTTATCAAC
TGTCGATGGTAGTTTATATGACTACCATGGTTGTAACGGGTAACGGAGAATCAGGGTTTGACTCCGGAGAGGGAGCCTGAGAAACGGCT
ACCACA

>TF4_SSU150234
ATGCATGTCTAAGCATGAGCCCTATAATGGTGAAGCCGCGAATGGCTCATTACAACAGCCGTTGTTTCTTGGATCTTGATTCACTTGGA
TAACTGTGGTAATTCTAGAGCTAATACATGCAACTCAGCTCCAACTGAGAAGAAGGAGCGCATTTATTTGACCAAAACTGACTAGGTTT
CGACCTAAAACTTGGTGACTCTGAATAACTCTGCTGATCACACAGTCTTGCACTGGTGACATATCTTTCAAGTGTCTGCCCTATCAACT
TTCGATGGTAGTTTATATGCCTACCGTGGTTGTAACGGGTAACGGAGAATAAGGGTTCGACTCCGGAGAGGGAGCCTGAGAAATGGCTA
CCACA
>TF5_SSU410031
ATGCATGTCTAAGCAGAAGCCGCAATACGGTGAAACCGCGAATAGCTCATTACAACAGCCTTAGTTTCTTGGATCTTCAACAGTTACTT
GGATAACTGTGGTAATTCTAGAGCTAATACACGCATGAAAATCTCGGCTTTGCGGTTGGGATGCATTTATTAGTACAAAACCAATCGAG
CGCGAGCTCGTTGTTGGTGAATCTGAATAACTTTGCTGATCGCACGCTCTTCGTAGTGGCGACGCATCTTTCAAGTGTCTGGCTTATCA
ACTTTCGATGGTAGTTTATACGACTACCATGGTTGTAACGGGTAACGGAGAATAAGGGTTCGACTCCGGAGAGGGAGCCTGAGAAACGG
CTACCACA
>TF5_SSU419519
ATGCATGTCTAAGCACAAGCCGAAAATGGTAAAGCCGCGAATGGCTCATTACAACAGCCTTTGTTTATTTGATCTTGAAATCCTACTTG
GATAACTGTGGTAATTCTAGAGCTAATACACGCATCTAAACTCGAGACGCCCGTCACGAGTGCATTTATTAGTACAAAACCAATCGGGC
TTGCCCGTTCTTTGGTGAATCTGAATAACTGAGCCGATCGCACGGTCATTGTACCGGCGACGTATCTTTCAAGTGTCTGCCCTATCAAC
TTTCGATGGTAGTTTATATGACTACCATGGTTATAACGGGTAACGGAGAATAAGGGTTCGACTCCGGAGAAGCAGCCTGAGAAACGGCT
ACCACA
>TF5_SSU430294
ATGCATGTCTAAGCATAAACTAAACTAAAGTGAAGCCGCGAATAGCTCATTACAACAGCCGTTGTTTCTTGGATCTCCGTATTACTTGG
ATAACTGTGGTAATTCTAGAGCTAATACACGCAATCGAGCTCCGACCTTACGGGACGAGCGCATTTATTAGATCAAAACCAATCAGTGC
TTGCACTGTAGTCTGGTGAATCTGAATAACTGAGCAGATCGCTTCGGTCTTGGTACCGGCGACATATCCTTCAAGTGTCTGCCTTATCA
ACTTTCGATGGTAGTTTATGTGCCTACCATGGTTGTAACGGGTAACGGAGAATTAGGGTTCGACTCCGGAGAGGGAGCCTGAGAAACGG
CTACCACA
>TF5_SSU437076
ATGCATGTGTAAGTACAAGCTTTTAGAACTGTGAAACCGCGAATGGCTCATTAGATCAGTTAATATTTATTAGATCGTAGAAAGTTACT
TGGATAACTGTGGTAATTCTAGAGCTAATACACGCCTTGAAGCTCTGACCTTCGGGGACGAGCGCATTTATTAGAACAAAACCAATGGG
GTTTGCCCCTCGGTTGGTGACTCTGAATAACTACTCCGATCGCACGGTCTCGCACCGGCGACGCATCTTTCAAGTGTCTGCCTTATCAA
CTGTCGATGGTAGTTTATGTGCCTACCATGGTTGTAACGGGTAACGGAGAATAAGGGTTCGACTCCGGAGAGGGAGCCCGAGAAACGGC
TACCACA
>TF5_SSU444034
ATGCATGTCTAAGCACAAGCCGCTTGATGGTAAAGCCGCGAATGGCTCATTACAACAGCTATTGTTTATTAGATCTTACCATCCTACTT
GGATAACTGTTGTAATTCTAGAGCTAATACACGCATCAAAAACGGGACCTAAGGGAACCGTTGCATTTATTAGAACAAAACCAATCGGG
CTTCGGCCCGTCATTTGGTGAATCTGAATAACTCTGCCGATCGCACGGTCCACGAACCGGCGACGCATCTTTCAAATGTCTGCCTTATC
AACTTTCGATGGTAGTTTATGCGCCTACCATGGTTGTAACGGGTAACGGAGAATCAGGGTTTGACTCCGGAGAGGGAGCCTGAGAAACG
GCTACCACA
>TF5_SSU446087
ATGCATGTCTAAGCAGAAGCCGAACAATGGCAAAGCCGCGAATGGCTCATTACAACAGCTGTTGTTTATTTGATCTTGAATTCCTACAT
GGATAACTGTGGTAATTCTAGAGCTAATACATGCTTACAAACTCATTTCCTTGGATCTGAGTGCATTTATTAGAACAAAACCAATCGGG
CTTGCCCGTTCATTGGTGAATCTGAATAACTATGCCGATCGCACGGTCTTCGCACCGGCGACGTATCTTTCAAGTGTCTGCCCTATCAA
CTTTCGATGGTAGTTTATGTGACTACCATGGTTATAACGGGTAACGGAGAATAAGGGTTCGACTCCGGAGAAGCAGCCTGAGAAATGGC
TACTACA
>TF5_SSU453472
ATGCATGTCTCAGCACAAGCCAATATATGGTAAAGCCGCGAATGGCTCATTACAACAGCCACTGTTTATTAGATCATCCTTCTTACTTG
GATAACTGTGGAAAAGCTAGAGCTAATACATGCTACAAGCTCTGACCTTACGGAAGGAGCGCATTTATTAGAACAAAAACAATCGGACT
TTGTCCGTCATTTGGTGAATCTGAATAACTATGCAGATCGCACGGTCTTAGAACCGGCGACATATCTTTCAAATGTCTGCCCTATCAAC
TTTCGACGGTATGTGATATGCTTACCGTGGTTGCAACGGGTAGCGGGGAATCAGGGTTCGATTCCGGAGAGGGAGCATGAGAAACGGCT
ACCACA
>TF5_SSU457543
ATGCATGTGTAAGCATAAGCCGATTAAATGGTGAAGCCGCGAATGGCTCATTACAACAGCCATAGTTTATTATATTTTTTCTTTTACTT
GGATAACTGAGGTAATTCTTGAGCTAATACACGCATCAAAGCTTCGACCTTACGGAAGGAGCGCATTTATTAGAACAAAACCAATCGGA
CTTCGGTCCGTTACTTGGTGACTCAGAATAACTCTGTGGATCGCATGGTCTTAGCACCGGCGACGTATCTTTCAAGTGTCTGCCTTATC
AACTTTCGTTGGTAATTTATGTGATTACCAAGGTTGTAACGGGTAACGGAGAATCAGGGTTTGATTCCGGAGAGGGAGCCTGAGAAATG
GCTACCACA
>TF5_SSU459305
ATGCATGTCTTAGTACAGACTATCTCACAGTGAAACTGCGAATGGCTCATTAAATCAGCTAAGGTTCCTTAGATCGTACAATCCTACAT
GGATACCTGTGGTAATTCTACAGCTAATACACGCATCAAAACCCAACCTTACGGTGGGGTGCGTTTGTTACTTCAAACCAATCGGACTT
CGGTCTGAAATCAAGTGATATTGAACAATTTAGCTGATCGCACGGTCTGAGAACCGGCGACATATCCTTCAAATGTCTGCCTTATCAAC
TTTCGATGGTAGATTACGCGCCTACCATGGTTGTAACGGGTAACGGAGAATCAGGGTTTGATTCCGGAGAGGGAGCCTGAGAAACGGCT
ACCACA
>TF5_SSU466315
ATGCATGTCTAAGCATAAACTAATCTAAAGTGAAGCCGCGAATAGCTCATTACAACAGCCGTTGTTTCTTGGATCTCCGTCTTACTTGG
ATAACTGTGGTAATTCTAGAGCTAATACATGCAGCTGAGATCTGACTTTACAGGAAGATCGCATTTATTAGATCAAAACCAATCGGCTT
CGGCCGTTGCTGGTGAATCTGAATAACTACGCAGATCGCTTAGGTTTTATACCGGCGACGTGTCCTTCAAGTGTCTGCCTTATCAACTT
TCGATGGTAGTTTCTACGCCTACCATGGTTGCGACGGGTAACGGAGAATCAGGGTTCGATTCCGGAGAGGGAGCCTGAGAAACGGCTAC
CACA
>TF6_SSU33463
ATGCATGTGTAAGCATGAACCATTTAATGGTGAAGCCGCGAATGGCTCATTACAACAGCCATAGTTTATTAGATTTCTTTTTTACTTGG
ATAACTGTGGTAATTCTAGAGCTAATACACGCAGCAAAGCTTCGACCTTACGGAAGGAGCGCATTTATTAGAACAAGACCAATCGTACT
TCGGTACGTATTTTGGTGAATCTGAATAACTTAGTCGATCTCATGGTCTTAGCACCGGAGACGCATCTTTCAAGTGTCTGCCTTATCAA
CTTTCGATGGTAGTTTATGCGCCTACCATGGTTGTAACGGGTAACGGAGAATCAGGGTTTGATTCCGGAGAGGGAGCCTGAGAAATGGC

```
TACCACA
>TF6_SSU33935
ATGCATGTCTAAGCAGAAGCCGAACAATGGCAAAGCCGCGAATAGCTCATTACAACAGCCATTGTTTACTTGATCTTGAAATCCTACTT
GGATAACTGTGGTAATTCTAGAGCTAATACACGCAATAAGCTCCGACCTCAGGGGAGGAGTGCATTTATTAGAACAAAACCAATCAGAC
CTCGGTCTGTCTCTTGGTGAATCAGAATAACTCAGCTGATTGCACAGTCTTGTACTGGCGACGTATCTTTCAAGTGTCTGCCTTATCAA
CTGTTGATGGTAGTTTATGCGACTACCATGGTTGTAACGGGTAACGGAGAATAAGGGTTCGACTCCGGAGAGGGAGCCTGAGAAACGGC
TACCACA
>TF6_SSU36442
ATGCATGTCTAAGCACAAGCCGAATATGGTGAAGCCGCGAATGGCTCATTACAACAGCCTTTGTTTATTTGATCTTGAAATCCTACTTG
GATAACTGTGGTAATTCTAGAGCTAATACACGCAACAAAACTTTGAACGTAAGTTCTTGGTGCATTTATTAGTACAAAACCTTCCGGAC
TTCGGTTCGTAAACTGGTGAATCTGAATAAATTAGCCGATCGCATGGCCTTCGCGCTGGCGACGTATCTTTCAAGTGTCTGCCCTATCA
ACTTTCGATGGTAGTTTATATGACTACCATGGTTATAACGGGTAACGGAGAATAAGGGTTCGACTCCGGAGAAGCAGCCTGAGAAACGG
CTACTACA
>TF6_SSU37421
ATGCATGTGTAAGTACAAGCTTTTAGAACGGTGAAACCGCGAATGGCTCATTAGATCAGTTAATATTTATTAGATCGTAGAAAGTTACT
TGGATAACTGTGGTAATTCTAGAGCTAATACATGCAACCAAGCTCCAACCTTTCGGAAGGAGCGCATTTATTATACCAAGACCAATCGG
GCTTTGCCCGCTATCTGGTGACTCTGAATAACTTTGCTGATCACATGGTCATAGTACCGGTGACATATCTTTCAAGTGTCTGCCCTATC
AACTTTCGATGGTAGTTTATGTGCCTACCATGGTTGTAACGGGTAACGGAGAATAAGGGTTCGACTCCGGAGAGGGAGCCTGAGAAATG
GCTACCACA
>TF6_SSU41803
ATGCATGTCTAAGTACAAGCCTCATTAAGGTGAAACCGCGAATGGCTCATTAAATCACACCTAATATACTGGATAGTATACAGTTACTT
GGATAACTGCGGTAATTCTGGAGCTAATACATGCAGCTGAGATCTGACTTTACAGGAAGATCGCATTTATTAGATCAAAACCAATCGGC
TTCGGCCGTTGCTGGTGAATCTGAATAACTACGCAGATCGCTTAGGTTTTATACCGGCGACGTGTCCTTCAAGTGTCTGCCTTATCAAC
TTTCGATGGTAGTTTCTACGCCTACCATGGTTGCGACGGGTAACGGAGAATTAGGGTTCGACTCCGGAGAGGGAGCCTGAGAAACGGCT
ACAACA
>TF6_SSU47996
ATGCATGTCTAAGCACAAACTGATTAATTGTGAAGCCGCGAATGGCTCATTACAACAGCCATTGTTTACTGGATATCTCATTACTACAT
GGATAACTGTGGTAATTCTACAGCTAATACACGCATCAAAACCCCAACTTTCGAAGGGGTGCGTTTGTTACTTCAAATCAATCGGACTT
CGGTCTGGTTTCAACTGAGATTGAACAATTTAGCTGATCGCACGGTCTAAGAACCGGCGACATATCCTTCAAACGTCTGCCTTATCAAC
TTTCGATGGTAGATTATGCGCCTACCATGGTTGTTACGGGTAACGGAGAATCAGGGTTTGATTCCGGAGAGGGAGCCTGAGAAACGGCT
ACCACA
>TF6_SSU48167
ATGCATGTCTAAGCAGAAGCCGCACAATGGTAAAGCCGCGAATGGCTCATTACAACAGCCGTCGTTTCTTGGATCTCTTGTTTTACTTG
GATAACTGTGGTAATTCTAGAGCTAATACACGCACTAAAGCTCTGACCTTACGGGACGAGCGCATTTATTAGAACAAAACCAATCGGGT
TTCGGCCCGTCCGTTGGTGACTCTGAATAACTACGCCGATCGCACGGTCTCGTACCGGCGACGTATCTTTCAAGTGTCTGCCTTATCAA
CTTTCGATGGTAGTTTATGTGCCTACCATGGTTGTAACGGGTAACGGAGAATAAGGGTTCGACTCCGGAGAGGGAGCCTGAGAAACGGC
TACCACA
>TF6_SSU53456
ATGCATGTCTAAGTATGAACTATCTATTGTGAAACCGCGAATGGCTCATTACAACAGCCATAGTTTACTGGATATATTCCTTTACATGG
ATAACTGTGGTAATTCTACAGCTAATACACGCATCAAAGCCCCGACTTTATGAAGGGGCGCGTTAGTTACTTCAAACCGATCGGTCTTC
GGACTGTATCCAAGTGAGATTGAACTATTTAGCTGAGCGCACGGTCTAAGCACCGGCGCCATATCCTTCAAATGTCTGCCTTATCAACT
TTCGATGGTAGATTATGCGCCTACCATGGTTGTAACGGGTAACGGAGAATCAGGGTTTGATTCCGGAGAGGGGGCCTGAGAAATGGCCA
CCACA
>TF6_SSU54250
ATGCACGTTTCAATACAAGCCTTACTAAGGTGAAATCGCGAATGGCTCATTACAACAGCCATTGTTTCTTGGATCTTATCTTTTACTTG
GATAACTGTGGTAATTCTAGAGCTAATACGTGCCACCAATCCCGACGCAAGAAGGGATGCATTTATTAGAACTAAACCGACCGGGCTCA
GCCCGCTGTTTGGTGAATCTGAATAACTCTGCAGATCGCATGGTCTCGCACCGGCGACATATCCTTCAAGTGTCTGCCTTATCAACTTT
CGATGGTAGTTTATGTGACTACCATGGTTGTCACGGGTAACGGAGAATTAGGGTTCGACTCCGGAGAGGGAGCCTGAGAAACGGCTACC
ACA
>TF6_SSU58877
ATGCATGTCTAAGCACAAGCCGAAAATGGTAAAGCCGCGAATGGCTCATTACAACAGCCTTTGTTTATTTGATCTTGAAATCCTACTTG
GATAACTGTGGTAATTCTAGAGCTAATACAAGCGATTAAACTCCAACCTTTTGGAAGGAGTGCATTTATTAGTACAAAACCAATCGGGG
TAAAACCCGTGGTTTGGTTAATCTGAATAACTCTGTCGATCGCACGGTCTTTGTACCGGCGACATATCTTTCAAGTGTCTGCCCTATCA
ACTGTCGATGGTAGTTTATATGACTACCATGGTTGTAACGGGTAACGGAGAATAAGGGTTCGTCTCCGGAGAAGCAGCCTGAGAAACGG
CTACTACA
>TF6_SSU74955
ATGCATGTCTAAGCATGAGCCATCAAATGGTGAAGCCGCGAACAGCTCATTACAACAGCCATAGTTCATTGGACTTCTCTCAATACTTG
GATATCTGTAATAATTTTAGAGCTAATATACGCAAGCAAACTCCAATCTCACGAGCGGAGTGCATTTATTAGAACAAAACCAATCGGGC
TTGCCTGTGCGTTTGGTGAATCTGAATAACTCAGCTGATCGCACAGTCTAGCACTGGCGACGTATCTTTCAAATGTCTGCCCTATCAAC
TTTCGACGGTATGTGATATGCTTACCGTGGTTGCAACGGGTAGCGGGGAATCAGGGTTCGATTCCGGAGAGGGAGCATGAGAAACGGCT
ACCACA
>TF6_SSU82210
ATGCATGTCTATGCATAAGCCAATTTATGGTGAAGCCGCGAATGGCTCATTACAACAGCTTTCATTTCTTGGGTGTCATTTTACTTGGA
TAACTTTGTCAACCCAGAGCTAATACATGCACAAAAGCCTTGACTTATGGAAAGGCGCAGTTATTTGATCAAAACCAATCAGGCTTGCC
TGAATTTGATGACTCTGAGTAAACTTGCAGATCGCACAGTCCTAGTACTGGCGACAAACTCTTCAAGTGTCTGCCTTATCAACTTTCGA
TGGTAGTTTCAGTGCCTACCATGGTTACAATGGGTAACGGAGAATAAGGGTTCGACTCCGGAGAGGGAGCCTGAGAAACGGCTACCACA
>TF6_SSU84268
ATGCATGTCTCAGTGCAAGTCCATTCAGGACGAAACCGCGAACGGCTCATTACAACAGCTATAATCTACGGGGAGTTTCCATACATGGA
TAACTCTGTCAACCCAGAGCTAATACATGCACAAAAGCCTCACCTCATGGCTGGGCGCATCTATTATACCAAAACCAACCGGGCTCTGC
CCGAGTCTTGGTGACTCTGAATAGAGAGTTAATGACGCAGTCTTAGTACTGGTCATGTTCCCCACGAGTGTCTGCCTTATCAACTTTCG
ATGGTAGTTTATGTGCCTACCATGGTTGCAACGGGTAACGGAGAATAAGGGTTCGACTCCGGAGAGGGAGCCCGAGAAACGGCTACCAC
```

A
>TF6_SSU98667
ATGCATGTTCCAGCAGAAACTGAATATAGTGAAGCCGCGAATAGCTCATTACAACAGCCATTGTTCATTGGAGTTGAAGACAATTTGGA
TAACCCTGTTAAATCAGAGCTAATACACGCAAAAAAGCTTGTCCTTTGCGGTTTAAGTGCATTTATTAGAACAAAATCATCCGGCCTTT
GGCCGTATTTGGTGAATCTGAATAACTTTGCTGATCGCACGGTCTAGTACTGGCGACGTATCTTTCAAGTGTCTGCCTTATCAACTTTC
GACGGTAGTTTATGTGCCTACCGTGGTTATAACGGGTAACGGAGAATTAGGGTTCGACTCCGGAGAGGGAGCCTGAGAAATGGCTACCA
CA
>TS1_SSU270885
ATGCATGTCTAAGCACAAGCTGAAAATGGTGAAGCTGCGAATGGCTCATTACAACAGCCTTTGTTTATTTGATCTTGAAATCCTACTTG
GATAACTGTGGTAATTCTAGAGCTAATACATGCTATCAAACTGAAGGCTCTGCCTTCAGTGCATTTATTAGTACAAAACCAATCAGGTT
TTACCTGCCAATTGGTGAATCAGAATAACTGTGCAGATCACATAGCCTATGAGCTGGTGACATATCTTTCAAGTGTCTGCCCTATCAAC
TTTCGATGGTAGTTTATATGACTACCATGGTTATAACGGGTAACGGAGAATAAGGGTTCGACTCCGGAGAAGCAGCCTGAGAAATGGCT
ACTACA
>TS1_SSU284163
ATGCATGTCTAAGCACAAGCCGAATATGGTGAAGCCGCGAATGGCTCATTACAACAGCCTTTGTTTATTTGATCTTGAAATCCTACTTG
GATAACTGTGGAAATTCTAGAGCTAATACACGCAAAAGAACTCGGAACGTAGGTTCCGGGTGCATTTATTAGTACAAGACCATCAGGGC
TCGTCCCTTCCAATGGTGAATCTGAATAACTGAGCCGATCGCATGGTCTCGCACCGGCGACGTATCTTTCAAATGTCTGCCCTATCAAC
TTTCGACGGTATGTGATATGCTTACCGTGGTTGCAACGGGTAGCGGGGAATAAGGGTTCGATTCCGGAGAGGGAGCATGAGAAACGGCT
ACCACA
>TS2_SSU821962
ATGCAAGTCTAAGCACAAGCCGTACAATGGTAAAGCCGCGAATAGCTCATTACAACAGCCATAGTTTATTAGATAGTTCTTTACTACAT
GGATAACTGTGGTAATTCTAGAGCTAATACATGCTACAAACGGCGACCTTTAGTTAGGAAGCCGCGCTTTTATTAGAACAAAACCAATC
GCTCTTTCGGGAGCGTCAATTGGTGAGTCTAAATAACATAGCAGATCGCACGGTCTGGGCACCGGCGACGAATCTTTCAAATGTCTGCC
TTATCAACTTTCGATGGTAGTTTACATGCCTACCATGGTGATAACGGGTAACAGAGAATTAGGGTTTGACTCTGGAGAGGCAGCCTGAG
AGACGGCTACCACA
>TS2_SSU823349
ATGCATGTCTAAGTACAGACTATATCACAGTGAAACTGCGAATGGCTCATTAAATCAGCTAAGGTTCCTTAGATCGTACAATCCTACAT
GGATAACTGTGGTAATTCTAGAGCTAATACATGCTACAAACGGCGACCTTTAGTTAGGAAGCCGCGCTTTTATTAGAACAAAACCAATC
GCTCTTTCGGGAGCGTCAATTGGTGAGTCTAAATAACATAGCAGATCGCACGGTCTTGGCACCGGCGACGAATCTTTCAAATGTCTGCC
TTATCAACTTTCGATGGTAGTTTACATGCCTACCATGGTGATAACGGGTAACAGAGAATTAGGGTTTGACTCTGGAGAGGCAGCCTGAG
AGACGGCTACCACA
>TS3_SSU475561
ATGCATGTCTAAGCACAAACTGAAATAAAGTGAAGCCGCGAATAGCTCATTACAACAGCCATTGTTTACTTGATCTTGAAATCCTACTT
GGATAACTGTGGTAATTCTAGAGCTAATACACGCAATTAAGCTCTGATCCTTTTGGTGACGAGTGCATTTATTAGAACAAAACCAATCG
GACTTCGGTCTGTTGTTGGTGAATCTGAATAACTCAGCTGATCGCACGGTCTTGTACCGGTGACGCATCTTTCAAGTGTCTGCCCTATC
AACTTTCGACGGTATGTGATATGCTTACCGTGGTTGCAACGGGTAGCGGGGAATCAGGGTTCGATTCCGGAGAGGGAGCATGAGAAACG
GCTACCACA
>TS3_SSU489684
ATGCATGTCTAAGCACAAGCCGAATATGGTGAAGCCGCGAATGGCTCATTACAACAGCCTTGGTTTATTTGATCTTGAAATCCTACATG
GATAACTGTGGTAATTCTAGAGCTAATACACGCATACAAGCTTCAGCCTTACGGCTGTGAGCGCATTTATTAGTACAAAACCAATCGGG
CCTTGCCCGTTGTTTGGTGAATCTGAATAACTGAGCCGATCGCATGGTCTCTGTACCGGCGACGTATCATTCAAGTGTCTGCCCTATCA
ACTTTCGATGGTAGTTTATGTGACTACCATGGTTACAACGGGTAACGGAGAATAAGGGTTCGACTCCGGAGAAGCAGCCTGAGAAACGG
CTACTACA
>TS3_SSU503133
ATGCAAGTCTAAGCACAAGCCGTTAAATGGTAAAGCCGCGAATAGCTCATTACAACAGCCATAGTTTATTAGATAGTTCCTTACTACAT
GGATAACTGTGGTAATTCTAGAGCTAATACATGCTTAAAAGAACGACCTCGCAAGAGGTTGTTCTGCACTTATTAGAACAAAACCAATC
GCGCTGCGGCAACGTAGTGCGTTATTTGGTGAATCTGAATAACTTGGCGGATCGCACGGTCTAGTACCGGCGACGCATCTTTCAAATGT
CTGCCTTATCAACTGTCGATGGTAGTTTACATGCCTACCATGGTGATAACGGGTAACAGAGAATAAGGGTTTGACTCTGGAGAGGGAGC
CTGAGAGACGGCTACCACA
>TS3_SSU508400
ATGCAGGTCTGAGCACGAGCTCAAGAAAATAATGGATGGATAATTGGATAACTGTGGCAATTCTAGAGCTAATACATGCATTGAAGCTC
CGACCAGCTTGCTGGGAAGAGCGCATTTATTAGAACAAAACCAATCGGACTTCGGTTCGTTATTGGTGACTCTGAATAACTCAGTCGAT
CGCACGGTCTTGTACCGGCGACAGATCATTCAAGTGTCTGCCTTATCAACTGTTGATGGTAGTTTATGTGACTACCATGGTTGCAACGG
GTAACGGAGAATAAGGGTTCGTCTCCGGAGAGGGAGCCTGAGAAATAGCTACCACA
>TS4_SSU543236
ATGCATGTCTCAGCACATACCAATATATGGCAAAGCCGCGAATGGCTCATTACAACAGCCGCATTTTATTAGATAATCCTATTTACTTG
GATAACTGTGGAAAAACTAGAGCTAATACATGTCATAAGCTTCGACCTTACGGAAGAAGTTCATTTATTAGAACAAAACCAATCGGACT
TTGTCCGTTACTTTGTTGACTCTGATTAACTTCATGATCGCACGGTCATAGAACCGGCGACATATCTTTCAAATGTCTGCTTTATCAAC
TTTCGATGGTAGTTTATGCGCCTACCATGGTTGTAACGGGTAACGAAGAATCAGGGTTTGATTTCGGAGAGGGAGCCTGAGAAACGGCT
ACCACA
>TS4_SSU544032
ATGCATTTCTAAGCACAAGCCGAATATGGTGAAGCCGCGAATGGCTCATTACAACAGCCTTTGTTTATTTGATCTTGAAATCCTACTTG
GATAACTGTGGTAATTCTAGAGCTAATACACGCAAAAGAACTCGGAACGTAGGTTCCGGGTGCATTTATTAGTACAAGACCATCAGGGT
TCGTCCCTTCCAATGGTGAATCTGAATAACTGAGCCGATCGCATGGTCTCGCACCGGCGACGTATCTTTCAAGTGTCTGCCCTATCAAC
TTTCGATGGTAGTTTATATGACTACCATGGTTATAACGGGTAACGGAGAATAAGGGTTCGACTCCGGAGAAGCAGCCTGAGAAACGGCT
ACTACA
>TS5_SSU874117
ATGCATGTATCAGCACAAGCCGTAATATGGTGAAGCCGCGAATAGCTCATTATAACAGTCGTAGTTTATTAGAAAGTCTGTACTGGATA
ACTGTGGTAATTCCAGAGCTAATACATGTTCCAAGCCCCAACTAACGAAGGGGTGCATTTATTAGAACAAGGCCGATCAGACTTTGTCT
GTCTCAGGTTGACTCTGAATAACTTTGCTAATCGCACAGTCTTTGCACTGGCGATGTATCTTTCAAATGTCTGCCCTATCAACTTTCGA
CGGTATGTGATATGCTTACCGTGGTTGCAACGGGTAGCGGGGAATCAGGGTTCGATTCCGGAGAGGGAGCATGAGAAACGGCTACCACA

>TS5_SSU875407
ATGCATGTCTAAGTACAGACTATATCACAGTGAAACTGCGAATGGCTCATTAAATCAGCTAAGGTTCCTTAGATCGTACAATCCTACTT
GGATAACTGTGGTAATTCTAGAGCTAATACACGCCTTGAAGCTCTGACCTTCGGGGACGAGCGCATTTATTAGAACAAAACCAATGGGG
TTCGCCCCTCGGTTGGTGACTCTGAATAACTACTCCGATCGCACGGTCTCGCACCGGCGACGCATCTTTCAAGTGTCTGCCTTATCAAC
TGTCGATGGTAGTTTATGTGCCTACCATGGTTGTAACGGGTAACGGAGAATAAGGGTTCGACTCCGGAGAGGGAGCCCGAGAAACGGCT
ACCACA
>TS5_SSU881546
ATGCATGTCTAAGCAGAAACTATTTTAAAGTGAAGCCGCGAAAAGCTCATTACAACAGCCGTCGTTTCTTGGGTCTCCGAATTTACTTG
GATAACTGTGGTAATTCTAGAGCTAATACATGCAATCGAGCTCTGAACGTAAGTGATGGGCGCATTTATTAGTAACAAAACCAATCGGT
TGCTTGCAACCGTGGTTTGGTGAATCTGAATAACTTAGCAGATCGCTTCGGTCTTTGTACCGGCGACATATCTTTCAAATGTCTGCCCT
ATCAACTTTCGATGGTACGTTATGCGCCTACCATGGTCGTAACGGGTAACGGAGAATCAGGGTTCGATTCCGGAGAGGGAGCCTGAGAA
ATGGCTACCACA
>TS5_SSU900338
ATGCATGTCTATGCACAAGCCGATAAATGGCAAAGCCGCGAATGGCTCATTACAACAGCCACTGTTCACTTGATCTGTATCATATCCTA
CTTGGATAACTGTGGTAATTCTAGAGCTAATACACGCACCCATTCTCCGACCGCAAGGGACGAGAGCATTTATTAGAACAAAACCAATC
GGCTTCGGTCGTTCGTTTGTGACTCTGAATAACTTTGCTGATCGTACGGTCTTTGTACCGACGACGCATCTTTCAAGTGTCTGCCTTAT
CAACTTTCGATGGTAAGTTCCTTGCTTACCATGGTTGTAACGGGTAACGGAGAATCAGGGTTCGACTCCGGAGAGGGAGCCTGAGAAAC
GGCTACCACA
>TS5_SSU901243
ATGCATGTCTATGCACACGCCGATTAATGGTAAAGCCGCGAATAGCTCATTACAACAGCCTCTGTTTATTAGATCTTTTTATCCTACTT
GGATAACTGTGGCAATTCTAGAGCTAATACACGCACCAAAACTCCGACCTTGCGGAAGGAGTGCATTTATTAGACCAAAACCAATGCAG
GCTTGTCCTGTTACTTTGGTGAATCTGAATAACCTTTGCCGAGTGAATGGTCTTTGAACCGTCACCATATCTTTCAAATGTCTGCCTTA
TCAACTTTCGACGGTAGTTTATATGACAACCGTGGTTGTAACGGGTAACGGGGAATCAGGGTTCGATTCCGGAGAGGGAGCATGAGAAA
CGGCTACCACA
>TS6_SSU559765
ATGCATGTCTTAGTACAGACTATCTCACAGTGAAACTGCGAATGGCTCATTAAATCAGCTAAGGTTCCTTAGATCGTACAATCCTACTT
GGATAACTGTAGTAATTCTAGAGCTAATACATGCTTAAAAGAACGACCTCGCAAGAGGTTGTTCTGCACTTATTAGAACAAAACCAATC
GCGCTGCGGCAACGTAGTGCGTTATTTGGTGAATCTGAATAACTTGGCGGATCGCACGGTCTAGTACCGGCGACGCATCTTCCAAATGT
CTGCCTTATCAACTGTCGATGGTAGTTTACATGCCTACCATGGTGATAACGGGTAACAGAGAATAAGGGTTTGACTCTGGAGAGGGAGC
CTGAGAGACGGCTACCACA
>TS6_SSU570763
ATGCATGTCTAAGTACAGACTATATCACAGTGAAACTGCGAATGGCTCATTAAATCAGCTAAGGTTCCTTAGATCGTACAATCCTACTT
GGATAACTGTAGTAATTCTAGAGCTAATACACGCATCAAGCTCTGACCTCATGGAATGAGCGCATTTATTAGAACAAAAACCAATCAGG
CTATGCCTGTTTTTGGTGGATCTGAATAACTCAGCTGACCGTATGCTCTCGTAGCGACGGCGATTCCTCCAAGTATCTGCCTTATCAAC
TGTTGATGGTAGTTTATGTGACTACCATGGTTGTAACGGGTAACGGAGAATAAGGGTTCGACTCCGGAGAGGGAGCCTGAGAAACGGCT
ACCACA
>TS6_SSU587229
ATGCATGTCTAAGCATAAATGAATTCATAGTGAAGCCGCGAATAGCTCATTACAACAGCCATCGTTTAATGGATATATTTTTACATGGA
TAACTGTGGTAATTCTACAGCTAATACACGCATCAAAACCCAACCTTACGGTGGGGTGCGTTTGTTACTTCAAACCAATCGGACTTCGG
TCTGAAATCAAGTGATATTGAACAATTTAGCTGATCGCACGGTCTGAGAACCGGCGACATATCCTTCAAATGTCTGCCTATCAACTTTC
GATGGTAGATTACGCGCCTACCATGGTTGTAACGGGTAACGGAGAATCAGGGTTTGATTCCGGAGAGGGAGCCTGAGAAACGGCTACCA
CA
>HE6_SSU372021
ATGCATGTCTAAGCACAAACTTTATGAGTGAAGCCGCGAAAAGCTCATTACAACAGCCGTCGTTTCTTGGATCTCCATTCTCTACTTGG
ATAACTGTGGTAATTCTAGAGCTAATACATGCGATCAAGCTCCGAACTCACGTGACGAGTGCATTTATTAGAACAAGACCATCCGGCTT
CGGCCGTTCTTTGGTGACTCTGAATAACTACGCGAATCACATGGTCCTCGCACCGGTGATGTATCTTTCAAGTGTCTGCCTTATCAACT
TCCGATGGTAGTTTATGTGCCTACCATGGTTGTAACGGGTAACGGGGAATTAGGGTTCGATTCCGGAGAGGGAGCATGAGAAACGGCTA
CCACA